# SCALING BEHAVIOR OF CIRCULAR COLLIDERS DOMINATED BY SYNCHROTRON RADIATION

Richard Talman
Laboratory for Elementary-Particle Physics Cornell University

*Abstract*

The scaling formulas in this paper—many of which involve approximation—apply primarily to electron colliders like CEPC or FCC-ee. The more abstract "radiation dominated" phrase in the title is intended to encourage use of the formulas—though admittedly less precisely—to proton colliders like SPPC, for which synchrotron radiation begins to dominate the design in spite of the large proton mass.

Optimizing a facility having an electron-positron Higgs factory, followed decades later by a p,p collider in the same tunnel, is a formidable task. The CepC design study constitutes an initial "constrained parameter" collider design. Here the constrained parameters include tunnel circumference, cell lengths, phase advance per cell, etc. This approach is valuable, if the constrained parameters are self-consistent and close to optimal. Jumping directly to detailed design makes it possible to develop reliable, objective cost estimates on a rapid time scale.

A scaling law formulation is intended to contribute to "ground-up" stage in the design of future circular colliders. In this more abstract approach, scaling formulas can be used to investigate ways in which the design can be better optimized. Equally important, by solving the lattice matching equations in closed form, as contrasted with running computer programs such as MAD, one can obtain better intuition concerning the fundamental parametric dependencies. The ground-up approach is made especially appropriate by the seemingly impossible task of simultaneous optimization of tunnel circumference for both electrons and protons. The fact that both colliders will be radiation dominated actually simplifies the simultaneous optimization task.

All GeV scale electron accelerators are "synchrotron radiation dominated", meaning that all beam distributions evolve within a fraction of a second to an equilibrium state in which "heating" due to radiation fluctuations is canceled by the "cooling" in RF cavities that restore the lost energy. To the contrary, until now, the large proton to electron mass ratio has caused synchrotron radiation to be negligible in proton accelerators. The LHC beam energy has still been low enough that synchrotron radiation has little effect on beam dynamics; but the thermodynamic penalty in cooling the superconducting magnets has still made it essential for the radiated power not to be dissipated at liquid helium temperatures. Achieving this has been a significant challenge. For the next generation p,p collider this will be even more true. Furthermore, the radiation will effect beam distributions on time scales measured in minutes, for example causing the beams to be flattened, wider than they are high [1] [2] [3]. In this regime scaling relations previously valid only for electrons will be applicable also to protons.

This paper concentrates primarily on establishing scaling laws that are fully accurate for a Higgs factory such as CepC. Dominating everything is the synchrotron radiation formula

$$\Delta E \propto \frac{E^4}{R}, \qquad (1)$$

relating energy loss per turn $\Delta E$, particle energy $E$ and bend radius $R$. [1] This is the main formula governing tunnel circumference for CepC because increasing $R$ decreases $\Delta E$.

The same formula will possibly dominate future proton colliders as well. But the strong dependence of cost on superconducting magnetic field causes the optimization of SPPC to be more complicated. Nevertheless scaling laws previously applicable only to electron rings, will apply also to SPPC. In particular, like CepC, as the SPPC helium cooling cost becomes fractionally more important, its proportionality to $1/R$ favors increased ring circumference.

With just one exception (deconstructing Yunhai Cai's intersection region (IR) design) this paper makes no use of any accelerator design code such as MAD. Apart from the fact that none is necessary, this is to promote the attitude that scaling from existing facilities (mainly LEP in this case) is more reliable than "accurate" numerical investigation.

This paper is intended to promote "ground up optimization" (as contrasted with "constrained parameter design") of future circular colliders, especially the Higgs factory. But not to perform an optimization. To the extent the investigation has been started, the main suggestion is that far longer cells, than have been used in existing studies, are favored, and should be investigated. From scaling relations, the optimal cell length is approximately 200 m, which is several times greater than some current designs. Scaling relations also suggest that an IP vertical beta function $\beta_y^* \approx 1$ mm, assumed by some, is unachievably small.

---

[1] Scaling formulas in this paper are indicated by a broad bar in the left margin. In some cases a constant of proportionality is included.

# CONTENTS

# 1 INTRODUCTION

## 1.1 Organization of the Paper

For a cicular e+e- colliding beam storage ring to serve as a "Higgs Factory" its single beam energy has to be significantly higher than the $E_\text{max} \approx 100\,\text{GeV}$ LEP energy. At such high energies, as Telnov [4] has pointed out, beamstrahlung limits the performance with a severity that increases rapidly with increasing $E_\text{max}$.

The basis for most of this paper, (the first formula in Appendix A, "Synchrotron Radiation Preliminaries") gives $U_1$, the energy loss per turn, per electron, as a function of ring radius $R$, and electron beam energy $E$;

$$U_1\ [\text{GeV}] = C_\gamma \frac{E^4}{R}, \quad (2)$$

where, for electrons, $C_\gamma = 0.8846 \times 10^{-4}\,\text{m/GeV}^3$. For protons $C_\gamma = 0.7783 \times 10^{-17}\,\text{m/GeV}^3$, For proton colliders preceeding LHC synchrotron radiation (SR) was always negligibly small owing to the large proton mass. For the LHC, SR influenced the design only through the efforts needed to avoid dissipating the radiated energy at liquid Helium temperature. The post-LHC future circular collider will be the first for which beam dynamics and ring optimization will be dominated by SR. This has always been true for electron colliders.

In this paper a simulation program is used to calculate the achievable luminosity for bend radii ranging up to four times the LEP/LHC ring radius, for single-beam energies from 100 GeV to 300 GeV. There are three phenomena giving luminosity limits: $\mathcal{L}^{RF}$, *RF-power limitation*; $\mathcal{L}^{bs}$, *beamstrahlung limitation*; and $\mathcal{L}^{bb}$, *beam-beam interaction limitation*, all of which have complicated dependencies on ring parameters. **Since the achievable luminosity is equal to the smallest of these limits, the optimal choice of parameters requires them all to be equal.** To be specially exploited is a scaling law to be obtained according to which the optimized luminosity is a function only of the product $RP^{RF}$, tunnel-radius multiplied by RF power.

Because a next-generation p,p collider is to follow the Higgs factory in the same tunnel, it is necessary to optimize the tunnel for both purposes. The proton energy for such a collider will be so large that, like all electron rings (in a first for protons) the performance will be dominated by synchrotron radiation and (for this reason) luminosity will depend primarily on a radius, power product $RP^\text{wall}$, much like the similar electron ring radius, power product. This differs from the electron case in that the power goes into refrigeration rather than into RF generation, but the impact on luminosity optimization will be similar.

Increasing either $R$ or $P$ is expensive but many accelerator costs are more sensitive to $P$ than to $R$. This can lead to the (possibly counter-intuitive) result that the Higgs luminosity can be increased, at modest increase in cost, by increasing $R$ and decreasing $P$. This will be even more true for the p,p collider. But it already suggests that **the excess cost incurred in tunnel circumference needed for eventual p,p operation at energy approaching 100 TeV (over and above what could be minimally adequate for the Higgs factory) may not be exhorbitant.**

Most of the paper consists of formulations largely specific to CepC, with emphasis on their scaling behavior. The remainder consists of more technical appendices. Some of the material has been copied with little change from reports produced over the last two years. As such the parameters in the various appendices are in some cases outdated and may be mutually inconsistent. This should be no more disconcerting than the fact that many important parameters, in fact, remain seriously uncertain. In any case it is the dependencies, not the numerical values, that are important.

The luminosity depends importantly on an inconveniently large number of collider parameters, and it is the detailed absolute numerical values of luminosity that are important. For that reason I have attempted to make all luminosity predictions self-consistent, even when the tables appear in different sections of the paper.

On the other hand, the ring parameters assumed in different sections may not be consistent. The most extreme instance of this is the "Single Beam Multibunch Operation and Beam Separation" section, in which the beam separation scheme assumes the quite short cell length adopted by the CEPC lattice design. To ease the electric separation challenge a multiple separator scheme was introduced, consistent with the CEPC parameters of early 2014 (and up to the present). But the "Lattice Optimization for Top-Off Injection" section advocates much longer cells, for which fewer, longer, electric separators may be adequate. So these two sections are mutually inconsistent. The "Scaling Law Dependence of Luminosity on Free Space $L^*$ also assumes an unaccountably small CEPC cell length in its semi-quantitative luminosity estimation.

Another open question concerns chromatic compensation of the interaction region optics. Scaling laws in the "Achromatic Higgs Factory Intersection Region Optics" and "Scaling Law Dependence of Luminosity on Free Space $L^*$" appendices assume local chromatic compensation internal to the intersection region optics. Though the $L^*$ scaling law is not in question, the example optics used to "derive" it are by no means settled. Problems associated with introducing strong bends within the intersection region optics may favor "old fashioned" chromatic compensation in the arcs, as in LEP.

The subject matter of the appendices can be discerned from the Table of Contents. They contain little original material, but are justified by the fact that the original sources are scattered and hard to merge consistently.

## 1.2 CEPC, then SPPC in the Same Tunnel

The quite low Higgs particle mass makes a circular electron collider an effective Higgs factory. Furthermore, just as LHC followed LEP in the same tunnel, building first an electron collider, and later a proton collider in the same tunnel, represents a natural future for elementary particle physics. Though this paper is almost entirely devoted to the circular electron Higgs factory, it is appropriate to consider the extent to which the parameter choices for the Higgs factory can be biased to improve the ultimate proton collider. A possible modest initial cost increase can be far more than compensated by the improvement in ultimate proton collider performance.

**The parameter most implicated in this discussion is, of course, the ring circumference.** Once fixed this choice will constrain the facility for its entire, at least half century, life. Furthermore this choice needs to be made before any of the many remaining design decisions have to be made.

To focus the discussion one can attempt to define a range of circumferences running from $C^{\min}$, a circumference such as 70 km, large enough to define a guaranteed frontier role in high energy physics through the life of the facility, to $C^{\max}$, such as 100 km, a circumference small enough to hold the project cost low enough to make project approval likely. Different high energy communities (all represented at the IAS Program) are differently qualified to determine these limits. $C^{\max}$ depends largely on cost; it is the project directorate (with the assistance of accelerator scientists) who are most qualified to assess the cost of the facility. But the cost is not the only factor influencing the approval probability. It is the particle theorists and experimentalists who are most qualified to establish the high level of enthusiasm without which project approval would be unlikely. Certainly there is a circumference value $C^{\min}$ below which their enthusiasm for the project would decline significantly.

Discussing and determining this regrettably subjective range of possible tunnel circumferences can be one theme of the IAS program, certainly with the hope that $C^{\min} < C^{\max}$.

Important ingredients of this formulation are bottom-up designs of the relation between tunnel circumference and beam energy for both electrons and protons. The natural order of construction is electron ring first, proton ring second. This is because the electron ring itself is much less challenging, much less expensive, and much less dependent on continuing improvements in superconducting magnet technology than is the proton ring.

Minimizing the initial cost (and thereby improving the approval likelihood) makes optimizing the electron ring design more urgent than optimizing the proton ring design. In fact, since the ideal circumference for protons is surely greater than for electrons, **what is needed is to maximize the electron ring circumference while minimizing its cost—a seemingly impossible task.**

**The thesis of this paper is that this optimization is not as hard as it seems.** More concretely, it will be shown that making the electron ring circumference "unnecessarily large" (from the point of view of minimally adequate Higgs particle production) can increase its cost less than proportionally, if at all, provided the RF power is reduced proportionally. This argument relies on a scaling law according to which the optimized luminosity is a function only of the product of circumference times RF power.

The "unnecessarily large" qualifying phrase in the previous paragraph may, itself, be unnecessarily conservative. Depending on initial discoveries with the Higgs factory, it may well be found appropriate in a second Higgs phase, to increase the RF power in order to increase the e+e- luminosity proportionally. This would be valuable not only to produce more Higgs particles, but also, for example, to increase the beam energy well above the Higgs production threshold to study other Higgs particle production channels.

### 1.3 Single Beam Parameters

Three important parameters of a circular colliding beam storage ring are $N^*$, the number of collision points, $N_b$, the number of circulating bunches, and $N_p$ the number of electrons or positrons in each bunch, assumed equal for the two beams. The optimal values for these parameters depend on the beam energy $E$ and on cost considerations. The case $N^* = 4$, $N_b = 4$ is illustrated in Figure 1. At the highest energy the RF power limit may fix the number of bunches to $N_b = 1$ or, for 4 detectors, to $N_b = 2$.

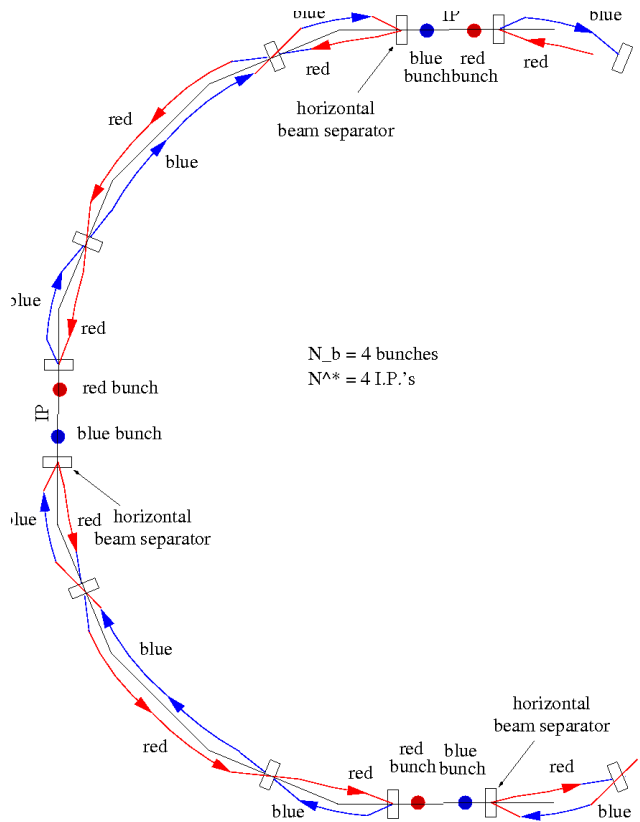

Figure 1: Schematic partial view of Higgs factory. As shown there are $N_b = 4$ bunches colliding at $N^* = 4$ collision points. To maximize luminosity, RF power considerations require $N_b$ to decrease as beam energy $E$ increases. The bunches are separated horizontally at the four quadrant arc midpoints. In the one ring design the beams share the same vacuum chamber, lattice elements and RF cavities.

Parameters for large ranges of single beam parameters are given in Table 1. Note: the $R = 3$ km entries are roughly appropriate for LEP; the maximum LEP RF voltage was 7.9 MV/m, making up the (approximate) $U_1 \approx 3$ GeV energy loss shown. For the proposed Higgs factory a maximum RF energy drop $eV_{\rm rf}^{\rm max} = 65$ GV is assumed, and "excess voltage" is defined by $eV_{\rm excess} = eV_{\rm rf}^{\rm max} - U_1$. The RF voltage is set to such a high value in order to be sufficient to compensate for $U_1$, even at quite high beam energies $E$. As a result the taulated RF over-voltages $eV_{\rm excess}$ are highly extravagant for low beam energies. For actual running at $E = 300$ GeV a higher value would be required. Subsequent tables in the paper simply extend the rows of this table for specific $\beta_y^*$ values, for example $\beta_y^* = 0.004$ m, $\beta_y^* = 0.006$ m, and $\beta_y^* = 0.008$ m, in order to establish trends. **According to the simulation model, the optimum is near $\beta_y^* \approx 5$ mm at the Higgs energy.** Numerical examples in the text are usually taken from the shaded rows.

Like Table 1, Table 2 contains single beam parameters, but specialized to 100 km circumference, with rows limited to physically significant energies; namely "Z" for the $Z_0$ resonance energy, "W" for the W-pair threshhold, "LEP" as the "nominal" LEP" beam energy, "H" for the Higgs production threshhold, and "tt" as the top-pair threshhold, representing high energy Higgs production channels.

Even in quite favorable cases the energy loss per turn $U_1$ is as much as several percent of the total energy. To keep the energy within 1% will then require a dozen or more RF accelerating sections. Because of its high energy loss, the Higgs factory will actually resemble a slowly curving linac. Nevertheless, it represents an economy, relative to a linear collider, to retain electrons along with most of their energy and restore their radiated energy every turn, rather than discarding and replacing them, as is required in a linear collider.

### 1.4 Optimization Considerations

This paper pays special attention to the beamstrahlung limitation pointed out by Telnov [4], and proceeds to quantify the limitation by a "beamstrahlung penalty" $\mathcal{P}_{\rm bs}$. This penalty turns out to be so severe, and its onset (with increasing beam energy $E$) so sudden (see Figure 14) that a sensible strategy is to fix parameters so that $\mathcal{P}_{\rm bs}$ remains just barely consistent with the capability to replenish the lost particles.

In previous, lower energy e+e- colliders, it has been customary to keep the r.m.s. bunch length $\sigma_z$ comparable to the vertical beta function $\beta_y$, in order to minimize the hourglass effect. Because of the beamstrahlung effect this strategy may no longer be optimal for a Higgs factory. Rather it may be more optimal to accept a higher hourglass penalty in order to reduce the beamstrahlung penalty. Lengthening the bunch "softens" the x-ray spectrum proportionally, which strongly reduces the likelihood of emission of a single photon of energy high enough for the radiating electron to be lost. As well as softening the beamstrahlung spectrum, increasing the bunch length also has the beneficial effect of reducing wall impedance effects. It must be kept in mind, however, that the bunch length is largely determined by the lattice design, and is not easily changed, for example as the beam energy is changed.

According to the simulation result given in Eq. (80), the saturated tune shift value $\xi^{\rm sat.}$ is proportional to $\sqrt{r_{yz}}$ where $r_{yz} = \beta_y/\sigma_z$. Figure 21 shows the hourglass correction factor $H(r_{yz})$ to be quite accurately equal to $\sqrt{r_{yz}}$. The product of $\xi^{\rm sat.} \cdot H(r_{yz})$ appearing, for example, in Eq. (93) is therefore proportional to $1/\sigma_z$ (for fixed $\beta_y$). This tends to frustrate efforts to increase luminosity by increasing bunch length for the purpose of decreasing beamstrahlung.

In all cases the luminosity is limited by available RF power per beam. Following recent designs that have adopted $P_{\rm rf} = 50$ MW as a kind of nominal choice, some tables in this paper use this value. **Other tables reflect my recommendation to reduce power to $P_{\rm rf} = 25$ MW while doubling the ring circumference.** Fixing $P_{\rm rf}$ fixes the maximum total number $N_{\rm tot}$ of particles stored in each beam. At pre-LEP beam energies all other parameters would then have been adjusted to "saturate the beam-beam tune shift [5]". At Higgs factory energies the RF power limitation, in conjunction with the beamstrahlung constraint, could make this impossible which will limit the luminosity accordingly.

Total power is not the only significant RF parameter. For some tables this paper, I also choose the maximum voltage drop to be $V_{\rm rf} = 65$ GV, which is almost certainly much higher than will actually be provided for initial operation; it is about 20 times higher than the maximum voltage in terminal LEP operation. A given value of $V_{\rm rf}$ sets an absolute maximum beam energy. To have non-zero luminosity at $E_{\rm max} = 300$ Gev, which is the highest energy appearing in Table 1, $V_{\rm rf}$ has to be at least 60 GV, for the maximum bend radius considered in the table.

At sub-LEP energies there will be ample RF power to saturate the vertical tune shift and and the luminosity can be further increased with multiple bunches.

It will be shown that the total Higgs particle production, summed over all detectors, increases with the number of detectors $N^*$. However, following the tentative CEPC design, this paper usually assumes $N^* = 2$.

As numerous authors [6] have pointed out, "topping-off" injection is highly favorable for a Higgs factory. There are various reasons for this, but the one that is crucial for the multi-detector approach is that topping-off injection can be expected to remove any limitation on the total beam-beam tune shift $\sum_{i=1}^{N^*} \xi_i$, where $\xi_i = \xi$ is the vertical tune shift in collision point $i$. Nonlinear dynamics will limit the value $\xi_i$ to the same maximum value at every intersection point. But, once stable circulating beams have been established, I assume that $N^*\xi$ can be arbitrarily high, even greater than 1, for example.

Weiren Chou and Tenaji Sen have pointed out that the topping off repetition rate has to be quite high. For example, if the beam lifetime (without topping off) is 30 minutes, the beam has to be topped off on the order of once per minute to keep the beam current constant to better that one percent.

The exact sequence of operations by which the stable, high current steady state is obtained will not be easy, nor will it be discussed here. Once in this state, there will be a far more easily-met requirement that the variation of $N^*\xi$ over the steady-state filling sequence remain less than the distance to the nearest destructive resonance.

| $E$ | $C$ | $R$ | $f$ | $U_1$ | $eV_{\text{excess}}$ | $n_1$ | $U_1/(D/2)$ | $\delta = \alpha_4$ | $u_c$ | $\epsilon_x$ | $\sigma_x^{\text{arc}}$ |
|---|---|---|---|---|---|---|---|---|---|---|---|
| GeV | km | km | KHz | GeV | GeV | elec./MW | MV/m | | GeV | nm | mm |
| 100 | 28 | 3.0 | 10.60 | 3.0 | 62 | 2.00e+11 | 0.626 | 0.0074 | 0.00074 | 6.354 | 0.523 |
| 150 | 28 | 3.0 | 10.60 | 14.9 | 50 | 3.94e+10 | 3.169 | 0.0249 | 0.00249 | 14.297 | 0.784 |
| 200 | 28 | 3.0 | 10.60 | 47.2 | 18 | 1.25e+10 | 10.016 | 0.0590 | 0.00591 | 25.417 | 1.05 |
| 250 | 28 | 3.0 | 10.60 | 115.2 | -50 | 5.11e+09 | 24.453 | 0.1152 | 0.01155 | 39.715 | 1.31 |
| 300 | 28 | 3.0 | 10.60 | 239.0 | -1.7e+02 | 2.46e+09 | 50.707 | 0.1991 | 0.01995 | 57.189 | 1.57 |
| 100 | 57 | 6.0 | 5.30 | 1.5 | 64 | 7.98e+11 | 0.157 | 0.0037 | 0.00037 | 3.177 | 0.37 |
| 150 | 57 | 6.0 | 5.30 | 7.5 | 58 | 1.58e+11 | 0.792 | 0.0124 | 0.00125 | 7.149 | 0.554 |
| 200 | 57 | 6.0 | 5.30 | 23.6 | 41 | 4.99e+10 | 2.504 | 0.0295 | 0.00296 | 12.709 | 0.739 |
| 250 | 57 | 6.0 | 5.30 | 57.6 | 7.4 | 2.04e+10 | 6.113 | 0.0576 | 0.00577 | 19.857 | 0.924 |
| 300 | 57 | 6.0 | 5.30 | 119.5 | -54 | 9.85e+09 | 12.677 | 0.0996 | 0.00998 | 28.595 | 1.11 |
| 100 | 75 | 8.0 | 3.98 | 1.1 | 64 | 1.42e+12 | 0.088 | 0.0028 | 0.00028 | 2.383 | 0.32 |
| 150 | 75 | 8.0 | 3.98 | 5.6 | 59 | 2.80e+11 | 0.446 | 0.0093 | 0.00094 | 5.361 | 0.48 |
| 200 | 75 | 8.0 | 3.98 | 17.7 | 47 | 8.87e+10 | 1.409 | 0.0221 | 0.00222 | 9.532 | 0.64 |
| 250 | 75 | 8.0 | 3.98 | 43.2 | 22 | 3.63e+10 | 3.439 | 0.0432 | 0.00433 | 14.893 | 0.8 |
| 300 | 75 | 8.0 | 3.98 | 89.6 | -25 | 1.75e+10 | 7.131 | 0.0747 | 0.00748 | 21.446 | 0.96 |
| 100 | 94 | 10.0 | 3.18 | 0.9 | 64 | 2.22e+12 | 0.056 | 0.0022 | 0.00022 | 1.906 | 0.286 |
| 150 | 94 | 10.0 | 3.18 | 4.5 | 61 | 4.38e+11 | 0.285 | 0.0075 | 0.00075 | 4.289 | 0.429 |
| 200 | 94 | 10.0 | 3.18 | 14.2 | 51 | 1.39e+11 | 0.901 | 0.0177 | 0.00177 | 7.625 | 0.573 |
| 250 | 94 | 10.0 | 3.18 | 34.6 | 30 | 5.68e+10 | 2.201 | 0.0346 | 0.00346 | 11.914 | 0.716 |
| 300 | 94 | 10.0 | 3.18 | 71.7 | -6.7 | 2.74e+10 | 4.564 | 0.0597 | 0.00599 | 17.157 | 0.859 |
| 100 | 113 | 12.0 | 2.65 | 0.7 | 64 | 3.19e+12 | 0.039 | 0.0018 | 0.00018 | 1.589 | 0.261 |
| 150 | 113 | 12.0 | 2.65 | 3.7 | 61 | 6.31e+11 | 0.198 | 0.0062 | 0.00062 | 3.574 | 0.392 |
| 200 | 113 | 12.0 | 2.65 | 11.8 | 53 | 2.00e+11 | 0.626 | 0.0148 | 0.00148 | 6.354 | 0.523 |
| 250 | 113 | 12.0 | 2.65 | 28.8 | 36 | 8.17e+10 | 1.528 | 0.0288 | 0.00289 | 9.929 | 0.653 |
| 300 | 113 | 12.0 | 2.65 | 59.7 | 5.3 | 3.94e+10 | 3.169 | 0.0498 | 0.00499 | 14.297 | 0.784 |

Table 1: Ring parameters for rings of various bending radii, assuming 2/3 fill factor, with half of total straight section length $D$ taken up by RF. The $U_1/(D/2)$ column therefore indicates the minimum required energy gain per meter to be supplied by the RF. $u_c$ is the critical energy of the synchrotron radiation energy spectrum. $\alpha_4$ is the appropriate damping decrement for $N^* = 4$ interaction points.

| name | $E$ | $C$ | $R$ | $f$ | $U_1$ | $eV_{\text{excess}}$ | $n_1$ | $\delta = \alpha_2$ | $u_c$ | $\epsilon_x$ † | $\sigma_x^{\text{arc}}$ |
|---|---|---|---|---|---|---|---|---|---|---|---|
| | GeV | km | km | KHz | GeV | GeV | elec./MW | | GeV | nm | mm |
| Z | 46 | 100 | 10.6 | 3.00 | 0.04 | 20 | 5.81e+13 | 0.00020 | 0.00002 | 0.573 | 2 |
| W | 80 | 100 | 10.6 | 3.00 | 0.34 | 20 | 6.08e+12 | 0.00107 | 0.00011 | 1.771 | 1.19 |
| LEP | 100 | 100 | 10.6 | 3.00 | 0.83 | 19 | 2.49e+12 | 0.00209 | 0.00021 | 2.767 | 0.972 |
| H | 120 | 100 | 10.6 | 3.00 | 1.73 | 18 | 1.20e+12 | 0.00361 | 0.00036 | 3.984 | 0.824 |
| tt | 175 | 100 | 10.6 | 3.00 | 7.83 | 12 | 2.66e+11 | 0.01119 | 0.00112 | 8.473 | 0.585 |

Table 2: Single beam parameters, assuming 100 km circumference. The second last column (†) lists the value of $\epsilon_x$ appropriate only for $\beta_y^* = 5$ mm. Though determined by arc optics, $\epsilon_x$ has to be adjusted, according to the value of $\beta_y^*$, to optimize the beam shape at the IP. Other cases can be calculated from entries in other tables. $U_1$ is the energy loss per turn per particle. $u_c$ is the critical energy for bending element synchrotron radiation. $\delta$ is the synchrotron radiation damping decrement.

It is certainly optimistic to permit arbitrarily large coherent beam-beam tune shift. But, from the point of view of linear lattice dynamics, the crossing points are indistinguishable from idealy thin lenses, focusing in both planes. They are not, however, "high quality" lenses, in that they have significant octupole, and higher multipole components, which may or may not limit the dynamic aperture. The "optimistic" aspect of our optimistic conjecture being made is that the dynamic aperture limitation does not worsen with multiple collision points. This is tantamount to assuming that a phasor sum need not have magnitude greater than any one of its terms.

The simulation model [5] used to establish saturated tune shift conditions, though ten years old, has been fired up again for the Higgs factory study. Operationally, saturation is specified, only semi-quantitatively, by the tune shift value at which the quadratic dependence of luminosity on beam current transitions from quadratic to linear. In the simulation model this transition point is quantitatively precise. Described in Appendix C, the model predicts the dependence of luminosity on damping decrement $\delta$, vertical beta function $\beta_y$, bunch length $\sigma_z$, the three tune values $Q_x$, $Q_y$ and $Q_s$, and the three bunch dimensions, $\sigma_x$, $\sigma_y$, and $\sigma_z$. Some of these dependencies are exhibited in figures in Appendix C.

Requiring beamstrahlung to be "barely acceptable" in the sense described so far influences the task of fixing the bend radius $R$, the total length of straight sections $D$, and the total circumference $C$. From a cost perspective these are the most important parameters. Before these parameters can be determined, the maximum energy $E_{max}$ has to be specified. Certainly the optimal ring size and cost increase more than proportionally with increasing $E_{max}$.

Apart from its reduced cost compared to a linear collider (which is due to the surprisingly low mass the Higgs particle has been found to have) the greatest advantage of a circular collider is its well-understood behavior and correspondingly small risk. The only significant uncertainty concerns the parameter $\eta$ which gives the fractional energy at which a beamstrahlung radiation causes the radiating electron to be lost. For numerical estimates in this report, following Telnov, I have adopted the value $\eta = 0.015$. That this value can be significantly larger than in existing rings is both because of specialized lattice design and the extremely strong betatron damping in a Higgs factory. An urgent design challenge is to confirm the validity of this assumption by detailed lattice design and tracking simulations.

## 2 RING CIRCUMFERENCE AND TWO RINGS VS ONE RING

The main Higgs factory cost-driving parameter choices include: tunnel circumference $C$, whether there is to be one ring or two, what is the installed power, and what is the "Physics" for which the luminosity deserves to be maximized. This section discusses some of the trade-offs among these choices, and attempts to show that the optimization goals for the Higgs factory and the later p,p collider are consistent.

**A good way to fix the circumference $C$ is to simply extrapolate from earlier colliding beam rings as is done in Figure 2.** Choosing $E = 300$ GeV to be the nominal beam energy yields circumference $C \approx 100$ km. Nothing in this paper is incompatible with this choice.

### 2.1 General Comments

The quite low Higgs mass (125 GeV) makes a circular e+e- collider (FCC-ep) ideal for producing background-free Higgs particles. There is also ample physics motivation for planning for a next-generation proton-proton collider with center of mass energy approaching 100 TeV. This suggests a two-step plan: first build a circular e+e- Higgs factory; later replace it with a 100 TeV pp collider (or, at least, center of mass energy much greater than LHC). This paper is devoted almost entirely to the circular Higgs factory step, but keeping in mind the importance of preserving the p,p collider potential. To illustrate this possibility the LHC and FCC-pp (scaled up from LHC based on radiation-dominated scaling) are also plotted on Figure 2.

The main Higgs factory cost-driving parameter choices include: tunnel circumference $C$, whether there is to be one ring or two, what is the installed power, and what are the physics priorities. From the outset I confess my prejudice towards a single LEP-like ring, optimized for Higgs production at $E = 120$ Gev, with minimum *initial* cost, and highest possible eventual p,p energy. This section discusses some of the trade-offs among these choices, and attempts to show that electron/positron and proton/proton optimization goals are consistent.

Both Higgs factory power considerations and eventual p,p collider favor a tunnel of the largest possible radius $R$. Obviously one ring is cheaper than two rings. For 120 GeV Higgs factory operation (and higher energies) it will be shown that one ring is both satisfactory and cheaper than two. But higher luminosity (by a factor of five or so) at the (45.6 GeV) $Z_0$ energy, requires two rings.

Unlike the $Z_0$, there is no unique "Higgs Factory energy". Rather there is the threshold turn-on of the cross section, shown, along with other applicable cross sections in Figure 3.

We arbitrarily choose 120 GeV per beam as the Higgs particle operating point and identify the single beam energy this way in subsequent tables. Similarly identified are the $Z_0$ energy (45.6 GeV), the W-pair energy of 80 GeV, the LEP energy (arbitrarily taken to be 100 GeV) and the $t\bar{t}$ energy of 175 GeV to represent high energy performance.

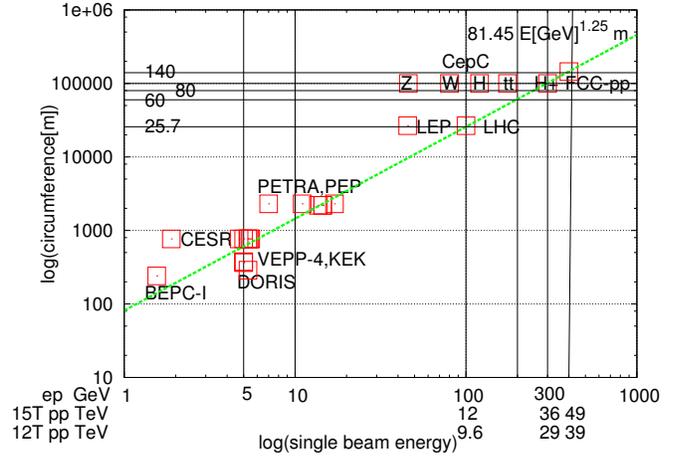

Figure 2: Relation between beam energy $E$ and circumference $C$ for numerous colliding beam rings. Because the FCC-pp collider will be synchrotron radiation dominated, its scaling up (from LHC) should follow the same trend. In this plot p,p colliders, LHC and FCC-pp are labeled on the right hand side of the linear fit. But the p,p horizontal axis depends on the assumed magnetic field value. The axis are labeled for 12 T and 15 T options. Like the earlier electron colliders, CESR, PEP, PETRA, and LEP, the dynamic ranges are about one octave in energy. The same will, presumeably, be true for the Higgs factory— from $Z_0$ at the low end, through $W$-pairs, Higgs, t-tbar, to associated Higgs production channels at the high energy end.

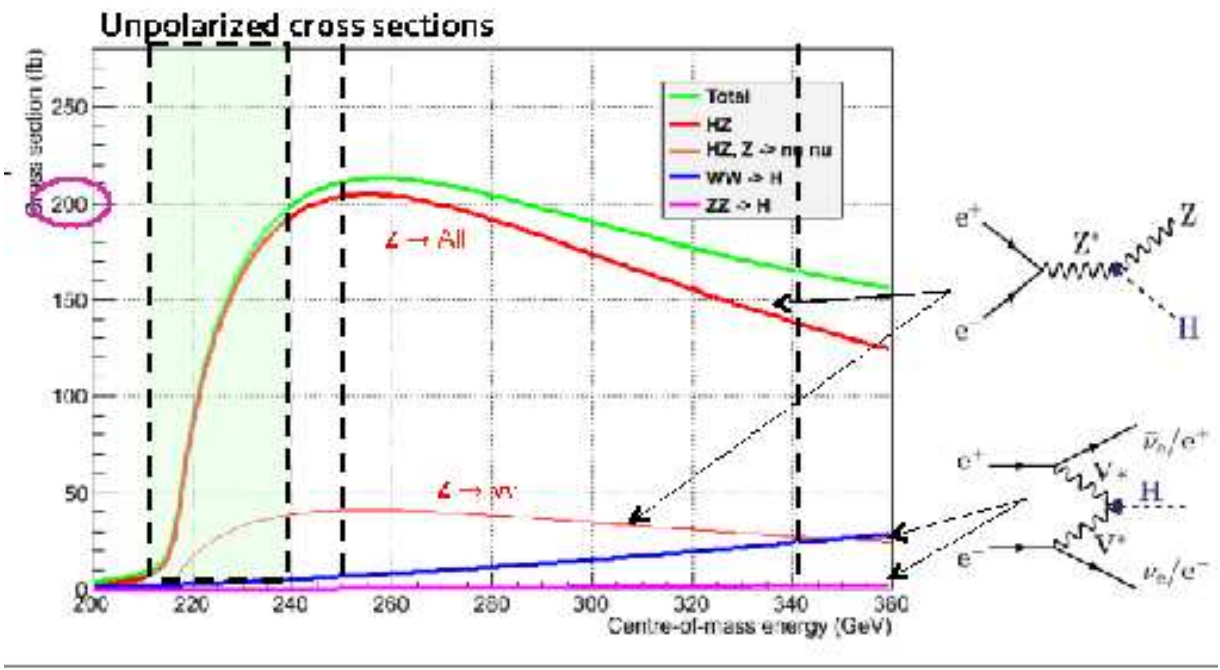

Figure 3: Higgs particle cross sections up to $\sqrt{s} = 0.3$ TeV (copied from Patrick Janot); $\mathcal{L} \geq 2 \times 10^{34}$ /cm$^2$/s, will produce 400 Higgs per day in this range.

## 2.2 Scaling up from LEP to Higgs Factory

**Radius × Power Scale Invariant.** Most of the conclusions in this paper are based on scaling laws, either with respect to bending radius $R$ or with respect to beam energy $E$. Scaling with bend radius $R$ is equivalent to scaling with circumference $C$. (Because of limited "fill factor", RF, straight sections, etc., $R \approx C/10$.)

Higgs production was just barely beyond the reach of LEP's top energy, by the ratio 125 GeV/105 GeV = 1.19. **This should make the extrapolation from LEP to Higgs factory quite reliable. In such an extrapolation it is increased radius more than increased beam energy that is mainly required.**

One can note that, for a ring three times the size of LEP, the ratio of $E^4/R$ (synchrotron energy loss per turn) is $1.19^4/3 = 0.67$—i.e. *less than final LEP operation*. Also, for a given RF power $P_{\rm rf}$, **the maximum total number of stored particles is proportional to $R^2$—doubling the ring radius cuts in half the energy loss per turn and doubles the time interval over which the loss occurs.** Expressed as a scaling law

$$\boxed{n_1 = \text{number of stored electrons per MW} \propto R^2.} \qquad (3)$$

This is boxed to emphasize its fundamental importance. Following directly from Eq. (1), it is the main consideration favoring large circumference for both electron and radiation-dominated proton colliders.

These comments should completely debunk a long-held perception that LEP had the highest energy practical for an electron storage ring.

There are three distinct upper limit constraints on the luminosity. As explained in Appendix D, "Luninosity Formulas", maximum luminosity results when the ring parameters have been optimized so the three constraints yield the same upper limit for the luminosity. For now we concentrate on just the simplest luminosity constraint $\mathcal{L}_{\rm pow}^{\rm RF}$, the maximum luminosity for given RF power $P_{\rm rf}$. With $n_1$ being the number of stored particles per MW; $f$ the revolution frequency; $N_b$ the number of bunches, which is proportional to $R$; $\sigma_y^*$ the beam height at the collision point; and aspect ratio $\sigma_x^*/\sigma_y^*$ fixed (at a large value such as 15);

$$\boxed{\mathcal{L}_{\rm pow}^{\rm RF} \propto \frac{f}{N_b}\left(\frac{n_1 P_{\rm rf}[{\rm MW}]}{\sigma_y^*}\right)^2.} \qquad (4)$$

Consider variations for which

$$\boxed{P_{\rm rf} \propto \frac{1}{R}.} \qquad (5)$$

Dropping "constant" factors, the dependencies on $R$ are, $N_b \propto R$, $f \propto 1/R$, and $n_1 \propto R^2$. With the $P_{\rm rf} \propto 1/R$ scaling of Eq. (5), $\mathcal{L}$ is independent of $R$. In other words, the luminosity depends on $R$ and $P_{\rm rf}$ only through their product $R P_{\rm rf}$. Note though, that this scaling relation *does not* imply that $\mathcal{L} \propto P_{\rm rf}^2$ at fixed $R$; rather $\mathcal{L} \propto P_{\rm rf}$.

In this paper this scaling law will be used in the form

$$\boxed{\mathcal{L}(R, P_{\rm rf}) = f(R P_{\rm rf}),} \qquad (6)$$

the luminosity depends on $R$ and $P_{\rm rf}$ as a function $f(R P_{\rm rf})$ of only their product.

This radius/power scaling formula can be checked numerically by comparing Tables 6 and 8. The comparison is only approximate since other parameters and the scalings from LEP are not exactly the same in the two cases.

**Parameter Scaling with Radius.** For simplicity, even if it is not necessarily optimal, let us assume the Higgs factory arc optics can be scaled directly from LEP values, which are: phase advance per cell $\mu_x = \pi/2$, full cell length $L_c = 79$ m. (The subscript "c" distinguishes the Higgs factory collider lattice cell length from injector lattice cell length $L_i$.)

Constant dispersion scaling formulas are given in Table 3. These formulas are derived in Section 4.2 "Lattice Optimization for Top-Off Injection". They are then applied to extrapolate from LEP to find the lattice parameters for Higgs factories of (approximate) circumference 50 km and 100 km, shown in Table 5.

| Parameter | Symbol | Proportionality | Scaling |
|---|---|---|---|
| phase advance per cell | $\mu$ | | 1 |
| collider cell length | $L_c$ | | $R^{1/2}$ |
| bend angle per cell | $\phi$ | $= L_c/R$ | $R^{-1/2}$ |
| quad strength $(1/f)$ | $q$ | $1/L_c$ | $R^{-1/2}$ |
| dispersion | $D$ | $\phi L_c$ | 1 |
| beta | $\beta$ | $L_c$ | $R^{1/2}$ |
| tunes | $Q_x, Q_y$ | $R/\beta$ | $R^{1/2}$ |
| Sands's "curly H" | $\mathcal{H}$ | $= D^2/\beta$ | $R^{-1/2}$ |
| partition numbers | $J_x/J_y/J_\epsilon$ | $= 1/1/2$ | 1 |
| horizontal emittance | $\epsilon_x$ | $\mathcal{H}/(J_x R)$ | $R^{-3/2}$ |
| fract. momentum spread | $\sigma_\delta$ | $\sqrt{B}$ | $R^{-1/2}$ |
| arc beam width-betatron | $\sigma_{x,\beta}$ | $\sqrt{\beta \epsilon_x}$ | $R^{-1/2}$ |
| -synchrotron | $\sigma_{x,synch.}$ | $D \sigma_\delta$ | $R^{-1/2}$ |
| sextupole strength | $S$ | $q/D$ | $R^{-1/2}$ |
| dynamic aperture | $x^{\max}$ | $q/S$ | 1 |
| relative dyn. aperture | $x^{\max}/\sigma_x$ | | $R^{1/2}$ |
| pretzel amplitude | $x_p$ | $\sigma_x$ | $R^{-1/2}$ |

Table 3: *Constant dispersion Constant dispersion* scaling is the result of choosing cell length $L \propto R^{1/2}$. The entry "1" in the last column of the shaded "dispersion" row, indicates that the dispersion is independent of $R$ when the cell length $L_c$ varies proportional to $\sqrt{R}$ with the phase advance per cell $\mu$ held constant.

## 2.3 Staged Optimization Cost Model.

For best likelihood of initial approval and best eventual p,p performance, the cost of the first step has to be minimized and the tunnel circumference maximized. Surprisingly, these requirements are consistent. Consider optimization principles for three collider stages:

- **Stage I, e+e-:** Starting configuration. Minimize cost at "respectable" luminosity, e.g. $10^{34}$. Constrain the number of rings to 1, and the number of IP's to $N^* = 2$.

- **Stage II, e+e-:** Maximize luminosity/cost for production Higgs (etc.) running. Upgrade the luminosity by some combination of: $P_{\rm rf} \to 2P_{\rm rf}$ or $4P_{\rm rf}$, one ring $\to$ two rings, increasing $N^*$ from 2 to 4, or decreasing $\beta_y^*$.

- **Stage III, pp:** Maximize the ultimate physics reach, i.e. center of mass energy, i.e. maximize tunnel circumference.

## 2.4 Scaling of Higgs Factory Magnet Fabrication

Unlike the rest of the paper, this section is conjectural and idiosyncratic. It contains my opinions concerning how best to construct the Higgs factory room temperature magnets. It does not pretend to understand the economics of superconducting magnet technology. But it is also not ruled out that similar arguments and conclusions may be applicable to the eventual p,p collider.

As a disciple of Robert Wilson, one cannot avoid approaching the Higgs factory design challenge by imagining how he would have. **Certainly Bob Wilson would have endorsed Nima Arkani-Hamed's attitude that we strive for 100 TeV collisions "because the project is big", rather than "in spite of the fact the project is big".**

"How would Bob do it?" also suggests unconventional design approaches. At the early design stage, based on good, but limited, understanding of the task, one of his principles can be stated as "It is better for the tentative parameter choices to be easy to remember than to be accurate". In the current context he would certainly have liked the round numbers in a statement such as "To obtain 100 TeV collisions we need a ring with 100 km circumference", especially because of (or, possibly, in spite of) the fact that the CERN FCC group favors just these values.

Another Wilson attitude was that, if a competent physicist (where he had himself in mind) could conceptualize an elegant solution to a mechanical design problem, consistent with the laws of physics, then a competent engineer (where he again had himself in mind) could certainly successfully complete the design.

In extrapolating the room temperature magnet design from LEP to CepC one must first acquire a prejudice as to the vacuum chamber bore diameter. Many of the scaling formulas in this paper are devoted to determining this, along with other self-consistent parameters. To make the subsequent discussion as simple as possible one can accept, as a first iteration, the choice of making the magnet bore the same as LEP, promising to later improve this choice, in a second, or third, iteration, as necessary. It is my guess that the first iteration will be close.

In round numbers, the 100 km Higgs factory ring magnet length is four times as great as LEP's, and the Higg's factory energy is greater than the maximum LEP energy in the ratio 120/100. The required Higgs factory magnetic field is therefore less than the LEP magnetic field in the ratio 1.2/4 = 0.3. The stored magnetic energy density scales as the square of this ratio. With the magnetic bore constant, the Higgs factory stored magnetic energy is less than for LEP in the ratio $4 \times 0.3^2 = 0.36$. Ferromagnetic magnets are often costed in Joules per cubic meter. If this were valid the Higgs factory magnet would be three times cheaper than the LEP magnet.

When one actually looks into magnet costs one finds the calculation in the previous paragraph to be entirely misleading. The actual costs tend to be dominated by end effects, fabrication, transportation and installation. Accepting these costs as dominant would, one might think, force one to accept the Higgs factory magnet cost being proportional to tunnel circumference; this would be the cost of simply replicating LEP magnets. One reason this might be too conservative is that, with the Higgs factory cell length being longer, the magnets could be longer. But this would also be misleading since the LEP magnets were already as long as economically practical (because of fabrication, transportation and installation costs).

To hold down magnet costs, the inescapable conclusion to be drawn from this discussion is that the magnets have to be built *in situ*, in their final positions in the Higgs factory tunnel. This is the only possible way to prevent the magnet cost from scaling proportional to the tunnel circumference, or worse. (The same is probably true for superconducting magnets in the later p,p phase of the project.)

It is not at all challenging to build the Higgs factory collider magnets in place. With top-off injection these magnets do not have to ramp up in field. As a result they have no eddy currents and therefore do not need to be laminated.

Regrettably the same is not true for the injector magnet, which will be more challenging, and may be more expensive, than the collider magnet.

An even more quixotic argument for building the magnet in place is to compare the arcs of the collider to high voltage electrical power lines, which carry vast amounts of power over vast distances. For example a $10^6$ V line, carrying $10^3$ A, carries $10^9$ W of power over a distance of 100 Km, with fractional energy loss of 1%. The arcs of the Higgs factory will similarly carry $10^{11}$ V at $10^{-2}$ A over a distance of 100 Km with fractional energy loss of 1%. Same power, same loss. One would not even think of building overland power lines in a factory before transporting them to where they are needed. The same should be true for accelerator magnets.

*SPPC:* For superconducting magnetic fields $B$ in the range from 4 to 7 Tesla the cost per unit volume [7] is roughly proportional $B^{2/3}$ but increasing "more than linearly for higher magnetic fields", perhaps proportional to $B$ at, say, 12 T. If true, at fixed bore diameter and fixed energy the magnet cost would be more or less independent of tunnel radius $R$, and there would be little need to worry about the tunnel circumference being "too big" from this point of view.

As discussed previously the synchrotron radiation heat load cost is proportional to $1/R^2$ at fixed $E$. In principle, none of the synchrotron radiation has to be stopped at liquid helium temperature but, in practice, this is very hard to achieve. As with electrons, the reduced synchrotron radiation power load can be exploited to increase the stored beam charge by increasing $R$. This has the further beneficial effect of increasing the beam burn-off (interaction) lifetime. Probably a more important interaction lifetime effect is that the stored charge can be proportional to $R$, causing the burn-off lifetime to be proportional to $R$.

## 2.5 Cost Optimization

Treating the cost of the 2 detectors as fixed, and letting $C$ be the cost exclusive of detectors, the cost can be expressed as a linearized fit, the sum of a term proportional to size and a term proportional to power;

$$C = C_R + C_P \equiv c_R R + c_P P_{\rm rf} \qquad (7)$$

where $c_R$ and $c_P$ are unit cost coefficients. As given by scaling formula (5), for constant luminosity, the RF power, luminosity, and ring radius, for small variations, are related by

$$P_{\rm rf} = \frac{\mathcal{L}}{k_1 R}. \qquad (8)$$

Minimizing $C$ at fixed $\mathcal{L}$ leads to

$$\boxed{R_{\rm opt} = \sqrt{\frac{1}{k_1}\frac{c_P}{c_R}\mathcal{L}}.} \qquad (9)$$

Conventional thinking has it that $c_P$ is universal world wide but, at the moment, $c_R$ is thought to be somewhat cheaper in China than elsewhere. If so, **the optimal radius should be somewhat greater in China than elsewhere.**

Exploiting $P_{\rm rf} \propto \mathcal{L}/R$, some estimated costs (in arbitrary cost units) and luminosities for Stage I and (Higgs Factory)Stage-II are given in Table 4.

The cost ratios in this table were originally extracted from the LEP "Pink Book" [8]. They have now been more reliably up-dated using values from the CEPC Pre-CDR design report [9]. The most significant finding is that doubling the circumference while cutting the power in half increases the cost by a factor of 1.4.

Being quite weak ferromagnets, the bending magnet costs could, in principle, be proportional to stored magnetic energy. For this assumption to be at all realistic the magnets have to be constructed *in situ* in the tunnel, in order to eliminate transportation and installation costs. I am confident that sophisticated engineering can accomplish this.

The luminosity estimates are from Table 8 and are explained in later sections, especially Section 4"Lattice Optimization for Top-Off Injection".

**Note that doubling the radius, while cutting the power in half, increases the cost only modestly, while leaving generous options for upgrading to maximize Higgs luminosity, as well as maximizing the potential p,p physics reach.** The shaded row in Table 4 seems like the best deal. Both Higgs factory and, later, p,p luminosities are maximized, and the initial cost is (almost) minimized. Of course this optimization has been restricted to a simple choice between 50 km and 100 km circumference.

|  | $R$ | $P_{\rm rf}$ | $C_{\rm tun}$ | $C_{\rm acc}$ | Phase-I cost | $\mathcal{L}^I$ (Higgs) | $\mathcal{L}^I$ ($Z_0$) |
|---|---|---|---|---|---|---|---|
|  | km | MW | arb. | arb. | arb. | $10^{34}$ | $10^{34}$ |
| 1 ring | 5 | 50 | 0.5 | 2.5 | 3.0 | 1.2 | 2.6 |
|  | 10 | 25 | 1.0 | 2.87* | **3.87** | 1.2 | 5.2 |
|  | 10 | 50 | 1.0 | 3.58 | 4.58 | 2.3 | 10.4 |
| 2 rings | 5 | 50 | 0.5 | 4.1† | 4.6 | 1.2 | 21 |
|  | 10 | 25 | 1.0 | 4.72 | 5.72 | 1.2 | 21 |
|  | 10 | 50 | 1.0 | 5.89 | 6.89 | 2.3 | 42 |

Table 4: Estimated costs, one ring in the upper table, two in the lower. $C_{\rm tun}$ is the tunnel cost, $C_{\rm acc}$ is the cost of the rest of the accelerator complex. Costs have been extrapolated from the CEPC pre-CDR proposal. *With one ring, changes $R \to 2R$ and $P \to P/2$ are estimated to increase the accelerator cost by a factor 1.15. †Changing from one ring to two rings with $R$ and $P$ held fixed is estimated to increase the cost by a factor 1.64. The ratio in the table, 3.87/2.87=1.35, is the cost ratio of doubling the ring while cutting the RF power in half. From an authoritative CEPC source, this ratio is more reliably calculated to be 1.40.

## 2.6 Luminosity Limiting Phenomena

**Saturated Tune Shift.** My electron/positron beam-beam simulation [5] dead reckons the saturation tune shift $\xi_{\rm max}$ which is closely connected to the maximum luminosity. For an assumed $R \propto E^{5/4}$ tunnel circumference scaling, $\xi_{\rm max}$ is plotted as a function of machine energy $E$ in Figure 4. This plot assumes that the r.m.s. bunchlength $\sigma_z$ is equal to $\beta_y^*$, the vertical beta function at the intersection point (IP).

The physics of the simulation assumes there is an equilibrium established between beam-beam heating versus radiation cooling of vertical betatron oscillations. Under ideal single beam conditions the beam height would be $\sigma_y \approx 0$. This would give infinite luminosity in colliding beam operation —*but this is unphysical*. In fact beam-beam forces cause the beam height to grow into a new equilibrium with normal radiation damping. It is parametric modulation of the vertical beam-beam force by horizontal betatron and longitudinal synchrotron oscillation that modulates the vertical force and increases the beam height. The resonance driving strength for this class of resonance is proportional to $1/\sigma_y$ and would be infinite if $\sigma_y=0$—*this too is unphysical*. Nature, "abhoring" both zero and infinity, plays off beam-beam emittance growth against radiation damping. However amplitude-dependent detuning limits the growth, so there is only vertical beam growth but no particle loss (at least from this mechanism). In equilibrium the beam height is proportional to the bunch charge. The simulation automatically accounts for whatever resonances are nearby.

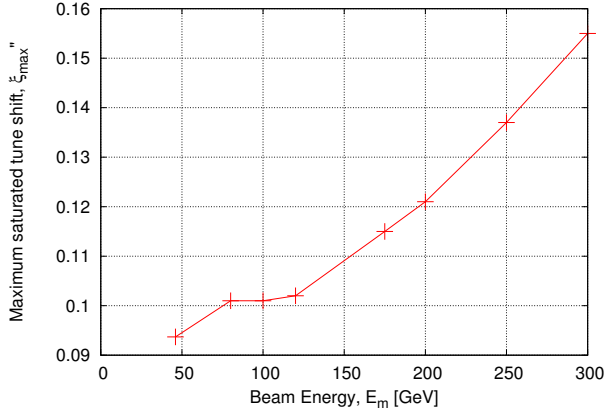

Figure 4: Plot of maximum tune shift $\xi_{\max}$ as a function of maximum beam energy for rings such that $E \propto R^{5/4}$. The non-smoothness has to be blamed on statistical fluctuations in the Monte Carlo program calculation. The maximum achieved tune shift parameter 0.09 at 100 GeV at LEP was less than shown, but their torturous injection and energy ramping seriously constrained their operations.

To estimate Higgs factory luminosity the tune plane is scanned for various vertical beta function values and bunch lengths, as well as other, less influential, parameters. The resulting ratio $(\xi^{\text{sat}}/\beta_y^*)$ is plotted in Figure 5. The ratio $\xi^{\text{sat.}}/\beta_y^*$ determines the beam area $A_{\beta_y}$ just sufficient for vertical saturation according to the formula, (see Eq. (10) for this and other of the following formulas),

$$A_{\beta_y} = \pi \sigma_x \sigma_y = \frac{N_p r_e}{2\gamma} \frac{1}{(\xi^{\text{sat.}}/\beta_y)}. \qquad (10)$$

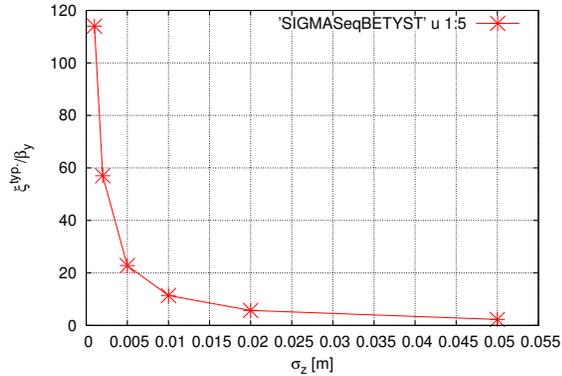

Figure 5: Plot of $\xi^{\text{typ.}}/\beta_y$, the "typical" tune shift value $\xi^{\text{typ.}}$ inversely weighted by $\beta_y$, as a function of $\sigma_z$, with $\beta_y = \sigma_z$, $\delta = 0.00764$, and synchrotron tune advance between collisions $Q_s = 0.0075$. Points from this plot can be substituted into Eq. (10) to obtain the transverse beam area $A$ just small enough for the tune shift to be saturated with the available number of electrons $N_p$ in each beam bunch.

This fixes the tune-shift-saturated charge density (per unit transverse area). It is only the product $\sigma_x \sigma_y$ that is fixed but there is a broad optimum in luminosity for aspect ratio $a_{xy} = \sigma_x/\sigma_y \approx 15$. To within this ambiguity all transverse betatron parameters are then fixed. $\beta_x^*$ is adjusted to make horizontal and vertical beam-beam tune shifts approximately equal. The lattice optics is adjusted so that the (arc-dominated) emittance $\epsilon_x$ gives the intended aspect ratio $a_{xy}$; $\epsilon_x = \sigma_x^2/\beta_x^*$.

(Incidentally, it will not necessarily be easy to optimize $\epsilon_x$ for each beam energy. Section "Lattice Optimization for Top-Off Injection" discusses tailoring cell length $L_c$ to adjust $\epsilon_x$. Unfortunately other considerations influence the choice of $L_c$ and, in any case, once optimized for one energy, $L_c$ remains fixed at all energies.)

**Beamstrahlung.** "Beamstrahlung" is the same as synchrotron radiation, except that it occurs when a particle in one beam is deflected by the electric and magnetic fields of the other beam. The emission of sychrotron radiation x-rays is inevitable and the lost energy has to be paid for. Much worse is the occasional radiation of a single photon (or, by chance, the sum of two) of sufficiently high energy that the reduction in momentum causes the particle itself to be lost. This magnifies the energy loss by the ratio of the x-ray energy lost to the energy of the circulating electron by some two orders of magnitude. It is this process that makes beamstrahlung so damaging. It contributes directly to the so-called "interaction lifetime". The damage is quantified by the beamstrahlung-dominated beam lifetime $\tau_{\text{bs}}$.

The important parameter governing beamstrahlung is the "critical energy" $u_c^*$ which is proportional to 1/bunch-length $\sigma_z$; beamstrahlung particle loss increases exponentially with $u_c^*$. To decrease beamstrahlung by increasing $\sigma_z$ also entails increasing $\beta_y^*$ which reduces luminosity. A favorable compromise can be to increase charge per bunch along with $\beta_y^*$.

**Reconciling the Luminosity Limits.** The number of electrons per bunch $N_p$ is itself fixed by the available RF power and the number of bunches $N_b$. For increasing the luminosity $N_b$ needs to be *reduced*. To keep beamstrahlung acceptably small $N_b$ needs to be *increased*. The maximum achievable luminosity is determined by this compromise between beamstrahlung and available power.

Three limiting luminosities can be defined: $\mathcal{L}_{\text{pow}}^{\text{RF}}$ is the RF power limited luminosity (introduced earlier to analyse constant luminosity scaling); $\mathcal{L}_{\text{sat}}^{\text{bb}}$ is the beam-beam saturated luminosity; $\mathcal{L}_{\text{trans}}^{\text{bs}}$ is the beamstrahlung-limited luminosity. Single beam dynamics gives $\sigma_y = 0$ which implies $\mathcal{L}_{\text{pow}}^{\text{RF}} = \infty$? Nonsense. Recalling the earlier discussion, the resonance driving force, being proportional to $1/\sigma_y$ would also be infinite. As a result the beam-beam force expands $\sigma_y = 0$ as necessary. *Saturation is automatic* (unless the single beam emittance is already too great for the beam-beam force to take control—it seems this condition was just barely satisfied in highest energy LEP operation [10]). Formulas for the luminosity limits are:

$$\mathcal{L}_{\text{pow}}^{\text{RF}} = \frac{N^*}{N_b} H(r_{yz}) \frac{1}{a_{xy}} \frac{f}{4\pi} \left(\frac{n_1 P_{\text{rf}}[\text{MW}]}{\sigma_y}\right)^2, \quad (11)$$

$$N_{\text{tot}} = n_1 P_{\text{rf}}[\text{MW}], \quad (12)$$

$$A_{\beta_y} = \pi \sigma_x \sigma_y = \frac{N_p r_e}{2\gamma} \frac{1}{(\xi^{\text{sat.}}/\beta_y)} = \pi \sigma_x \sigma_y, \quad (13)$$

$$\mathcal{L}_{\text{sat}}^{\text{bb}} = N^* N_b N_p H(r_{yz}) f \frac{\gamma}{2r_e} (\xi^{\text{sat.}}/\beta_y), \quad (14)$$

$$\mathcal{L}_{\text{trans}}^{\text{bs}} = N^* N_b H(r_{yz}) a_{xy} \sigma_z^2 f \left(\frac{\sqrt{\pi}}{28.0\,\text{m}} \frac{1.96 \times 10^5}{\sqrt{2/\pi}}\right)^2$$
$$\times \frac{1}{r_e^2 \widetilde{E}^2} \left(\frac{91\eta}{\ln\left(\frac{1/\tau_{\text{bs}}}{f n_{\gamma,1}^* \mathcal{R}_{\text{unif.}}^{\text{Gauss}}}\right)}\right)^2, \quad (15)$$

$$N_b = \sqrt{\frac{\mathcal{L}_{\text{sat}}^{\text{bb}}}{\mathcal{L}_{\text{trans}}^{\text{bs}}}}. \quad (16)$$

Here $H(r_{yz})$ is the hourglass reduction factor. If $\mathcal{L}_{\text{trans}}^{\text{bs}} < \mathcal{L}_{\text{sat}}^{\text{bb}}$ we must increase $N_b$. But $\mathcal{L}_{\text{trans}}^{\text{bs}} \propto N_b$, and $\mathcal{L}_{\text{pow}}^{\text{RF}} \propto 1/N_b$. We accept the better of the compromises $N_{b,\text{new}}/N_{b,\text{old}} = \mathcal{L}_{\text{sat}}^{\text{bb}}/\mathcal{L}_{\text{trans}}^{\text{bs}}$ or $N_{b,\text{new}}/N_{b,\text{old}} = \sqrt{\mathcal{L}_{\text{sat}}^{\text{bb}}/\mathcal{L}_{\text{trans}}^{\text{bs}}}$ as good enough.

Parameter tables, scaled up from LEP, are given for 100 km circumference Higgs factories in Tables 6 and 8. The former of these tables assume the number of bunches $N_b$ is unlimited. The latter table derates the luminosity under the assumtion that $N_b$ cannot exceed 200. Discussion of the one ring vs two rings issue can therefore be based on Table 8.

Some parameters not given in tables are: Optimistic=1.5 (a shameless excuse for actual optimatization), $\eta_{\text{Telnov}}$=0.01 (lattice fractional energy acceptance), $\tau_{\text{bs}}$=600 s, $R_{\text{GauUnif}}$= 0.300, $P_{rf}$ = 25 MW, Over Voltage=20 GeV, aspect ratio $a_{xy}$=15, $r_{yz} = \beta_y^*/\sigma_z$=1, and $\beta_{\text{arc max}}$=198.2 m.

With the exception of the final table, which is specific to the single ring option, the following tables apply equally to single ring or dual ring Higgs factories. The exception relates to $N_b$, the number of bunches in each beam. With $N_b$ unlimited (as would be the case with two rings) all parameters are the same for one or two rings (at least according to the formulas in this paper).

### 2.7 One Ring or Two Rings?

With one ring, the maximum number of bunches is limited to approximately ≤ 200. (I have not studied crossing angle schemes which may permit this number to be increased.) For $N_b > 200$ the luminosity $\mathcal{L}$ has to be de-rated accordingly; $\mathcal{L} \to \mathcal{L}_{\text{actual}} = \mathcal{L} \times 200/N_b$. This correction is applied in Table 8. This table, whose entries are simply drawn from Table 6, makes it easy to choose between one and two rings. Entries in this table have been copied into the earlier Table 4. When the optimal number of bunches is less than (roughly) 200, single ring operation is satisfactory, and hence favored. When the optimal number of bunches is much greater than 200, for example at the $Z_0$ energy, two rings are better.

Note though, that the $Z_0$ single ring luminosities are still very healthy. In fact, with $\beta_y^*$=10 mm, which is a more conservative estimate than most others in this paper and in other FCC reports, the $Z_0$ single ring penalty is substantially less.

Luminosities and optimal numbers of bunches in a second generation scaled-up-luminosity Higgs factory running are shown in Figure 6.

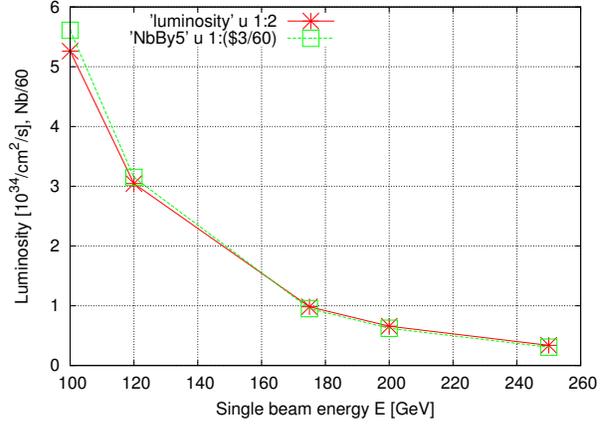

Figure 6: Dependence of luminosity on single beam energy (after upgrade to Stage II luminosity). The number of bunches (axis label to be read as $N_b/60$) is also shown, confirming that (as long as the optimal value of $N_b$ is 1 or greater) the luminosity is proportional to the number of bunches. There is useful luminosity up to $E = 500$ GeV CM energy.

### 2.8 Predicted Luminosities

With one 100 km circumference ring, the maximum number of bunches is limited to about 200. For $N_b < 200$ the luminosity $\mathcal{L}$ has to be reduced proportionally. $\mathcal{L} \to \mathcal{L}_{\text{actual}} = \mathcal{L} \times N_b/200$. Luminosities in the 100 km, 25 MW case are given in Section "Ring Circumference and Two Rings vs One Ring". Here, for comparison, and to more nearly match the separation scheme shown in Figure 11, the circumference is assumed to be $C$=50 km, the RF power 50 MW per beam, and the number of bunches $N_b$=112. The results are shown in Table 7 (unlimited $N_b$) and Table 9 (with $N_b$=112).

The values of parameters not shown in the tables are $\eta_{\text{Telnov}}$=0.01, $\beta_y^*$ =5 mm, $xi^{\text{typ.}}/\beta_y^*$=22.8, $\tau_{\text{bs}}$=600 s, Optimistic= 1.5, $R_{\text{Gau-unif}}$=0.30, $eV_{\text{rf}}$=20 GeV, $OV_{\text{req.}}$=20 GV, $a_{xy}$=15, $r_{yz}$=1, $\beta_{x,\text{arcmax}}$=120 m.

### 2.9 Reconciling the Luminosity Formulas

Several formulas have been given for the luminosity. The luminosity actually predicted is the smallest of the entries intries in the three luminosity columns, for example in

| Parameter | Symbol | Value | Unit | Energy-scaled | Radius- | scaled |
|---|---|---|---|---|---|---|
| bend radius | $R$ | **3026** | m | **3026** | 5675 | 11350 |
| | $R/3026$ | | | 1 | 1.875 | 3.751 |
| Beam Energy | $E$ | **45.6/91.5** | GeV | **120** | **120** | **120** |
| Circumference | $C$ | 26.66 | km | 26.66 | 50 | 100 |
| Cell length | $L_c$ | | m | 79 | 108 | 153 |
| Momentum compaction | $\alpha_c$ | 1.85e-4 | | 1.85e-4 | 0.99e-4 | 0.49e-4 |
| Tunes | $Q_x$ | 90.26 | | 90.26 | 123.26 | 174.26 |
| | $Q_y$ | 76.19 | | 76.19 | 104.19 | 147.19 |
| Partition numbers | $J_x/J_y/J_\epsilon$ | 1/1/2 | | 1/1.6/1.4 ! | 1/1/2 | 1/1/2 |
| Main bend field | $B_0$ | 0.05/0.101 | T | 0.1316 | 0.0702 | 0.0351 |
| Energy loss per turn | $U_0$ | 0.134/2.05 | GeV | 6.49 | 3.46 | 1.73 |
| Radial damping time | $\tau_x$ | 0.06/0.005 | s | 0.0033 | 0.0061 | 0.0124 |
| | $\tau_x/T_0$ | 679/56 | turns | 37 | 69 | 139 |
| Fractional energy spread | $\sigma_\delta$ | 0.946e-3/1.72e-3 | | 0.0025 | 0.0018 | 0.0013 |
| Emittances (no BB), x | $\epsilon_x$ | 22.5/30 | nm | 21.1 | 8.2 | 2.9 |
| y | $\epsilon_y$ | 0.29/0.26 | nm | 1.0 | 0.4 | 0.14 |
| Max. arc beta funcs | $\beta_x^{max}$ | 125 | m | 125 | 171 | 242 |
| Max. arc dispersion | $D^{max}$ | 0.5 | m | 0.5 | 0.5 | 0.5 |
| Beta functions at IP | $\beta_x^*, \beta_y^*$ | 2.0,0.05 | m | 1.25/0.04 | N/Sc. | N/Sc. |
| Beam sizes at IP | $\sigma_x^*, \sigma_y^*$ | 211, 3.8 | $\mu$m | 178/11 | N/Sc. | N/Sc. |
| Beam-beam parameters | $\xi_x, \xi_y$ | 0.037,0.042 | | 0.06/0.083 | N/Sc. | N/Sc. |
| Number of bunches | $N_b$ | 8 | | 4 | N/Sc. | N/Sc. |
| Luminosity | $\mathcal{L}$ | 2e31 | cm$^{-2}$s$^{-1}$ | 1.0e32 | N/Sc. | N/Sc. |
| Peak RF voltage | $V_{RF}$ | 380 | MV | 3500 | N/Sc. | N/Sc. |
| Synchrotron tune | $Q_s$ | 0.085/0.107 | | 0.15 | N/Sc. | N/Sc. |
| Low curr. bunch length | $\sigma_z$ | 0.88 | cm | $\frac{\alpha_c R \sigma_e}{Q_s E}$ | N/Sc. | N/Sc. |

Table 5: Higgs factory parameter values for 50 km and 100 km options. The entries are mainly extrapolated from Jowett's, 45.6 Gev report [12] [13], and educated guesses. "N/Sc." indicates (important) parameters too complicated to be estimated by scaling. Duplicate entries in the third column, such as 45.6/91.5 are from Jowett [12]; subsequent scalings are based on the 45.6 Gev values.

Table 6. For the middle shaded row the lowest value is $\mathcal{L} = 1.14 \times 10^{34}$ cm$^{-2}$s$^{-1}$.

Under ideal single beam conditions, the beam height $\sigma_y$ is vanishingly small and Eq. (11) predicts infinite luminosity, even for arbitrarily small RF power. Of course this is nonsense; nature "abhors" both zero and infinity. In fact, when in collision, the beam-beam force causes $\sigma_y$ to grow (as the simulation model assumes). In the current context this implies that it is *always* possible to saturate the tune shift operationally. In this circumstance Eq. (13) is applicable, and gives the beam area $A_{\beta_y}$ small enough for the tune shift to be saturated with the available number of electrons, which is given by Eq. (13). Tentatively we assume $N_b = 1$ and, therefore, $N_p = N_{tot}$. Then

$$\sigma_y = \sqrt{\frac{A_{\beta_y}}{\pi a_{xy}}}, \quad \text{and} \quad \sigma_x = a_{xy}\sigma_y. \quad (17)$$

With the beam aspect ratio $a_{xy}$ being treated as if known, this permits the bunch height and width to be determined. But this determination is only preliminiary since the number of bunches $N_b$ is not yet fixed. Then, for a tentatively adopted value of bunch length $\sigma_z$, with $(\xi^{sat.}/\beta_y)$ read from Figure 5, Eq. (14) gives the predicted luminosity with all the beam in one bunch.

But this has neglected the beamstrahlung limitation; Eq. (15) gives the maximum luminosity allowed by beamstrahlung. (Factors have not been collected in this embarrassingly-cluttered formula so they can be traced from earlier formulas.) This beamstrahlung-limited luminosity will usually be less than the beam-power limited luminosity. The only recourse in this case is to split the beam into $N_b$ bunches. Changing $N_b$ does not change $\mathcal{L}_{sat}^{bb}$, because $N_b N_p$ is fixed, but it increases $\mathcal{L}_{trans}^{bs}$, and it decreases $\mathcal{L}_{pow}^{RF}$ by the same factor. Unfortunately, not yet definitively knowing $\sigma_y$, we cannot yet reckon the optimal value of $N_b$. As a compromise we use the square-rooted ratio in Eq. (16) to fix $N_b$. This increases $\mathcal{L}_{trans}^{bs}$ and decreases $\mathcal{L}_{pow}^{RF}$ by the same factor (assuming $N_b > 1$).

A more agressive approach is to replace Eq. (16) by $N_b = \mathcal{L}_{sat}^{bb}/\mathcal{L}_{trans}^{bs}$. This is justifiable, since $\mathcal{L}_{sat}^{bb}$ depends only on $N_{tot.}$ and is unaffected by changing $N_b$ This has not been done for the tables since it leads to unacchievably large values of $N_b$ at low beam energy. It would, however, give luminosity more than twice as great in some cases.

For $N^* = 1$ or $N^* = 2$, if Eq. (16) gives a value of $N_b$ less than 1 it means that $N_b = 1$ is optimal. For $N^* = 4$, if Eq. (16) gives a value of $N_b$ less than 2 it means that $N_b = 2$

| name | $E$ | $\epsilon_x$ | $\beta_y^*$ | $\epsilon_y$ | $\xi_{sat}$ | $N_{tot}$ | $\sigma_y$ | $\sigma_x$ | $u_c^*$ | $n_{\gamma,1}^*$ | $\mathcal{L}^{RF}$ | $\mathcal{L}_{trans}^{bs}$ | $\mathcal{L}^{bb}$ | $N_b$ | $\beta_x^*$ | $P_{rf}$ |
|---|---|---|---|---|---|---|---|---|---|---|---|---|---|---|---|---|
| | GeV | nm | mm | pm | | $10^{12}$ | $\mu$m | $\mu$m | GeV | | $10^{34}$ | $10^{34}$ | $10^{34}$ | | m | MW |
| Z | 46 | 0.949 | 2 | 63.3 | 0.094 | 1500 | 0.356 | 5.34 | 0.000 | 2.01 | 52.5 | 103 | 52.5 | 65243 | 0.03 | 25 |
| W | 80 | 0.336 | 2 | 22.4 | 0.101 | 150 | 0.212 | 3.17 | 0.001 | 2.10 | 9.66 | 17.2 | 9.6 | 10980 | 0.03 | 25 |
| LEP | 100 | 0.223 | 2 | 14.9 | 0.101 | 62 | 0.172 | 2.59 | 0.002 | 2.13 | 4.95 | 8.46 | 4.94 | 5421 | 0.03 | 25 |
| H | 120 | 0.159 | 2 | 10.6 | 0.102 | 30 | 0.146 | 2.19 | 0.003 | 2.17 | 2.86 | 4.74 | 2.86 | 3044 | 0.03 | 25 |
| tt | 175 | 0.078 | 2 | 5.33 | 0.118 | 6.6 | 0.103 | 1.55 | 0.006 | 2.24 | 0.923 | 1.43 | 0.92 | 920 | 0.03 | 25 |
| Z | 46 | 17.2 | 5 | 1140 | 0.094 | 1500 | 2.39 | 35.89 | 0.001 | 2.16 | 21 | 35.1 | 21. | 3605 | 0.075 | 25 |
| W | 80 | 6.11 | 5 | 408 | 0.101 | 150 | 1.43 | 21.42 | 0.003 | 2.26 | 3.86 | 5.83 | 3.86 | 602 | 0.075 | 25 |
| LEP | 100 | 4.07 | 5 | 271 | 0.101 | 62 | 1.16 | 17.47 | 0.005 | 2.31 | 1.98 | 2.86 | 1.97 | 296 | 0.075 | 25 |
| H | 120 | 2.92 | 5 | 195 | 0.102 | 30 | 0.987 | 14.80 | 0.008 | 2.35 | 1.15 | 1.6 | 1.14 | 166 | 0.075 | 25 |
| tt | 175 | 1.47 | 5 | 98.1 | 0.118 | 6.6 | 0.7 | 10.51 | 0.017 | 2.43 | 0.369 | 0.479 | 0.37 | 49 | 0.075 | 25 |
| Z | 46 | 155 | 10 | 10300 | 0.094 | 1500 | 10.2 | 152.3 | 0.002 | 2.29 | 10.5 | 15.5 | 10.5 | 400 | 0.15 | 25 |
| W | 80 | 55.4 | 10 | 3690 | 0.101 | 150 | 6.08 | 91.17 | 0.007 | 2.41 | 1.93 | 2.55 | 1.93 | 66 | 0.15 | 25 |
| LEP | 100 | 37.0 | 10 | 2470 | 0.101 | 62 | 4.97 | 74.48 | 0.011 | 2.46 | 0.989 | 1.25 | 0.99 | 32 | 0.15 | 25 |
| H | 120 | 26.6 | 10 | 1770 | 0.102 | 30 | 4.21 | 63.15 | 0.016 | 2.50 | 0.573 | 0.696 | 0.57 | 18.3 | 0.15 | 25 |
| tt | 175 | 13.5 | 10 | 898 | 0.118 | 6.6 | 3.0 | 44.94 | 0.036 | 2.60 | 0.185 | 0.207 | 0.19 | 5.5 | 0.15 | 25 |

Table 6: The major factors influencing luminosity, assuming 100 km circumference and 25 MW/beam RF power. The predicted luminosity is the smallest of the three luminosities, $\mathcal{L}^{RF}$, $\mathcal{L}_{trans}^{bs}$, and $\mathcal{L}^{bb}$. All entries in this table apply to either one ring or two rings, except where the number of bunches $N_b$ is too great for a single ring.

| name | $E$ | $\epsilon_x$ | $\beta_y^*$ | $\epsilon_y$ | $\xi_{sat}$ | $N_{tot}$ | $\sigma_y$ | $\sigma_x$ | $u_c^*$ | $n_{\gamma,1}^*$ | $\mathcal{L}^{RF}$ | $\mathcal{L}_{trans}^{bs}$ | $\mathcal{L}^{bb}$ | $N_b$ | $\beta_x^*$ | $P_{rf}$ |
|---|---|---|---|---|---|---|---|---|---|---|---|---|---|---|---|---|
| | GeV | nm | mm | pm | | | $\mu$m | $\mu$m | GeV | | $10^{34}$ | $10^{34}$ | $10^{34}$ | | m | MW |
| Z | 46 | 0.916 | 2 | 61.1 | 0.094 | 7.3e+14 | 0.35 | 5.24 | 0.000 | 1.97 | 52.5 | 96.8 | 52.513 | 33795 | 0.03 | 50 |
| W | 80 | 0.323 | 2 | 21.6 | 0.101 | 7.6e+13 | 0.208 | 3.12 | 0.001 | 2.06 | 9.66 | 16.2 | 9.661 | 5696 | 0.03 | 50 |
| LEP | 100 | 0.215 | 2 | 14.3 | 0.101 | 3.1e+13 | 0.169 | 2.54 | 0.002 | 2.10 | 4.95 | 8 | 4.947 | 2814 | 0.03 | 50 |
| H | 120 | 0.153 | 2 | 10.2 | 0.102 | 1.5e+13 | 0.143 | 2.15 | 0.003 | 2.13 | 2.86 | 4.48 | 2.863 | 1581 | 0.03 | 50 |
| tt | 175 | 0.077 | 2 | 5.12 | 0.118 | 3.3e+12 | 0.101 | 1.52 | 0.006 | 2.19 | 0.923 | 1.35 | 0.923 | 478 | 0.03 | 50 |
| Z | 46 | 16.5 | 5 | 1100 | 0.094 | 7.3e+14 | 2.35 | 35.21 | 0.001 | 2.12 | 21 | 33.2 | 21.005 | 1872 | 0.075 | 50 |
| W | 80 | 5.88 | 5 | 392 | 0.101 | 7.6e+13 | 1.4 | 20.99 | 0.003 | 2.22 | 3.86 | 5.52 | 3.864 | 313 | 0.075 | 50 |
| LEP | 100 | 3.91 | 5 | 261 | 0.101 | 3.1e+13 | 1.14 | 17.12 | 0.005 | 2.26 | 1.98 | 2.71 | 1.979 | 154 | 0.075 | 50 |
| H | 120 | 2.80 | 5 | 187 | 0.102 | 1.5e+13 | 0.966 | 14.50 | 0.007 | 2.30 | 1.15 | 1.52 | 1.145 | 86 | 0.075 | 50 |
| tt | 175 | 1.41 | 5 | 94 | 0.118 | 3.3e+12 | 0.686 | 10.28 | 0.016 | 2.38 | 0.369 | 0.455 | 0.369 | 26 | 0.075 | 50 |
| Z | 46 | 149 | 10 | 9900 | 0.094 | 7.3e+14 | 9.95 | 149.28 | 0.002 | 2.24 | 10.5 | 14.7 | 10.503 | 208 | 0.15 | 50 |
| W | 80 | 53.1 | 10 | 3540 | 0.101 | 7.6e+13 | 5.95 | 89.26 | 0.007 | 2.36 | 1.93 | 2.42 | 1.932 | 34 | 0.15 | 50 |
| LEP | 100 | 35.4 | 10 | 2360 | 0.101 | 3.1e+13 | 4.86 | 72.88 | 0.011 | 2.41 | 0.989 | 1.19 | 0.989 | 17 | 0.15 | 50 |
| H | 120 | 25.4 | 10 | 1700 | 0.102 | 1.5e+13 | 4.12 | 61.78 | 0.016 | 2.45 | 0.573 | 0.663 | 0.573 | 9.5 | 0.15 | 50 |
| tt | 175 | 12.9 | 10 | 857 | 0.118 | 3.3e+12 | 2.93 | 43.92 | 0.035 | 2.54 | 0.185 | 0.198 | 0.185 | 2.9 | 0.15 | 50 |

Table 7: Luminosity influencing parameters and luminosities with unlimited number of bunches $N_b$, assuming 50 km circumference ring and 50 MW per beam RF power.

| $E$ | $\beta_y^*$ | $\xi_{sat}$ | $\mathcal{L}_{actual}$ | $N_{b,actual}$ | $P_{rf}$ |
|---|---|---|---|---|---|
| GeV | m | | $10^{34}$ | | MW/beam |
| 46 | 0.002 | 0.094 | 0.161 | 200 | 25 |
| 80 | 0.002 | 0.1 | 0.176 | 200 | 25 |
| 100 | 0.002 | 0.1 | 0.182 | 200 | 25 |
| 120 | 0.002 | 0.1 | 0.188 | 200 | 25 |
| 175 | 0.002 | 0.12 | 0.200 | 200 | 25 |
| 46 | 0.005 | 0.094 | 1.165 | 200 | 25 |
| 80 | 0.005 | 0.1 | 1.282 | 200 | 25 |
| 100 | 0.005 | 0.1 | 1.334 | 200 | 25 |
| 120 | 0.005 | 0.1 | 1.145 | 166 | 25 |
| 175 | 0.005 | 0.12 | 0.369 | 50 | 25 |
| 46 | 0.010 | 0.094 | 5.247 | 200 | 25 |
| 80 | 0.010 | 0.1 | 1.932 | 66.5 | 25 |
| 100 | 0.010 | 0.1 | 0.989 | 32.7 | 25 |
| 120 | 0.010 | 0.1 | 0.573 | 18.3 | 25 |
| 175 | 0.010 | 0.12 | 0.185 | 5.5 | 25 |

Table 8: Luminosites achievable with a single ring for which the number of bunches $N_b$ is limited to 200, assuming 100 km circumference and 25 MW/beam RF power. Entries in this table have been distilled down to include only the most important entries in Table 6, as corrected for the restricted number of bunches. The luminosity entries in Table 4 have been obtained from this table.

is optimal, since $N_b = 2$ is the minimum number of bunches with 4 collision points.

| $E$ | $\beta_y^*$ | $\xi_{sat}$ | $\mathcal{L}_{actual}$ | $N_{b,actual}$ | $P_{rf}$ |
|---|---|---|---|---|---|
| GeV | m | | $10^{34}$ | | MW |
| 46 | 0.002 | 0.094 | 0.174 | 112 | 50 |
| 80 | 0.002 | 0.1 | 0.190 | 112 | 50 |
| 100 | 0.002 | 0.1 | 0.197 | 112 | 50 |
| 120 | 0.002 | 0.1 | 0.203 | 112 | 50 |
| 175 | 0.002 | 0.12 | 0.216 | 112 | 50 |
| 46 | 0.005 | 0.094 | 1.256 | 112 | 50 |
| 80 | 0.005 | 0.1 | 1.380 | 112 | 50 |
| 100 | 0.005 | 0.1 | 1.434 | 112 | 50 |
| 120 | 0.005 | 0.1 | 1.145 | 86.6 | 50 |
| 175 | 0.005 | 0.12 | 0.369 | 26.1 | 50 |
| 46 | 0.010 | 0.094 | 5.644 | 112.0 | 50 |
| 80 | 0.010 | 0.1 | 1.932 | 34.7 | 50 |
| 100 | 0.010 | 0.1 | 0.989 | 17.1 | 50 |
| 120 | 0.010 | 0.1 | 0.573 | 9.5 | 50 |
| 175 | 0.010 | 0.12 | 0.185 | 2.9 | 50 |

Table 9: Luminosity influencing parameters and luminosities with the number of bunches limited to $N_b = 112$, assuming 50 km circumference ring and 50 MW per beam RF power.

Most of the beam parameters have been, or can now be, determined. In principle this optimization procedure can be iterated, but I have not attemped this. If all luminosity measures are comparable, the parameters are probably close to optimal.

## 2.10 Qualitative Comments

The nature of the required compromises can be inferred from the luminosity formulas. Except for beam-beam effects the beam height can, in principle, be arbitrarily small. (This assumes perfect decoupling and zero vertical dispersion. Both of these conditions are adequately achieved in low energy rings, but neither was persuasively achieved at LEP.) Because the Higgs factory will be operationally simpler than LEP, we assume that $\sigma_y$ will actually be driven by the beam-beam effect. In this case Eq. (11) implies that the tune shift can be saturated irrespective of the beam power. At too high energy this will surely cease to be applicable, and the formulas will have to be modified accordingly.

For now we assume that saturated operation is always possible. Then, except for possible beamstrahlung limitation, the luminosity given by Eq. (14) is, in principle, achievable, and the beam area $A_{\beta_y}$ is given by Eq. (13). (Note, though, that a graph such as Figure (18) assumes $\sigma_z = 0.01$ m which is not the result of any optimization. Also, repeating what has already been implied, the luminosity will be very small if the area $A_{\beta_y}$ has to be made excessively small to achieve saturation.)

Starting from some given parameter set, to increase luminosity by reducing beamstrahlung favors increasing $a_{xy}$ and $\sigma_z$. But increasing $a_{xy}$ reduces $\mathcal{L}_{\text{pow}}^{\text{RF}}$ and inreasing $\sigma_z$ reduces $\mathcal{L}_{\text{sat}}^{\text{bb}}$—it can be seen by multiplying by $(\beta_y/\sigma_z)/r_{yz} \equiv 1$. Based on a cursory preliminary investigation there seem to be broad optima roughly centered on $a_{xy} = 15$ and $r_{yz} = 0.6$ and I have adopted values close to these for all results in the present paper. Though disappointing that the luminosity cannot be "peaked up" using these parameters, a corollary is that they can be moved significantly without reducing the luminosity greatly. For example, increasing $r_{yz}$ by a factor as great as ten (which may very well be demanded by higher mode considerations) decreases the luminosity by less than a factor of three.

To compare predicted luminosities let us take, for example, the case with $R = 10$ km, $E = 250$ GeV, which is a higher energy than exhibited in tables of the present paper. I have found luminosity 0.11 compared, for comparable parameters, to Telnov's storage ring value of 0.087 (in the final column of Telnov's Table II [4], which is his most nearly comparable case).

A much-debated question (prior to the surprisingly low Higgs mass discovery) concerned the relative effectiveness of circular and linear colliders at very high beam energies. My optimized luminosity value of $1.0 \times 10^{34}$ cm$^{-2}$s$^{-1}$ (summed over 4 IP's) is comparable with the advertised ILC luminosity value of about 1.8 at 250 GeV [14]. This suggests that, as well as being based on routine technology, circular storage rings can remain competitive with linear colliders up to almost three times the LEP energy. [2]

Telnov's number of bunches per beam is $N_b = 31$; whereas my example has only $N_b = 2$. My $\beta_y = 6$ mm is twenty times greater (and hence easier) than Oide's [15] $\beta_y = 0.26$ mm, for the same beam energy; (see the final column of Telnov's Table I). Oide assumes 2 bunches per beam and obtains luminosity $0.024 \times 10^{34}$ m$^{-2}$s$^{-1}$, much less than my value.

For simplicity my RF acceleration has been patterned after the final-configuration LEP design, which had total voltage drop $V_{\text{rf}} = 3.63$ GV. My value of 65 GV is 18 times greater than this; I chose this value just big enough to avoid having all tabulated luminosity values at $E = 250$ GeV from vanishing. To actually run at $E = 300$ GeV would require the maximum RF voltage to be increased. Conversely, during a first phase of operation, a far smaller value would be adequate. During final LEP operation there were 44 klystrons, with each having power dissipation of roughly 1 MW. My RF power dissipation is only two or three times this great. This reflects the fact that I assume that, as in all previous circular electron rings, there is a substantial "overvoltage" compared to the average energy loss per turn.

An exhaustive investigation of the parameter space has not been attempted. However, within my model, it seems the parameters in the tables in this report are fairly close to optimal.

Fortunately most of the entries in the tables for the larger $R$ values show performance ranging from respectable to excellent. This is true in spite of the fact that (except for $eV_{\text{rf}}$ and $P_{\text{rf}}$) none of the parameters seem to be especially challenging. Even a ring only twice the circumference of LEP (except for possibly unachievably low $\beta_y$) could serve as a respectable Higgs factory. For various reasons a larger circumference would be more conservative short term, and have far greater long term potential.

Parameters in the tables have been "tuned" for the shaded row, $E = 250$ GeV, twice the energy for maximum Higgs production. The large range from minimum to maximum luminosities at $E = 125$ GeV imply that, after retuning, a luminosity significantly higher than $10^{34}$ cm$^{-1}$s$^{-1}$ will be possible, and the 200 per day Higgs particle production increased proportionally. This too improves with increased circumference.

Especially satisfying is the finding that bunch lengths $\sigma_z \approx 1$ cm will be appropriate (rather than the much shorter bunches some designs have assumed, with the large higher mode losses to be expected.) My bunch lengths are ten times greater than the LBNL/SLAC design [17] and yet my luminosity at 100 GeV in a LEP-sized ring is comparable. I also have a much smaller number of bunches $N_b$.

---

[2] A heuristically plausible mnemonic: circular colliders are superior as long as their energy loss per turn is an order of magnitude less than their full energy. At this point the number of bunches in the collider cannot be further reduced while still serving all of the intersection regions.

Also satisfying is that vertical emittance values less than $\epsilon_y \approx 0.1$ nm seem not to be required. This is comparable with values already achieved at LEP [18] [19] [11] under more challenging operating conditions. In discussions at the 2013 FNAL Higgs Factory Workshop it seemed to be taken for granted that Herculean efforts would be required to reduce $\epsilon_y$ to an acceptably small value. I find this not to be the case.

The intersection region optics seems to be manageable as well. In a presentation at the FNAL Higgs Factory Workshop in early 2013, Yunhai Cai [17] showed, for example, a final focus with $\beta_x = 50$ mm, $\beta_y = 1$ mm. These values are considerably more challenging than most of the configurations shown in tables in this paper, especially for the larger circumference rings.

### 2.11 Advantages of Vertical Injection and Bumper-Free, Kicker-Free, Top-Off Injection

I proposed kicker-free, septum free, vertical injection at Beijing in April 2014, and described it in paper SAT4A3, "Lattice Optimization for Top-Off Injection" at the 55th ICFA Advanced Beam Dynamics Workshop on High Luminosity Circular e+e- Colliders, in the WG 6 "Injection" working group for HF2014 October 11.

Handling the synchrotron radiation at a Higgs Factory is difficult and replenishing the power loss is expensive. Otherwise the RF power loss is purely beneficial, especially for injection. Betatron damping decrements $\delta$ (fractional amplitude loss per turn) are approximately half the energy loss per turn divided by the beam energy, (e.g. $\delta \approx 0.5 \times 2.96/120 = 1.25\%$.) Also the energy dependence is large enough for injection efficiency to improve significantly with increasing energy.

According to Liouville's theorem, increasing the beam particle density by injection is impossible for a Hamiltonian system. The damping decrement $\delta$ measures the degree to which the system is *not* Hamiltonian. Usually bumpers and kickers are needed to keep the already stored beam captured while the injected beam has time to damp. If $\delta$ is large enough one can, at least in principle, inject with no bumpers or kickers.

The most fundamental parameter limiting injection efficiency is the emittance of the injected beam. The vertical emittance in the booster accelerator can be very small, perhaps $\epsilon_y < 10^{-10}$ m. This may require a brief flat top at full energy in the booster. For injection purposes the beam height can then be taken to be effectively zero. The next most important injector parameter is the septum thickness. For horizontal injection this septum normally also has to carry the current to produce a horizontal deflection. Typically this requires the septum thickness to be at least 1 mm. For vertical injection, with angular deflection not necessarily required, the septum can be very thin, even zero. The remaining (and most important) injection uncertainty is whether the ring dynamic aperture extends out to the septum. If not, it may be possible to improve the situation by moving the closed orbit closer to the wall using DC bumpers. (However this may be disadvantageous for vertical injection as vertical bends contribute unwanted vertical emittance to the stored beams.)

A virtue of top-off injection is that, with beam currents always essentially constant, the linear part of the beam-beam tune shift can be designed into the linear lattice optics. One beam "looks", to a particle in the other beam, like a lens (focusing in both planes). Large octupole moments makes this lens far from ideal. Nevertheless, if the beam currents are constant the pure linear part can be subsumed into the linear lattice model. And the octupole component, though nonlinear, does not necessarily limit the achievable luminosity very severely.

With injection continuing during data collection there would be no need for cycling between injection mode and data collection mode. This could avoid the need for the always difficult "beta squeeze" in transitioning from injection mode to collision mode. Runs could then last for days, always at maximum luminosity. This would improve both average luminosity and data quality.

Kicker-free vertical injection is illustrated schematically in Figure 7. Let $n_{\text{inj.}}$ be a small integer indicating the number of turns following injection before the injected beam threatens to wipe out on the injection septum. The fractional shrinkage of the Courant-Snyder invariant after $n_{\text{inj.}}$ turns is $n_{\text{inj.}}.\delta$. By judicious choice of vertical, horizontal, and synchrotron tunes this shrinkage may be great enough that less than, say, 10 % of the beam is lost on the septum.

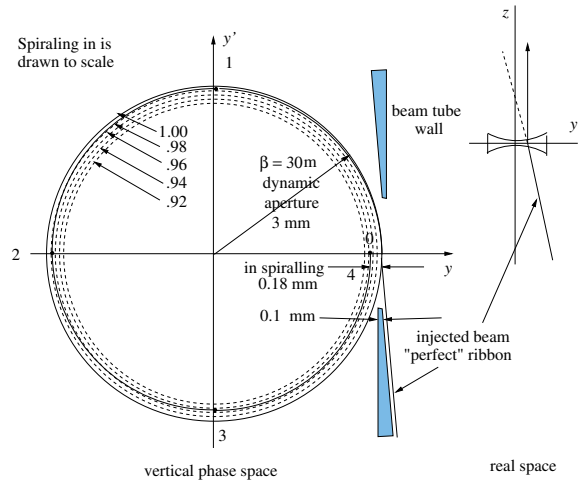

Figure 7: A cartoon of kicker-free, vertical injection. The dashed line shows the Courant-Snyder amplitude of the injected particle with the fractional shrinking per turn drawn more or less to scale.

# 3 SINGLE RING MULTIBUNCH OPERATION AND BEAM SEPARATION

As illustrated in Figure 8, the counter-circulating electrons and positrons in a circular Higgs Factory have to be separated everywhere except at the $N^*$ intersection points (IP). The separation has to be electric and, to avoid unwanted increase of vertical emittance $\epsilon_y$, the separation has to be horizontal. This section considers only head-on collisions at $N^* = 2$ IP's, with the beams separated everywhere else (but with nodes at RF cavities) by closed electric bumps.

## 3.1 Electric Bump Beam (Pretzel) Separation

**Operating Energies.** Typical energies for "Higgs Factory" operation are established by the cross sections shown in Figure 3. We arbitrarily choose 120 GeV per beam as the Higgs particle operating point and identify the single beam energy this way in subsequent tables. Similarly identified are the $Z_0$ energy (45.6 GeV), the W-pair energy of 80 GeV, the LEP energy (arbitrarily taken to be 100 GeV) and the $t\bar{t}$ energy of 175 GeV to represent high energy performance.

**Bunch Separation at LEP.** Much of the material in this section has been drawn from John Jowett's article "Beam Dynamics at LEP" [12]. When LEP was first commissioned for four bunches ($N_b$=4) and four IPs ($N^*$=4) operation, bunch collisions at the 45 degree points were avoided by vertical electric separation bumps. It was later realized that vertical bumps are inadvisable because of their undesirable effect on vertical emittance $\epsilon_y$, which needs to be minimized. We therefore consider only horizontal separation schemes.

Various horizontal pretzel separation schemes were tried at LEP. They were constrained by the need to be superimposed on an existing lattice. LEP investigations in the early 1990's mainly concentrated on what now would be called quite low energies, especially the $Z_0$ energy, $E = 45.6$ GeV. For a Higgs factory we need to plan for energies four or five times higher. The required product of separator length multiplied by electric separator field has to be greater by the same factor to obtain the same angular separation. Actually the factor may have to be somewhat greater than this because of the larger bunch separation needed with increased ring circumference.

Before continuing, allow me a brief digression concerning the etymology of the technical and metaphorical term "pretzel". The term was coined by Raphael Littauer, the inventor of the eventually workable pretzel beam separation scheme. The pretzel "idea" first came to Boyce "Mac" McDaniel, director of the Cornell Laboratory of Nuclear Studies at the time.

As first realized by McDaniel, instead of having closed bumps one can make do with a single separator. The effect of a single electric deflection is to make the closed orbits of the counter-circulating beams different *everywhere*. Even in this case there are periodic "nodes" at which the distorted orbits cross. To achieve the desired beam separation, one has only to arrange for the desired crossing points to be at nodes and the parasitic crossing points to be at "loops" of the respective closed orbits. Raphael Littauer introduced the picturesque term "pretzel" to distill the entire discussion of this paragraph into a single word.

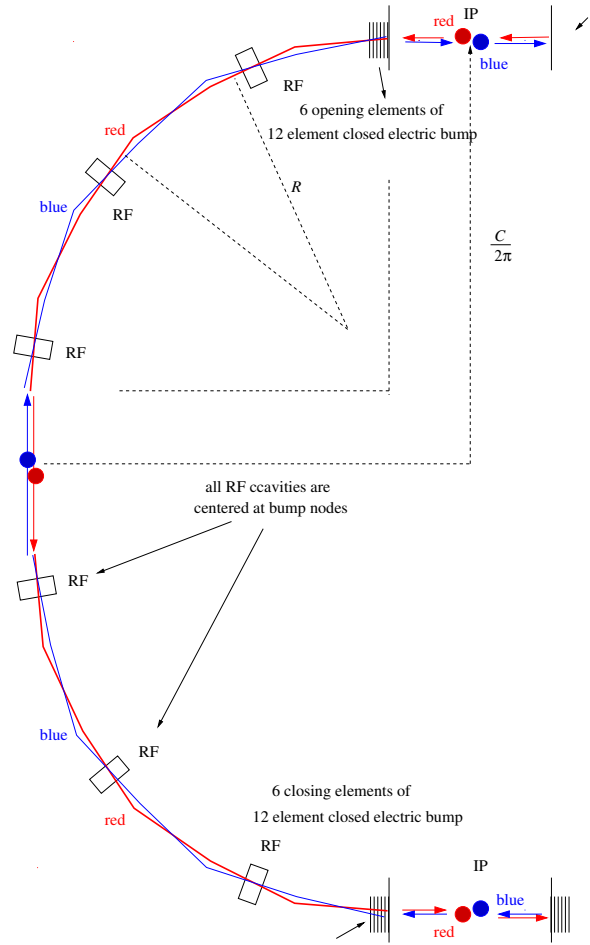

Figure 8: Cartoon illustrating beam separation in one arc of a Higgs factory. There are $N_b$=4 bunches in each beam and $N^*$=2 interaction points (IP). The bend radius $R$ is significantly less than the average radius $C/(2\pi)$; roughly $C = 3\pi R$. For scaling purposes $R$ and $C$ are taken to be strictly proportional. Far more separation loops and crossovers are actually needed than are shown.

The original pretzel conception was not good enough, however, since the crossing angle at the IP was damaging to the luminosity. It was Littauer who fleshed out the idea and led its successful implementation. It proved necessary to introduce four electric separators in matched pairs about the North and South IP's. This invalidated the original name "pretzel" since the scheme amounted to closed electric bump separation separately in the East and West arcs. Nevertheless, by that time the "pretzel" name had caught on and the scheme continued to be called pretzel separation in CESR and in all subsequent rings.

A disadvantage of the metaphorical terminology is that it conveys a picture of the whole ring being a single "pretzel",

obscuring the fact that the separation bumps are closed in each arc—two closed pretzels, if one prefers. To emphasize this point, for this paper only, I discuss closing electric multi-bumps, arc-by-arc. But what is being described is a pretzel separation scheme.

Separating the beam in a pre-existing ring is significantly more difficult than designing beam separation during the planning stage, as was done, for example, for the 45 degree separation points in the initial LEP ring. Obviously the separators have to be electric and therefore probably quite long. At CESR there was no free space long enough, so existing magnets had to be made shorter and stronger to free up space for electric separators. Even so, the required electric field was uncomfortably large.

With $N_b$ equally-spaced bunches in each of the counter-rotating beams the beams need to be separated at the $2N_b - N^*$ "parasitic" crossing points. Standard closed bumps are typically $\pi$-bumps or $2\pi$ bumps. But, with 4 deflectors, two at each end of a sector, bumps can easily be designed to be $n\pi$ bumps, where $n$ is an arbitrary integer matched to the desired number of separation points.

This discussion is illustrated pictorially in Figure 9 using a space-time plot introduced (in this context) by John Jowett. The beams are plotted as "world trajectories", whose crossings in space do not, in general, coincide with their crossings in time. Separated world events with the same label correspond to the same point at different times.

In the figure, associating point 4 with point 1 would correspond to the original McDaniel pretzel scheme in which the counter-circulating orbits are different everywhere in the ring. With the "closed pretzels" there is no such association. The separated beams are smoothly merged onto common orbits at both ends.

(With care) one can associate the transverse bump displacement pattern with the space-time diagram, interpreting the vertical axis as bump amplitude. A head-on collision occurs when two populated bunches pass through the same space-time event. To avoid parasitic crossings the minimum bunch separation distance is therefore twice the closed bump period.

Another separation scheme tried at LEP was local electric bumps close to the 4 IP's and angle crossing to permit "trains" with more than one bunch per train. This permitted as many as 4 bunches per train though, in practice, more than 3 were never used. For lack of time this option is not considered in this paper.

The primary horizontal separation scheme at LEP is illustrated in Jowett's clear, but complicated, Figure 3 [12]. The scheme used 8 primary separators and 2 trim separators with the separation bumps continuing through the 4 IP's, but with vanishing crossing angles at all 4 IP's. Starting from scratch in a circular collider that is still on the drawing board, one hopes for a simpler separation scheme.

**Separated Beams and RF Cavities.** By introducing slightly shortened, slightly strengthened, special purpose bending magnets to make space for electric separators, multi-element electric bumps can be located arbitrarily without seriously perturbing any existing lattice design. But there is an issue with separated orbits and RF cavities. Probably both beams should pass through the centers of the RF cavities. But it seems safe to assume that the closed orbit angles through the RF can be (symmetrically) different from zero. Otherwise, far more electric separators will be required, and far fewer bunches would be possible. RF cavities are therefore to be placed at beam separation nodes.

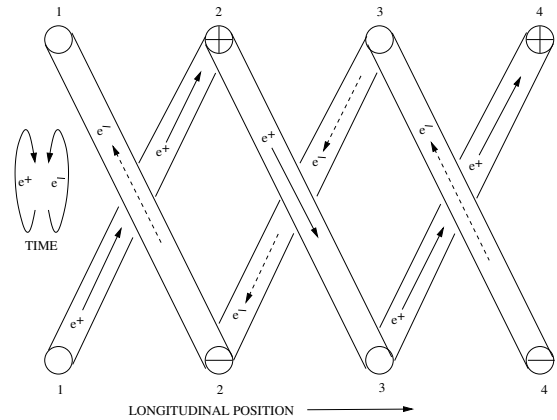

Figure 9: A minimal and modified "Jowett Toroidal Space-Time Beam Separation Plot" illustrating the separation of counter-circulating beams. Points with the same label at the top and the bottom of the plot are the same points (at different times). Though drawn to suggest a toroid the plot is purely two dimensional. The original McDaniel pretzel encompassed the whole ring—that is, in this figure, points 1 and 4 would also be identified. But this identification is not essential. It is important to interpret "toroid space-time separation" in topological terms, not as a synonym for "barber-pole separation" in which the orbits are actually toroidal in space. Though thinkable, such a separation scheme brings with it serious problems that almost inevitably lead to increased vertical emittance.

"Topping-off" injection is essential; especially to permit large beam-beam tune shifts. As long as the beam current is constant the beam-beam deflection is equivalent to a thin lens, focusing in both planes, though with strong octupole superimposed. The linear focusing part can be incorporated into the (linear) lattice optics. And the superimposed octupole is not necessarily very damaging. Strong damping makes bump-free, kicker-free, bunch-by-bunch, high-efficiency, vertical injection possible. Then steady-state, continuous operation without fill cycling may be possible.

Somewhat reduced beam separation at bump ends may be acceptable. With crossing angle the number of bunches may later be increased.

### 3.2  6 + 6 Element Closed Electric Bump

Bunches must not collide in arcs. They should be separated by at least 10 beam width sigmas when they pass. With both beams passing through the same RF, the path lengths between RF cavities probably have to be equal. A single ring is as good as dual rings if the total number of bunches can be limited to, say, less than 200. Here it is proposed to support only head-on collisions at each of the two IP's. The minimum bunch spacing will then be slightly greater than the total length of the intersection region (IR).

The half ring shown in Figure 8 shows a closed electric bump in the west arc. Orbits are common only in the two IR's. On the exit from each IR an electric bump is started and the bump is closed just before the next IR. These "bumps" are very long, almost half the circumference. As already explained, this is not "pretzel" beam separation, as that term was initially understood. Other than being horizontal rather than vertical and having multiple avoided parasitic collisions, these are much like the four separation bumps in the original LEP design.

Closed bumps require at least 3, or for symmetry, 4 controllable deflectors. Here a 12-bump scheme is described, with 6 electrostatic separators at the bump start and 6 at the bump stop. This scheme could be needed if the lattice cell lengths are too short to contain sufficiently strong electric separators. In Section "Lattice Optimization for Top-Off Injection", I show that the optimal collider cell length $L_i$ is significantly longer than was assumed when this separation scheme was designed. With longer cells a simpler 4 or 6 kicker bump may be adequate.

The design orbit spirals in significantly; this requires the RF acceleration to be distributed quite uniformly. Basically the ring is a "curved linac". The only betatron tune tunability is in the arcs. As the arc phase advances are changed (by a percent or so) the bumps have to be closed (very accurately) by tuning phase advance per cell and trim electric separators. As with beam separation in LEP, trim separators may be required.

Sketches and design formulas for a multi-element electric bump are shown in Figure 10. Figure 11 exhibits the separation of up to 112 bunches in a 50 km ring. Notice that, to avoid head-on parasitic collisions, the bunch separations are equal to two wavelengths of the electric bump pattern.

### 3.3 Bunch Separation Partition Number Shift

(Mangling Jowett's careful formulation [12] for brevity) the longitudinal partition number $J_\epsilon$ depends on focusing function $K_1$, dispersion $D$, and on fractional momentum offset, $\delta = \bar{\delta} + \delta_{s.o.}$ (where "s.o." stands for synchrotron oscillation) and separator displacement $x_p(s)$;

$$J_\epsilon(\delta, x_p) = 2 + \frac{2 \oint K_1^2 D^2 ds}{\oint (1/R^2) ds} (\bar{\delta} + \delta_{s.o.}) + \\ + \frac{2 \oint K_1^2 (D(s) - D_0(s)) x_p(s) ds}{\oint (1/R^2) ds}; \quad (18)$$

here $D/D_0$ are the separator-on/separator-off horizontal dispersion functions. The middle term here can be used to shift $J_\epsilon$ away from 2, as proved useful at LEP, but it does not depend on $x_p$; it is shown only as protection against confusing it with the final term.

Setting $\bar{\delta} + \delta_{s.o.} = 0$, and averaging, the separator-displaced partition number is

$$J_\epsilon(|x_p|) = 2 + \frac{2 < K_1^2(D - D_0) x_p >}{< 1/R^2 >}. \quad (19)$$

In spite of $x_p$ averaging to zero, there is a non-vanishing shift of $J_\epsilon(|x_p|)$ because $K_1$, $D$, and $x_p$ are correlated. At LEP this shift was observed to be significantly damaging and to be dominated by sextupoles. The factors in Eq. (19) scale as

$$x_p \propto \sigma_x \propto \frac{1}{R^{1/2}}, \quad K_1 = \frac{q}{l_q} \propto \frac{1/R^{1/2}}{R^{1/2}} \propto \frac{1}{R},$$
$$D - D_0 \propto S \propto \frac{1}{R^{1/2}}, \quad \Delta J_\epsilon(|x_p|) \propto \frac{1}{R}. \quad (20)$$

These scaling formulas (derived in Section "Lattice Optimization For Top-Off Injection") indicate that the seriousness of this partition number shift actually decreases with increasing $R$. Even if this were not true, should the partition number shift be unacceptably large, it can be reduced by increasing the quadrupole length $l_q$ to decrease $K_1$. The partition number shift is due to excess radiation in the quadrupoles. Since this radiation intensity is proportional to the square of the magnetic field, doubling the quadrupole length halves the radiation and the partition number shift.

### 3.4 Beam Separation in Injection-Optimized Collider Lattice

Section "Lattice Optimization for Top-Off Injection") describes the scaling of lattice parameters obtained after redesigning both injector and collider for efficient injection. The resulting collider cell length is $L_c = 213$ m. Because the cells are so long, there may be no need for multiple electrostatic separators. Instead one may use, for example, two or three electric kickers to launch each electric bump, with two or three matched kickers to terminate it. A large increase in cell length will surely also require a corresponding increase in longitudinal separation of circulating bunches. The single beam luminosity will be correspondingly reduced if the luminosity is already limited by the maximum number of bunches, as in the case of $Z_0$ production. The luminosity reduction should be little affected at the Higgs energy and above.

Irrespective of bunch separation schemes, the minimum bunch separation will still be at least equal to the length of the intersection region. For single ring operation this will probably be less restrictive than the bunch separation required for the separation scheme.

## 4 LATTICE OPTIMIZATION FOR TOP-OFF INJECTION

This section discusses Higgs factory injection. Full energy, top-off injection is assumed. Vertical injection seems

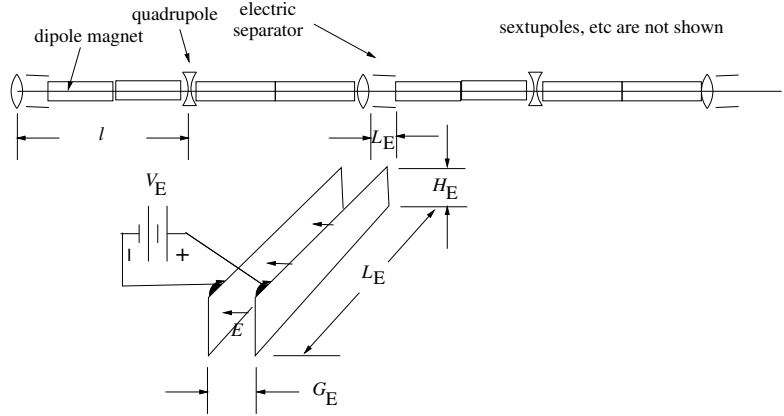

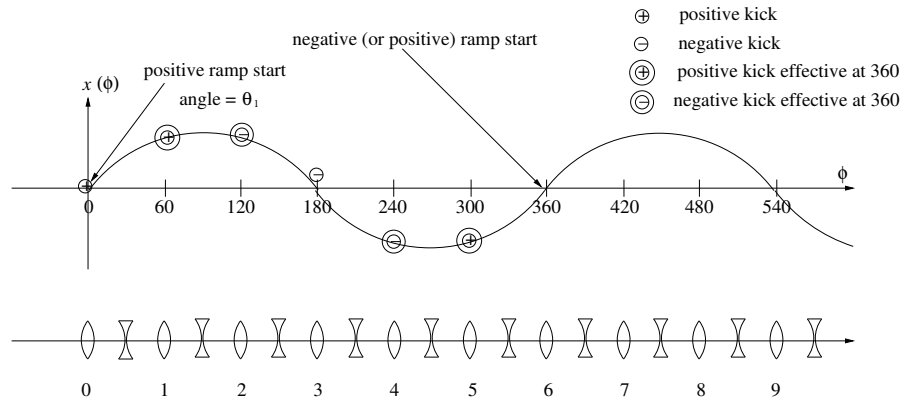

displacement at "j" due to kick angle $\theta_1$ at "k" = $\hat{\beta}_x \sin(\phi_j - \phi_k)\,\theta_1$

$N_1$ = number of effective kicks per half bump = 4 (for 60 degree lattice)

$N_2$ = number of accumulating bump stages

$x(120 N_2) = N_1 N_2 \hat{\beta}_x \sin(\phi_j - \phi_k)\,\theta_1$

$= 4 \times 69.3\text{m} \times \theta_1\, \dfrac{\sqrt{3}}{2}\, N_2$ ( for 60 degree lattice )

NOTE: deflections by arc quadrupoles are typically greater than electric separator deflections

Figure 10: Sketches and design formulas for a multi-element electric bump.

preferable to horizontal (as will be shown). Kicker-free, bunch-by-bunch injection concurrent with physics running may be feasible. Achieving high efficiency injection justifies optimizing injector and/or collider lattices for maximum injection efficiency. Stronger focusing in the injector and weaker focusing in the collider improves the injection efficiency. Scaling formulas (for the dependence on ring radius $R$) show injection efficiency increasing with increasing ring circumference. Scaling up from LEP, more nearly optimal parameters for both injector and collider are obtained. Maximum luminosity favors adjusting the collider cell length $L_c$ for maximum luminosity and choosing a shorter injector cell length, $L_i < L_c$, for maximizing injection efficiency.

## 4.1 Injection Strategy: Strong Focusing Injector, Weak Focusing Collider

**Introduction.** I take it as given that full energy top-off injection will be required for the FCC electron-positron Higgs factory. Without reviewing the advantages of top-off injection, one has to be aware of one disadvantage. The cost in energy of losing a full energy particle due to injection inefficiency is the same as the cost of losing a circulating particle to Bhabha scattering or to beamstrahlung or to any other loss mechanism. Injection efficiency of 50% is equivalent to doubling the irreducible circulating beam loss rate. To make this degradation unimportant one should therefore try to achieve 90% injection efficiency.

Achieving high efficiency injection is therefore sufficiently important to justify optimizing one or both of injector and collider lattices to improve injection. The aspect of this op-timization to be emphasized here is shrinking the injector beam emittances and expanding the collider beam acceptances by using stronger focusing in the injector than in the collider. What are the dynamic aperture implications? They will be shown to be progressively more favorable as the ring radius $R$ is increased relative to the LEP value. The dynamic-aperture/beam-width ratio increases as $R^{1/2}$ and is the same for injector and collider.

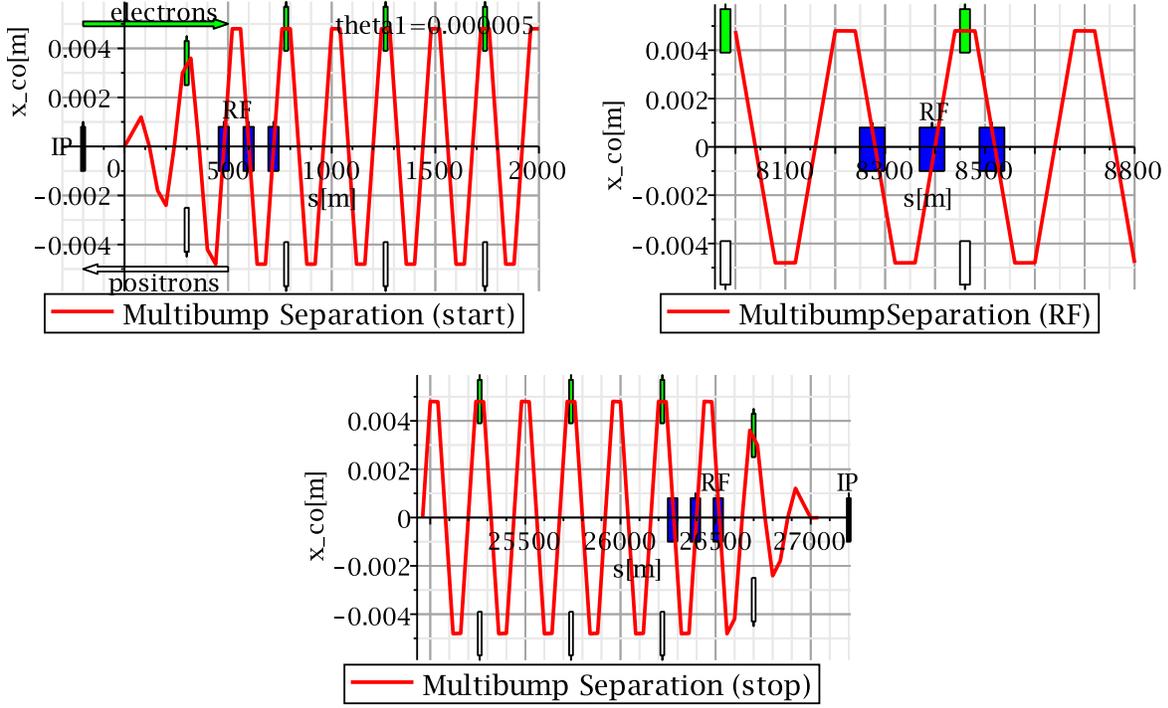

Figure 11: Short partial sections of the multibump beam separation are shown: one at the beginning, one at an RF location in the interior, and one at the far end of a long arc in Figure 8. The bunch separations are 480 m in a 50 km ring with cell length $L_c = 60$ m. IP's are indicated by vertical black bars, RF cavities by blue rectangles, electron bunches are green rectangles moving left to right, positron bunches are open rectangles moving from right to left. Counter-circulating bunches are separated at closed bump loop locations, and they must not pass through the nodes at the same time.

### 4.2 Constant Dispersion Scaling with $R$

**Linear Lattice Optics.** Most of the following scaling formulas come from Jowett [12] or Keil [20] or from reference [21]. The emphasis on parameter scaling is in very much the spirit emphasized by Alex Chao [22]. For simplicity, even if it is not necessarily optimal, assume the Higgs factory arc optics can be scaled directly from LEP values, which are: phase advance per cell $\mu_x = \pi/2$, full cell length $L_c = 79$ m. (The subscript "c" distinguishes the collider lattice cell length from the injector lattice cell length $L_i$.) At constant phase advance, the beta function $\beta_x$ scales as $L_c$ and dispersion $D$ scales as bend angle per cell $\phi = L_c/R$ multiplied by cell length $L_c$;

$$D \propto \frac{L_c^2}{R}. \tag{21}$$

(For 90 degree cells, the constant of proportionality in this formula is approximately 0.5, for the average dispersion $< D >$.) Holding $L_c$ constant as $R$ is increased would decrease the dispersion, impairing our ability to control chromaticity. Let us therefore *tentatively adopt the scaling*

$$L_c \propto R^{1/2}, \quad \text{corresponding to} \quad \phi \propto R^{-1/2}. \tag{22}$$

This is tantamount to holding dispersion $D$ constant, and is consistent with electron storage ring design trends over the decades.

These quantities and "Sands curly H" $\mathcal{H}$ then scale as

$$\beta_x \propto R^{1/2}, \quad D \propto 1, \quad \mathcal{H} \propto \frac{D^2}{\beta_x} \propto \frac{1}{R^{1/2}}. \tag{23}$$

Copied from Jowett [12], the fractional energy spread is given by

$$\sigma_\epsilon^2 = \frac{55}{32\sqrt{3}} \frac{\hbar}{m_e c} \gamma_e F_\epsilon, \quad \text{where}$$

$$F_\epsilon = \frac{< 1/R^3 >}{J_x < 1/R^2 >} \propto \frac{1}{R}, \tag{24}$$

and the horizontal emittance is given by

$$\epsilon_x = \frac{55}{32\sqrt{3}} \frac{\hbar}{m_e c} \gamma_e F_x, \quad \text{where}$$

$$F_x = \frac{<\mathcal{H}/R^3>}{J_x <1/R^2>} \propto \frac{1}{R^{3/2}}. \quad (25)$$

The betatron contribution to beam width scales as

$$\sigma_{x,\text{betatron}} \propto \sqrt{\beta_x \epsilon_x} \propto 1/R^{1/2}. \quad (26)$$

Similarly, at fixed beam energy, the fractional beam energy (or momentum) spread $\sigma_\delta$ scales as

$$\sigma_\delta \propto \sqrt{B} \propto 1/R^{1/2}. \quad (27)$$

**Scaling with $R$ of Arc Sextupole Strengths and Dynamic Aperture.** At this stage in the Higgs Factory design, it remains uncertain whether the IP-induced chromaticity can be cancelled locally, which promises more than a factor of two increase in luminosity, but would require strong bends close to the IP. For the time being I assume the IP chromaticity is cancelled in the arcs. Individual sextupole strengths can be apportioned as

$$S = S^{\text{arc chr.}} + S^{\text{IP chr.}} \quad (28)$$

The IP-compensating sextupole portion $S^{\text{IP chr.}}$ depends on the IP-induced chromaticity. A convenient rule of thumb has the IP chromaticity equal to the arc chromaticity. By this rule doubling the arc-compensating sextupole strengths cancels both the arc and the IP chromaticity.

With dispersion $D \propto 1$, quad strength $q \propto 1/R^{1/2}$, and $S^{\text{arc chr.}} \propto q/D$, one obtains the scaling of sextupole strengths and dynamic aperture;

$$S \propto \frac{1}{R^{1/2}}, \quad \text{and} \quad x^{\text{dyn. ap.}} \propto \frac{q}{S^{\text{arc chr.}}} \propto 1. \quad (29)$$

The most appropriate measure of dynamic aperture is its ratio to beam width,

$$\frac{x^{\text{dyn. ap.}}}{\sigma_x} \propto \frac{1}{1/R^{1/2}} \propto R^{1/2}. \quad (30)$$

The increase of this ratio with increasing $R$ would allow the IP optics to be more aggressive for the Higgs factory than for LEP. Unfortunately it is the chromatic mismatch between IP and arc that is thought to be more important in limiting the dynamic aperture than is the simple compensation of total chromaticity. The constant dispersion scaling formulas derived so far are given in Table 3.

### 4.3 Revising Injector and/or Collider Parameters for Improved Injection

What has been discussed so far has been "constant dispersion scaling". But, as already stated, we wish to differentiate the injector and collider optics such that the injector emittances are smaller and the collider acceptances are larger. This can be accomplished by shortening injector length $L_i$ and lengthening collider cell length $L_c$. The resulting $R$-dependencies and scaling formulas are shown in Table 10. They yield the lattice parameters in Table 11 for both the 50 km and 100 km circumference options.

**Implications of Changing Lattices for Improved Injection.** According to these calculations there is substantial advantage and little disadvantage to strengthening the injector focusing and weakening the collider focusing. This is achieved by shortening the injector cell length $L_i$ and increasing the collider cell length $L_c$. Weakening the collider focusing has the effect of increasing the equilibrium transverse beam dimensions.

As indicated in the caption to Table 11, the improvement in the injector emittance/collider acceptance ratio is probably unnecessaily large, at least for a 100 km ring, where the improvement in the injector/collider emittance ratio is a factor of seven.

Furthermore there is at least one more constraint that needs to be met. Maximum luminosity results only when the beam aspect ratio at the crossing point is optimal. Among other things this imposes a condition of the horizontal emittance $\epsilon_x$. At the moment the preferred method for controlling $\epsilon_x$ is by the appropriate choice of cell length $L_c$. With lattice manipulations other than changing the cell length it may be possible to increase, but probably not decrease $\epsilon_x$.

According to Table 2 of Section "Ring Circumference and Two Rings vs One Ring", with $\beta_y^* = 5$ mm the optimal choice of $\epsilon_x$ is 3.98 nm. According to Table 11 the actual value will be $\epsilon_x = 7.82$ nm. These values can be considered "close enough for now", or they can be considered different enough to argue that further design refinement is required (which is obvious in any case). But the suggestion is that the $L_c = 213$ m collider cell length choice in Table 11 may be somewhat too long.

Unfortunately the optimal value of $\epsilon_x$ depends strongly on the optimal value of $\beta_y^*$, which is presently unkown. These considerations show that the arc and intersection region designs cannot be separately optimized. Rather a full ring optimization is required.

| Parameter | Symbol | Proportionality | $L \propto R^{1/4}$ injector | $L \propto R^{1/2}$ | $L \propto R^{3/4}$ collider |
|---|---|---|---|---|---|
| phase advance per cell | $\mu_x$ | | 90° | 90° | 90° |
| cell length | $L$ | | $R^{1/4}$ | $R^{1/2}$ | $R^{3/4}$ |
| | | | 110 m | 153 m | 213 m |
| bend angle per cell | $\phi$ | $= L/R$ | $R^{-3/4}$ | $R^{-1/2}$ | $R^{-1/4}$ |
| momentum compaction | | $\phi^2$ | $R^{-3/2}$ | $R^{-1}$ | $R^{-1/2}$ |
| quad strength (1/f) | $q$ | $1/L$ | $R^{-1/4}$ | $R^{-1/2}$ | $R^{-3/4}$ |
| dispersion | $D$ | $\phi L$ | $R^{-1/2}$ | 1 | $R^{1/2}$ |
| beta | $\beta$ | $L$ | $R^{1/4}$ | $R^{1/2}$ | $R^{3/4}$ |
| tune | $Q_x$ | $R/\beta$ | $R^{3/4}$ | $R^{1/2}$ | $R^{1/4}$ |
| | | | 243.26 | 174.26 | 125.26 |
| tune | $Q_y$ | $R/\beta$ | $R^{3/4}$ | $R^{1/2}$ | $R^{1/4}$ |
| | | | 205.19 | 147.19 | 106.19 |
| Sands's "curly H" | $\mathcal{H}$ | $= D^2/\beta$ | $R^{-5/4}$ | $R^{-1/2}$ | $R^{1/4}$ |
| partition numbers | $J_x/J_y/J_\epsilon$ | 1/1/2 | 1/1/2 | 1/1/2 | 1/1/2 |
| horizontal emittance | $\epsilon_x$ | $\mathcal{H}/(J_x R)$ | $R^{-9/4}$ | $R^{-3/2}$ | $R^{-3/4}$ |
| fract. momentum spread | $\sigma_\delta$ | $\sqrt{B}$ | $R^{-1/2}$ | $R^{-1/2}$ | $R^{-1/2}$ |
| arc beam width-betatron | $\sigma_{x,\beta}$ | $= \sqrt{\beta \epsilon_x}$ | $R^{-1}$ | $R^{-1/2}$ | 1 |
| -synchrotron | $\sigma_{x,synch.}$ | $= D\sigma_\delta$ | $R^{-1}$ | $R^{-1/2}$ | 1 |
| sextupole strength | $S$ | $q/D$ | $R^{1/4}$ | $R^{-1/2}$ | $R^{-5/4}$ |
| dynamic aperture | $x^{da}$ | $q/S$ | $R^{-1/2}$ | 1 | $R^{1/2}$ |
| relative dyn. aperture | $x^{da}/\sigma_x$ | | $R^{1/2}$ | $R^{1/2}$ | $R^{1/2}$ |
| separation amplitude | $x_p$ | $\sigma_x$ | N/A | $R^{-1/2}$ | 1 |

Table 10: To improve injection efficiency (compared to constant dispersion scaling) the injector cell length can increase less, for example $L_i \propto R^{1/4}$, and the collider cell length can increase more, for example $L_i \propto R^{3/4}$. The shaded entries assume circumference $C=100$ km, $R/R_{\text{LEP}}=3.75$.

| Parameter | Symbol | LEP(sc) | Unit | Injector | | Collider | |
|---|---|---|---|---|---|---|---|
| bend radius | $R$ | **3026** | m | **5675** | **11350** | 5675 | 11350 |
| beam Energy | | 120 | GeV | 120 | 120 | 120 | 120 |
| circumference | $C$ | 26.7 | km | 50 | 100 | 50 | 100 |
| cell length | $L$ | 79 | m | 92.4 | 110 | 127 | 213 |
| momentum compaction | $\alpha_c$ | 1.85e-4 | m | 0.72e-4 | 0.25e-4 | 1.35e-4 | 0.96e-4 |
| tunes | $Q_x$ | 90.26 | | 144.26 | 243.26 | 105.26 | 125.26 |
| | $Q_y$ | 76.19 | | 122.19 | 205.19 | 89.19 | 106.19 |
| partition numbers | $J_x/J_y/J_\epsilon$ | 1/1.6/1.4 | | 1/1/2 | 1/1/2 | 1/1/2 | 1/1/2 |
| main bend field | $B_0$ | 0.1316 | T | 0.0702 | 0.0351 | 0.0702 | 0.0351 |
| energy loss per turn | $U_0$ | 6.49 | GeV | 3.46 | 1.73 | 3.46 | 1.73 |
| radial damping time | $\tau_x$ | 0.0033 | s | 0.0061 | 0.0124 | 0.0061 | 0.0124 |
| | $\tau_x/T_0$ | 37 | turns | 69 | 139 | 69 | 139 |
| fractional energy spread | $\sigma_\delta$ | 0.0025 | | 0.0018 | 0.0013 | 0.0018 | 0.0013 |
| emittances (no BB), $x$ | $\epsilon_x$ | **21.1** | nm | **5.13** | **1.08** | **13.2** | **7.82** |
| $y$ | $\epsilon_y$ | 1.0 | nm | 0.25 | 0.05 | 0.66 | 0.39 |
| max. arc beta functs | $\beta_x^{\max}$ | 125 | m | 146 | 174 | 200 | 337 |
| max. arc dispersion | $D^{\max}$ | 0.5 | m | 0.37 | 0.26 | 0.68 | 0.97 |
| quadrupole strength | $q \approx \pm 2.5/L_p$ | 0.0316 | 1/m | 0.027 | 0.0227 | 0.0197 | 0.0117 |
| max. beam width (arc) | $\sigma_x = \sqrt{2\beta_x^{\max}\epsilon_x}$ | $1.6\sqrt{2}$ | mm | $0.865\sqrt{2}$ | $0.433\sqrt{2}$ | $1.62\sqrt{2}$ | $1.62\sqrt{2}$ |
| (ref) sextupole strength | $S = q/D$ | 0.0632 | $1/m^2$ | 0.0732 | 0.0873 | 0.0290 | 0.0121 |
| (ref) dynamic aperture | $x^{da} \sim q/S$ | ~0.5 | m | ~0.370 | ~0.260 | ~0.679 | ~0.967 |
| (rel-ref) dyn.ap. | $x^{da}/\sigma_x$ | ~0.313 | | ~0.428 | ~0.600 | ~0.417 | ~0.621 |
| separation amplitude | $\pm 5\sigma_x$ | $\pm 8.0\sqrt{2}$ | mm | | | $\pm 8.1\sqrt{2}$ | $\pm 7.8\sqrt{2}$ |

Table 11: Lattice parameters for improved injection efficiency. This table is to be compared with Table 5 to assess the effect of lattice changes on injection efficiency. The shaded row indicates how successfully the injector emittance has been reduced relative to the collider emittance. The factor of seven improvement, 7.82/1.08, in this ratio for a 100 km ring, seems unnecessarily large, indicating that less radical scaling should be satisfactory. As it happens the 213 m collider cell length agrees almost perfectly with the cell length adopted for the FCC-pp collider, as reported by Schulte [16] at the 2015 FCC-week in Washington D.C.

# 5  $\mathcal{L} \times L^*$ LUMINOSITY × FREE SPACE INVARIANT

Yunhai Cai's intersection region design [17] is analysed in detail in Appendix E, "Deconstructing Yunhai Cai's IR Optics". For maximum operational convenience in changing IP beta functions, Yunhai's design was designed to be scalable. This makes the IR design ideal for using dimensional analysis to derive scaling law dependence on the free space length $L^*$, which is the length of the space left free for the particle collision reconstruction apparatus. This scaling law can be employed to investigate how the choice of free IP length $L^*$ affects the achievable luminosity. Yunhai's design is probably close to optimal. But, even if it is not, the same results, based purely on scaling behavior, will still be valid. This prescription *does not* establish the *absolute* luminosity but it *does* determine the *relative* luminosity under the plausible hypothesis that the luminosity maximum will be governed by the maximum $\beta$ functions (anywhere in the ring).

For convenience in extracting scaling laws, Yunhai's MAD lattice file was modified to include a scaling factor "FAC" such that FAC=1 results in no change. All lengths are multiplied by FAC and all quadrupole focal lengths are divided by FAC. (This means the quadrupole coefficients in the MAD file have to be divided by FAC$^2$ to account for the altered quadrupole lengths). Similarly sextupole coefficients have to be divided by FAC$^3$. After these changes, MAD runs produced the beta function plots shown in Figure 12 for the four parameter sets given in Table 12. Other than noting their identical shapes (confirming the scaling) only the maximum $\beta_y^{\max}$ values are extracted from the plots.

| FAC | $L^*$ m | $\beta_x^*$ mm | $\beta_y^*$ mm | $\beta_y^{\max}$ m | $L^*$ depend. | $\beta_y^*$ depend. |
|---|---|---|---|---|---|---|
| 1.0 | 2 | 200 | 2 | 4900 | ✓ | ✓ |
| 1.5 | 3 | 450 | 4.5 | 4900 | ✓ | |
| 1.0 | 2 | 400 | 4 | 2450 | | ✓ |
| 1.0 | 2 | 1000 | 10 | 990 | | ✓ |

Table 12: Parameter sets for dimensional analysis of Yunhai Cai Higgs factory intersection region design.

Results of the MAD runs are plotted in Figure 13. The smooth fitting function in the left plot of Figure 13 gives the scaling law

$$\beta_y^{\max} = \frac{10[\mathrm{m}^2]}{\beta_y^*} \Big(\frac{L^*}{L^{*\mathrm{nom}}}\Big)^p, \qquad (31)$$

where $L^{*\mathrm{nom}} = 2$ m is the nominal distance from IP to the entrance edge of the first quadrupole. The final factor (which is equal to 1 for the plot) has been included to allow power law dependence on $L^*$, with exponent $p$ to be determined later. The right plot of Figure 13 gives the scaling law

$$\beta_y^* = \beta_y^{*\,\mathrm{nom}} \Big(\frac{L^*}{L^{*\mathrm{nom}}}\Big)^2. \qquad (32)$$

For Eqs. (31) and (32) to be compatible requires $p = 2$. Then Eq. (31) becomes

$$\boxed{\beta_y^* = 2.5 \, \frac{L^{*2}}{\beta_y^{\max}}} \quad \stackrel{\mathrm{e.g.}}{=} \; 2.5 \, \frac{2^2}{4900} = 2 \, \mathrm{mm}. \quad \checkmark \qquad (33)$$

Using Eq. (33) the luminosity is given by

$$\boxed{\mathcal{L}^{\mathrm{static}} \stackrel{\mathrm{e.g.}}{=} 1.6 \times 10^{31} \mathrm{cm}^{-2} \mathrm{s}^{-1} \mathrm{m} \times \frac{\beta_y^{\max}}{L^{*2}}.} \qquad (34)$$

*The constant of proportionality in this equation has not been determined by the scaling formula. It has been chosen to match preliminary CEPC estimated luminosities.* For local chromaticity I.P. design (Yunhai Cai), lengths are scalable, quad strengths scale as 1/length, beta functions scale as length. Assuming some upper limit on $\beta_y^{\max}$ has been established, an upper limit for the luminosity times detector length has also been detemined;

$$\boxed{L^* \times \mathcal{L} \quad \text{product (upper limit) is fixed.}} \qquad (35)$$

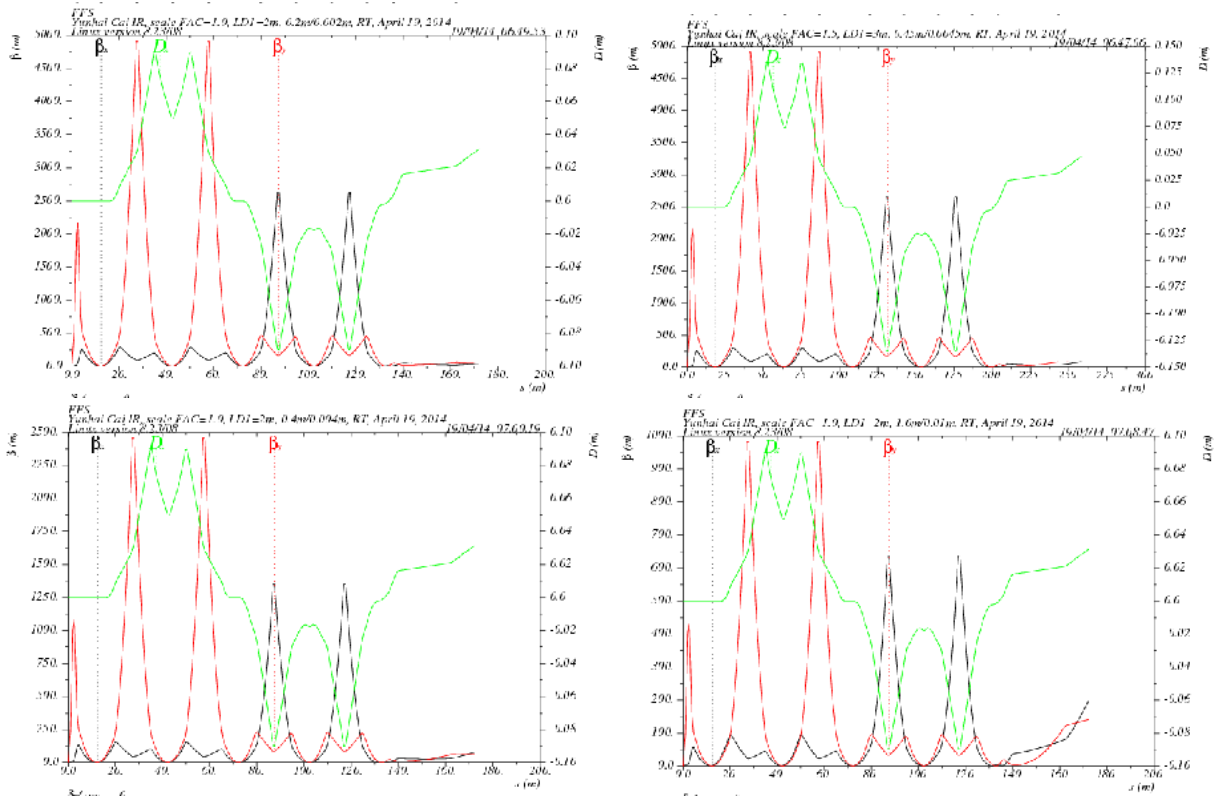

Figure 12: The upper plots differ by the choice $FAC = 1$ in the upper left plot and $FAC = 1.5$ in the upper right plot. So, for example, $L^*$ is 50% greater in the lower plot. The lower plots differ by the choice $\beta_y = 4$ mm in the lower left plot and $\beta_y = 10$ mm in the lower right plot. The important parameters are copied from the data line into the table.

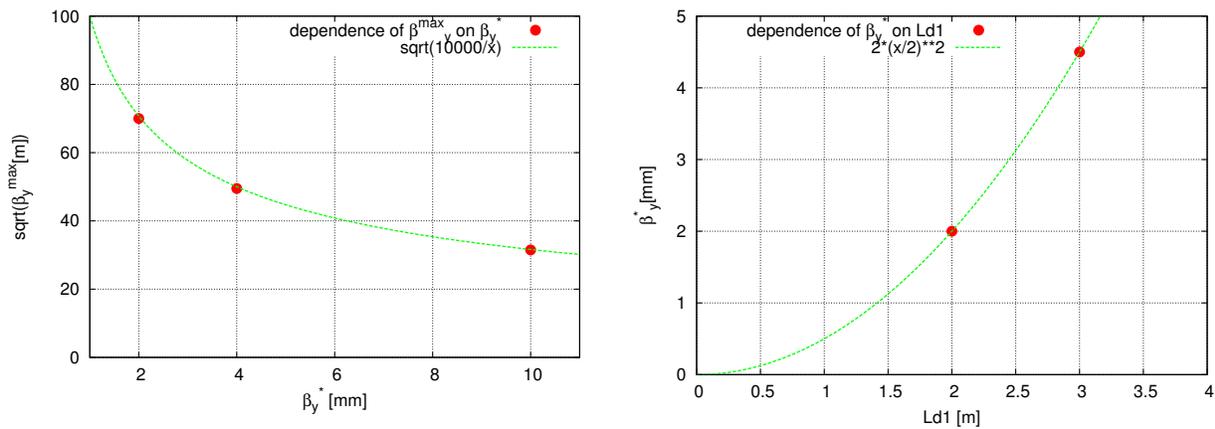

Figure 13: Parameter dependencies implied by the Yunhai Cai intersection region design. The left plots $\beta_Y^{max}$ (or rather its square root) versus $\beta_y^*$ with the longitudinal scale held constant. The right plots $\beta_Y^*$ versus $L^*$ with $\beta_Y^{max} = 4900$ m held constant.

# 6 TRANSVERSE SENSITIVITY LENGTH

## 6.1 Estimating $\beta_y^{\max}$ (and from it $\mathcal{L}$)

According to Eq. (34) the achievable luminosity $\mathcal{L}$ is proportional to the maximum achievable beta function value $\beta_y^{\max}$. This quantity is hard to determine, however, since it depends on unknown parameter uncertainty, such as misalignments and fringe fields. Here we attempt to determine $\beta_y^{\max}$ by scaling from operational experience with existing rings.

All of accelerator physics is based on a radial multipole expansion of the magnetic fields, with coefficients that are bend (constant), then quadrupole (linear), then sextupole (quadratic), and so on. The Courant-Snyder formalism is based on the first two (linear) terms and is considered to be perfectly reliable for small amplitudes. For various reasons, such as head-tail instability, the ring must be achromatic. This requires sextupoles; i.e. *nonlinearity*. The perturbative resonance-driving effect of any individual sextupole is proportional to $\beta^{3/2}$ (multiplied by the sextupole strength) and each higher multipole order brings in another power of $\beta^{1/2}$. Uncertain nonlinear multipole moments are magnified by large beta functions (raised to power $\beta^{3/2}, \beta^2, \beta^{5/2}, \dots$). For protons (but probably not electrons) because of their coil-dominated magnets, the very convergence of the multipole series is an issue. In the design stage there are too many unknown higher order multipoles to compensate reliably. Historically the dominant lattice defects can only be identified and cured in machine studies during the first year of operation and beyond.

The maximum tolerable beta function depends on unknown errors. For purposes of estimation one can guess that the most sensitive lattice element is the quadrupole situated at the location where $\beta_y$ is maximal, and that it produces an uncorrected angular deflection error $\Delta x' = q \Delta x$, proportional to the quadrupole strength $q$ and to an unknown "effective" displacement factor $\Delta x$. This error could be due to the quadrupole itself being displaced or due to other elements in the ring being displaced from their design locations. Dropping a factor of order 1 (or greater in case of resonance) that depends on betatron phase, the maximum displacement caused by this error would be $\beta_y^{\max} q \Delta x^{\text{tol.}}$, where $\Delta x^{\text{tol.}}$ is to be a phenomenologically determined "transverse displacement tolerance length". If the resulting error exceeds the dynamic aperture the particle will be lost. Using scaling equation (29) and $q \propto 1/L_c$, these quantities are given by

$$\Delta x \approx \beta_y^{\max} q \Delta x^{\text{tol.}},$$
$$x^{\text{dyn. ap.}} \propto \frac{q}{S} \propto D, \tag{36}$$

Using $q \propto 1/L_c$, setting these equal, and solving for $\Delta x^{\text{tol.}}$

$$\Delta x^{\text{tol.}} \propto \frac{D L_c}{\beta_y^{\max}}. \tag{37}$$

Dimensionally this quantity is a length. Though expressed unambiguously, for any actual ring it has to be multiplied by a factor depending on unknown transverse positioning imperfections, the absolute magnitude of which can only be inferred phenomenologically. To the extent the errors are due to positioning and field quality construction errors they may be expected to be quite similar in modern accelerators constructed with best practically achievable precision.

## 6.2 Why $\beta_y$ must not be too large

To get higher luminosity requires reducing $\beta_y^*$. Reducing $\beta_y^*$ increases $\beta_y^{\max}$, which invariably makes the collider more erratic, often unacceptably so. Sensitivity to beam-beam effects and other effects is greatly magnified by large $\beta$ anywhere in the ring. There are inevitable unknown transverse element displacement errors $\Delta y_{\text{transverse}}$. From the transverse sensitivity just discussed, the limitation imposed by a large $\beta_y^{\max}$ at one or a few points in the ring is expressed by a

$$\boxed{\text{transverse sensitivity length} = \frac{D L_C}{\beta^{\max}}.} \tag{38}$$

The optical deviation caused by $\Delta y$ will be negligible only in the limit

$$\Delta y \ll \text{transverse sensitivity length}. \tag{39}$$

### 6.3 Maximum $\beta_y$ Phenomenology Based on Transverse Orbit Sensitivity

The inverse of the sensitivity length is a "figure of demerit,"FOD" = $\frac{\beta_y^{\max}}{DL_c}$ that can be used to compare different rings, either proton or electron, independent of their beam energies. When $\beta_y^{\max}$ is large, it is always because $\beta_y^*$ is small. But the value of $\beta_y^*$ is irrelevant in assessing the dynamic aperture limitation caused by the large value of $\beta_y^{\max}$. Nevertheless $\beta_y^*$ values are given in the table. Note that $\beta_y^*$ tends to be "big" for the ancient rings toward the top of the table, and "small" toward the bottom. The two CepC rows assume identical IP optics with $\beta_y^* = 10$ mm, but different arc parameters. For the CepC$_1$ row the ring parameters are copied from the CEPC, CDR design. For the CepC$_2$ row the ring parameters are copied from Table 5. For the CEPC row $\beta_y^* = 1$ mm (which accounts for its hyper-transverse-sensitivity). CEPC and FCC-ee values differ due to different dispersion and different $L^*$ values (1.5 m for CEPC, 2.0 m for FCC-ee).

| $\beta_y^*$ | Ring | | $D$ | $L_c$ | $\beta_y^{\max}$ | $\frac{\beta_y^{\max}}{DL_c}$ |
|---|---|---|---|---|---|---|
| m | | | m | m | m | 1/m |
| 0.015 | CESR | exp. | 1.1 | 17 | 95 | 5.1 |
| 0.08 | PETRA | exp. | 0.32 | 14.4 | 225 | 49 |
| | HERA | exp. | 1.5 | 48 | 2025 | 28 |
| 0.05 | LEP | exp. | 0.8 | 79 | 441 | 7.0 |
| 0.007 | KEKB | exp. | 0.5 | 20 | 290 | 29 |
| | LHC | exp. | 1.6 | 79 | 4500 | 36 |
| 0.01 | CepC$_1$ | des. | 0.31 | 47 | 1225 | 84 |
| 0.01 | CepC$_2$ | des. | 1.03 | 153 | 1225 | 8.8 |
| 0.001 | CEPC | des. | 0.31 | 47 | 6000 | 410 |
| 0.001 | FCC-ee | des. | 0.10 | 50 | 9025 | 1805 |

Table 13: "Figures of Demerit", inverses of "transverse sensitivity lengths", are plotted for various low and high energy colliders, both proton and electron.

Compared in this way the transverse tolerances of KEKB and LHC are close in value, even though, as storage rings, they could scarcely be more disimilar; KEKB is a "small" electron collider, LHC is a large proton collider. The pessimistic behavior of LEP can be blamed on the absence of top-off injection, which led to the tortuous ramping and beta squeeze operations. This limited the $\beta_y^*$ to be not less than 5 cm. Entries in Table 13 suggest an empirically determined upper limit rule on $FOD_{\text{trans.sens.}}$,

$$FOD_{\text{trans.sens.}} < 40. \qquad (40)$$

CEPC exceeds this limit by a factor of 10, FCC-ee by a factor of 50. This is partly due to their way too short cell lengths.

This transverse sensitivity discussion has been only semi-quantitative but, at least, it is dimensionally consistent, and it provides a prescription for comparing performance of very different colliders. For the "transverse sensitivity length" to be a valid comparison gauge implicitly assumes that this length (dependent of survey and positioning precision) can be expected to be the same for accelerators of all sizes, and for both electrons and protons. The approach has been somewhat *ad hoc* however, and it depends on the validity of the scaling laws emphasized in this paper. Some length other than $DL_c/\beta_y^{\max}$ might provide a more valid comparison, though it would probably disrupt the good agreement between two modern rings, KEKB and LHC, in the last column of Table 13.

### APPENDICES

## 7 A. SYNCHROTRON RADIATION PRELIMINARIES

The dependence of ring radius $R$ on maximum electron beam energy $E_{\max}$ is dominated by the relation giving $U_1$, the energy loss per turn, per electron;

$$U_1 \text{ [GeV]} = C_\gamma \frac{E^4}{R}, \qquad (41)$$

where, for electons, $C_\gamma = 0.8846 \times 10^{-4}$ m/GeV$^3$. [3]

$C_\gamma$ can legitimately be referred to as the "Sands [23] constant"[4]. In these units the formula yields an energy in GeV units. Except as noted in a footnote[5] all energies in this report are expressed in GeV. Corresponding to energy loss $U_1$, with each beam having $N_b$ bunches of $N_p$ particles, the radiated power from each beam, in GW units, is given by

$$P_{\text{rad}} = U_1 N_{\text{tot}} f, \qquad (42)$$

where $N_{\text{tot}} = N_b N_p$ and $f = c/C$ is the revolution frequency.

The spectrum of radiated photons is characterized by a "critical energy" given by [24]

$$u_c \text{ [GeV]} = 10^{-9} \frac{3}{4\pi} hc \frac{\gamma^3}{R} = 2.96 \times 10^{-16} \text{ GeV.m} \frac{\gamma^3}{R} \qquad (43)$$

---

[3] For protons $C_\gamma = 0.7783 \times 10^{-17}$ m/GeV$^3$. Post-LHC p,p storage rings will, for the first time, be dominated by synchrotron radiation, in much the way that electron rings always have been. This is due to a combination of the energy ratio (to the fourth power) and the fact that some fraction of the synchrotron radiation energy is inevitably dissipated at liquid helium temperature, "amplifying" its cost in wall power.

[4] The convention in this report is for all quantities except energies to be expressed in SI units. Following Sands, energies are expressed in GeV. By this time an energy unit of 100 GeV would be more convenient. In this report, to avoid redefining the Sands constant $C_\gamma$, a dimensionless energy $\widetilde{E} = E/100$ GeV is introduced.

[5] When a factor occuring on the right hand side of an equation is given numerically (rather than symbolically) and it represents a quantity with units, the units are appended explicitly to the numerical value, as in the equation in the previous footnote. The units of quantities expressed symbolically are determined by convention but, for emphasis, the units may occasionally be be given in square brackets. The intention has been to make it possible to mentally check the dimensional consistency of every equation.

The factor of $10^{-9}$ in the first expression is included so that, when the remaining factor is evaluated using SI units, $u_c$ will come out in GeV. The factor $hc$ is equal to $12390\,\text{eV}.\text{Å}=1.239\times 10^{-6}\,\text{eV.m}$.

For numerical convenience, $\gamma$ can be written as

$$\gamma = \frac{10^{11}}{0.511\times 10^6}\left(\frac{E\,[\text{GeV}]}{100\,\text{GeV}}\right) \equiv 1.96\times 10^5\,\widetilde{E}, \quad (44)$$

where $\widetilde{E}$ is the (now dimensionless) beam energy divided by a "nominal" reference energy taken to be $100\,\text{GeV}$. For LEP operating at $100\,\text{GeV}$, which we will be using as a reference energy, $\widetilde{E}=1$. The beam energy itself, in GeV, is

$$E = 100\,\text{GeV}\,\widetilde{E}. \quad (45)$$

The $\widetilde{E}$-values to be considered will range from 1 to (at our most optimistic) 3. For radiated photons of energy $u$, the average energy value is related to the critical energy by $<u> = 0.31 u_c \approx u_c/3.2$. The number of radiated photons per electron per turn is then

$$n_{\gamma,1} = \frac{3.2\times 10^8\,\text{GeV}^4}{u_c}\,C_\gamma\,\frac{\widetilde{E}^4}{R}. \quad (46)$$

This number is invariably much greater than 1. Later, when comparing beamstrahlung to bremstrahlung, it will be convenient to use this relation in the form

$$U_1 = 0.31\,n_{\gamma,1}u_c, \quad (47)$$

even if this may seem, at present, to involve circular reasoning.

The horizontal ring emittance $\epsilon_x$ is established by the equilibrium between quantum fluctuation heating and synchrotron radiation damping. The important lattice parameter for this is the "Sands curly-H parameter" $\mathcal{H}$. I adopt, as a numerical value, (independent of $E_{\max}$ for simplicity) the value from Cai et al. [17];

$$\langle\mathcal{H}\rangle \approx \left\langle\frac{\eta_x^2 + (\beta_x\eta_x' - \beta_x'\eta_x/2)^2}{\beta_x}\right\rangle \approx 1.3\times 10^{-3}\,\text{m}. \quad (48)$$

The horizontal emittance is then given by [25]

$$\epsilon_x = \frac{1.323}{2J_x}\frac{u_c}{E}\langle\mathcal{H}\rangle, \quad (49)$$

where I will take $J_x = 1$ as the value of the horizontal partition number. From $\epsilon_x$ the r.m.s. beam width is given by

$$\sigma_x = \sqrt{\beta_x\epsilon_x}. \quad (50)$$

# 8 B. BEAMSTRAHLUNG

## 8.1 Determining Parameters to Suppress Beamstrahlung

As first explained by Telnov [4], with increasing beam energy, beamstrahlung has an even greater impact on the energy dependence of circular e+e- storage rings than does the energy loss per turn given by Eq. (41). Instead of being limited by power loss that can be restored by RF cavities, beamstrahlung causes reduced beam lifetime. This sets a luminosity limit based on the maximum beam current the injection system can provide. This section re-derives beamstrahlung formulas.

To quantify the lattice energy acceptance Telnov introduces a parameter $\eta \approx 0.015$, which is the maximum fractional energy loss an electron can suffer without being lost. Each beam bunch, looking like a quadrupole of strength $q$ to a particle in the opposing bunch, causes deflection

$$\Delta y' = -q\,y. \quad (51)$$

The effective quadrupole strength $q$ is related to the "beam-beam tune shift parameter" $\xi$;

$$\xi = \frac{\beta_y}{4\pi}\,q, \quad (52)$$

where $\beta_y$ is the vertical beta function at the interaction point. $\xi$ is given in terms of other beam parameters by [4]

$$\xi = \frac{1}{2\pi}\frac{1}{\gamma}N_p r_e\frac{\beta_y}{\sigma_x\sigma_y}, \quad (53)$$

where $r_e = 2.718\times 10^{-15}$ m is the classical electron radius. Note that there is no "hourglass" correction to the tune shift. For flat beams, by Gauss's law, the vertical electric field is proportional to the fractional excess charge below the test particle, and this fraction is independent of longitudinal coordinate $z$.

The first of two important aspects of beamstrahlung is the energy scale of the radiated x-rays (eventually, $\gamma$-rays). The vertical displacement of a typical electron is $\sigma_y$, the r.m.s. beam height, which is the approximate location where a particle is subject to the maximum deflection force from the other beam. Treating the other beam as a quadrupole of inverse focal length $q$, the slope change of the particle is

$$\Delta y' = -q\sigma_y \stackrel{\text{also}}{\approx} -\frac{\ell_z}{\rho^*}, \quad (54)$$

where $\ell_z$ is the "effective" bunch length of each of the beams. To treat the bunch length as uniform, with density per unit length equal to the actual charge density at the center of the bunch, while retaining the correct total charge, we take

$$\ell_z = \sqrt{\frac{\pi}{2}}\,\sigma_z, \quad (55)$$

where $\sigma_z$ is the r.m.s. bunch length. The formula will be used only to calculate the effective bending radius of the orbit for purposes of calculating the effective critical beamstrahlung energy $u_c^*$. This treats individual bunches as uniform, with distance from the back of the bunch to the front of the bunch as $2\ell_z$. The time duration of the deflection pulse is reduced by a factor of two by the relative speed being $2c$; which hardens the radiation. The hardening effect is being accounted for by reducing the effective bunch length by a factor of two in Eq. (54)[6].

The fact that the distribution is actually Gaussian, not uniform, softens the radiation. This softening effect is being neglected. This gives a (substantial) overestimate of $u_c^*$, a corresponding overestimate of the importance of beamstrahlung, and a corresponding underestimate of the luminosity.

Solving for $\rho^*$ yields

$$\rho^* = \frac{\ell_z}{\sigma_y} \frac{1}{q}. \tag{56}$$

Substituting from Eq. (52) into Eq. (56) produces

$$\rho^* = \frac{\ell_z}{\sigma_y} \frac{\beta_y}{4\pi\xi}. \tag{57}$$

The qualitative behavior of radiation from a short radiator depends on whether the radiator is technically "short" meaning the deflection angle is short compared to the radiation cone angle $1/\gamma$ or "long", in the opposite case. (Drawn from undulator radiation terminology) this distinction can be quantified by a strength parameter $K$ defined by

$$K = \Delta x' \gamma. \tag{58}$$

The nominal boundary between short and long bend element length is at $K = 1$. In our colliding beam beamstrahlung application the trend is from $K < 1$ at low beam energy to $K \gg 1$ at high beam energy. At the relatively low energies of pre-LEP storage rings, beamstrahlung was negligible. For post-LEP operation, the strong beam-beam force requires the "long target", $K \gg 1$, formulation. Fortunately this is not a calculational hardship since the formulas required are the same as needed to describe the effects of ordinary synchrotron radiation. But, because the beamstrahlung radiation is much "harder" than the synchrotron radiation, it is the ultrahard photons at the upper end of the beamstrahlung spectrum that need to be described.

The spectrum is most compactly charactized by the "critical photon wavelength" $\lambda_c^*$ [24];

$$\frac{1}{\lambda_c^*} = \frac{3}{4\pi} \frac{\gamma^3}{\rho^*}, \tag{59}$$

where $\rho^*$ is the radius of curvature of a radiating electron, assuming its orbit while passing through the other beam is a perfect circular arc of radius $\rho^*$. Corresponding to $\lambda_c^*$ is the "critical energy" $u_c^*$ of the radiation which, as in Eq. (43), is given by

$$u_c^* \text{ [GeV]} = 2.96 \times 10^{-16} \text{GeV.m} \frac{\gamma^3}{\rho^*} \tag{60}$$

---
[6] There is a similar cancelation in the tuneshift calculation. Because the relative speed of the test particle and the opposing beam bunch is $2c$, the pulse duration of the transverse force is reduced by a factor of two compared to the time interval between arrival time of head and tail at the origin. There is a compensating factor of two because the electric and magnetic forces are almost exactly equal. The beam-beam tune shift parameter $\xi$ accounts for both of these factors.

Expressed in terms of $\widetilde{E}$,

$$u_c^* = 28.0 \,\text{GeV.m} \, \sqrt{2/\pi} \frac{\sigma_y}{\sigma_z} \frac{\xi}{\beta_y} \widetilde{E}^3. \tag{61}$$

$u_c^*$ is the reduction in beam particle energy caused by radiation of one photon having energy equal to the critical energy. The fractional energy loss caused by a single emission at the critical energy is

$$\frac{u_c^*}{E} = 0.280 \,\text{m} \sqrt{2/\pi} \frac{\sigma_y}{\sigma_z} \frac{\xi}{\beta_y} \widetilde{E}^2. \tag{62}$$

When applied to LEP operation at 100 Gev, with $\xi = 0.083$, this works out to $u_c^*/E = 0.081 \times 10^{-3}$. This is in acceptable agreement, given the ambiguities, with the value of $0.09 \times 10^{-3}$ given in Telnov's Table I. Appendix B discusses the effect a Gaussian (as contrasted to uniform) longitudinal bunch distribution has on the beamstrahlung loss calculation.

The second beamstrahlung parameter of importance is the total energy $U_1^*$ radiated by a single electron in its $N^*$ passages through the counter-circulating beam as it completes one full turn. Treating the bunches as long and uniform, Eq. (41), multiplied by $N^* \ell_z / (2\pi \rho^*)$ to account for the fraction of the circumference,

$$U_1^* \text{[GeV]} \stackrel{?}{=} \mathcal{F}_1^* C_\gamma \frac{E^4}{\rho^{*2}} \frac{N^* \ell_z}{2\pi} = \mathcal{F}_1^* C_\gamma E^4 8\pi \frac{\xi^2 \sigma_y^2}{\beta_y^2} \frac{N^*}{\ell_z}. \tag{63}$$

Here the question mark serves as a warning that, in a quadrupole, only a fairly small fraction of the electrons are actually subject to the maximum deflecting force. Following Telnov [4] we estimate that about 10% of the electrons contribute importantly to the radiation. The factor $\mathcal{F}_1^*$, which we tentatively set equal to 0.1, accounts for this fraction. The fact that this estimate is so crude will ultimately have little effect on our results since the ring parameters will always be arranged to make beamstrahlung almost negligible.

As in Eq. (46), the number of beamstrahlung photons emitted per turn is

$$n_{\gamma,1}^* = \frac{3.2 \times 10^8 \,\text{GeV}^4}{u_c^*} \mathcal{F}_1^* C_\gamma \widetilde{E}^4 8\pi \frac{\xi^2 \sigma_y^2}{\beta_y^2} \frac{N^*}{\ell_z}. \tag{64}$$

Note (from Eq. (46)), that $n_{\gamma,1}^*$ has the same explicit $\widetilde{E}^4$ factor as $n_{\gamma,1}$.

What makes the synchrotron radiation damaging is the energy loss increase proportional to $\gamma^4$, making it expensive to restore the power loss. What makes beamstrahlung potentially even more damaging than synchrotron radiation at ultra-high energy is that photons near the upper end of the beamstrahlung spectrum can be energetic enough that, when an electron emits one such photon, the electron's energy is reduced to below the ring acceptance, causing the particle to be lost.

At a minimum, when an electron is lost due to beamstrahlung, its energy has to be made up in accelerating its replacement. If we neglect other complications, such as positron production and other acceleration and injection issues, the energy loss can be made commensurate with synchrotron radiation loss by introducing a "beamstrahlung penalty" defined by the energy loss ratio,

$$\mathcal{P}_{bs} = \frac{U^*_{1\,\text{effective}}}{U_1} = \frac{\mathcal{F}^*_2 0.31\, n^*_{\gamma,1} E}{0.31\, n_{\gamma,1} u_c}; \quad (65)$$

it is the replacement of $u^*_c$ by $E$ in the numerator that makes this penalty large. The factor $\mathcal{F}^*_2$ is the fraction of beamstrahlung photons energetic enough to cause the radiating electron to be lost. An accurate calculation of $\mathcal{F}^*_2$ will not be necessary since the proposed colliding beam design strategy will be to reduce beamstrahlung to unimportance.

Neglecting the energy loss from beamstrahlung photons of too low energy to cause particle loss, the numerator of Eq. (65) gives the energy loss due to beamstrahlung. The denominator gives the usual energy loss due to synchrotron radiation. Substituting from Eqs. (46) and (64)

$$\mathcal{P}_{bs} = 8\pi\,\mathcal{F}^*_1\, \frac{\xi^2 \sigma^2_y}{\beta^2_y}\, \frac{N^* R}{\ell_z} \left[ \frac{E}{u^*_c} \mathcal{F}^*_2\!\left(\frac{u^*_c}{E}, \eta\right) \right]. \quad (66)$$

This time the arguments of function $\mathcal{F}^*_2$ are given. Recall that $\mathcal{F}^*_2(\frac{u_c*}{E},\eta)$ is equal to the probability that radiation of a beamstrahlung photon will cause the radiatin electron to be lost. Alternatively, substituting from Eq. (61), the beamstrahlung penalty is

$$\mathcal{P}_{bs} = 8\pi\,\mathcal{F}^*_1 \mathcal{F}^*_2\!\left(\frac{u^*_c}{E},\eta\right) \frac{\xi \sigma_y}{\beta_y} \frac{N^* R}{0.28\,\text{m}} \frac{1}{E^2}. \quad (67)$$

## 8.2 Probability of Electron Loss Due to Beamstrahlung

The (normalized to 1) probability distribution of the (relative) energy $\zeta = u^*/u^*_c$ of a synchrotron-radiated photon, when the characteristic photon energy is $u^*_c$ is [28]

$$s(\zeta) = 1.0599\,\zeta^{0.275}\, e^{-0.965\zeta}. \quad (68)$$

$\zeta$ can be thought of as the photon energy in units of the characteristic energy. This formula is valid for all radiated energies in the range $0 < \zeta < \infty$, except in the (unimportant, both as regards normalization and physical effect) long wavelength, low energy limit. Defining a cumulative distribution function $S(\zeta) = \int_0^\zeta s(\zeta')\,d\zeta'$, its complement is $Sc(\zeta) = 1 - S(\zeta)$

$$Sc(\zeta_0) = \int_{\zeta_0}^\infty s(\zeta)\,d\zeta = \text{probability that }\zeta > \zeta_0. \quad (69)$$

The function $Sc(x)$ is plotted in Figure 14; it shows a rapidly falling loss probability for values of $x$ increasing beyond $x = 10$. The value $x = 20.05$ causes the loss probability to be $10^{-8}$.

To achieve a beamstrahlung lifetime $\tau_{bs}$ requires

$$\frac{1}{\tau_{bs}} = n^*_{\gamma,1} f \mathcal{R}^{\text{Gauss}}_{\text{unif.}}\, Sc\!\left(\frac{\eta E}{u^*_c}\right) \approx n^*_{\gamma,1} f \mathcal{R}^{\text{Gauss}}_{\text{unif.}}\, \exp\!\left(-0.91\,\frac{\eta E}{u^*_c}\right), \quad (70)$$

where the factor $\mathcal{R}^{\text{Gauss}}_{\text{unif.}}$ is explained in an appendix. Solving this for $u^*_c$ yields

$$u^*_c = \frac{-0.91\eta E}{\ln\!\left(\frac{1/\tau_{bs}}{n^*_{\gamma,1} f \mathcal{R}^{\text{Gauss}}_{\text{unif.}}}\right)}. \quad (71)$$

This determines the value of $u^*_c$ that would correspond to a beamstrahlung lifetime of $\tau_{bs}$.

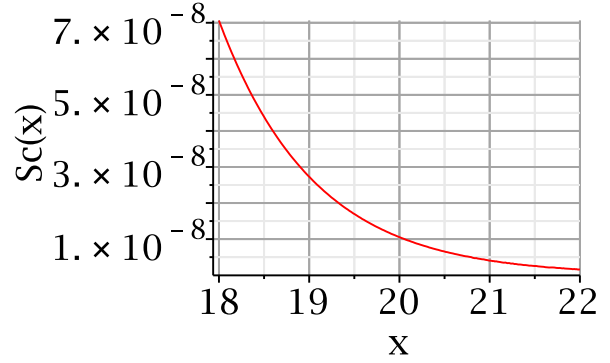

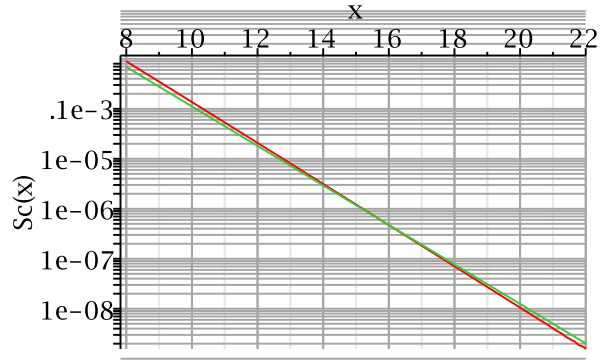

Figure 14: The loss probability per beamstrahlung-radiated photon plotted with linear and logarithmic scales. On the log plot both $Sc(x)$ and $\exp(-0.91 * x)$ are plotted, and can be seen to be approximately equal over the central range. Below some value such as $x \approx 15$, the beamstrahlung loss probability becomes unsustainable. But the loss probability falls rapidly as $x$ is increased.

## 8.3 Beamstrahlung from a Beam Bunch with Gaussian Longitudinal Profile

In this appendix, as in the body of the paper, beamstrahlung is treated with the same formulas as synchrotron radiation in the arcs, and all corresponding parameters are distinguished by asterisks. The charge density of the opposing bunch is now allowed to depend on longitudinal position $z$. As a result the beamstrahlung radius of curvature $\rho^*(z)$ and the critical energy $u^*_c(z)$ also depend on $z$. As in Eq. (63), the energy radiated in longitudinal interval $dz$ is

$$dU^* = \mathcal{F}^*_1 C_\gamma\, \frac{E^4}{\rho^*(z)}\, \frac{dz}{2\pi\rho^*(z)}, \quad (72)$$

and, as in Eq. (64), the number of photons radiated in the same range is

$$dn_\gamma^* = \frac{3.2}{u_c^*(z)\rho^*(z)} \mathcal{F}_1^* C_\gamma \frac{E^4}{\rho^*(z)} \frac{dz}{2\pi}. \quad (73)$$

The denominator factor $u_c^*(z)\rho^*(z)$ can be replaced using Eq. (60);

$$dn_\gamma^* = \frac{1}{2\pi} \frac{3.2}{2.96 \times 10^{-16} \text{ GeV.m}} \mathcal{F}_1^* C_\gamma \frac{E^4/\gamma^3}{\rho^*(z)} dz$$

$$\equiv \frac{\mathcal{K}}{\rho_0^*} \exp\left(\frac{-z^2}{2\sigma_z^2}\right) dz. \quad (74)$$

In the last step constant factors have been abbreviated by factor $\mathcal{K}$, the opposing beam has been taken to be Gaussian distributed longitudinally, and it subjects the beam being analysed to curvature $\rho_0^*$ at the origin.

It is only the occasional emission of a very hard beamstrahlung photon that can cause an electron to be lost. The distribution of photon energies $u$ can be expressed in terms of a universal (normalized to 1) probability distribution $s(\zeta)$ where $\zeta = u/u_c^*$ is the photon energy $u$ in units of the critical energy $u_c^*$ (which now depends on $z$). The cumulative probability distribution corresponding to $s(\zeta)$ is $S(\zeta_0)$ which is the probability of emissions for which $\zeta < \zeta_0$. The "complementary" cumulative distribution function $Sc(\zeta_0) = 1 - S(\zeta_0)$ gives the probability that $\zeta > \zeta_0$. These functions are discussed in greater detail in conjunction with Eqs. (68) and (69).

The fraction of the $dn_\gamma^*$ photons in Eq. (74) having energy in excess of $\eta E$ (the energy loss great enough for the electron to be lost) is given by

$$\int_{\eta E/u_c^*(z)}^{\infty} s(\zeta)\, d\zeta = Sc\left(\frac{\eta E}{u_c^*(z)}\right) = Sc\left(\frac{\eta E}{u_{c,0}^*} \exp\left(\frac{z^2}{2\sigma_z^2}\right)\right), \quad (75)$$

where $u_{c,0}^*$ is the critical energy at the origin and $u_c^*(z) = u_{c,0}^* \exp(-z^2/(2\sigma_z^2))$. Finally, integrating over $z$, the probability for an electron to be lost is equal to the number of beamstrahlung photons exceeding the loss threshold;

$$P_e^{\text{Gaussian}} = \frac{\mathcal{K}}{\rho_0^*} \int_{-\infty}^{\infty} dz \exp\left(\frac{-z^2}{2\sigma_z^2}\right) Sc\left(\frac{\eta E}{u_{c,0}^*} \exp\left(\frac{z^2}{2\sigma_z^2}\right)\right). \quad (76)$$

For a uniform longitudinal distribution the formula is similar, the integration is trivial, and the loss probability is

$$P_e^{\text{uniform}} = \frac{\mathcal{K}}{\rho_0^*} 2\sigma_z\, Sc\left(\frac{\eta E}{u_{c,0}^*}\right). \quad (77)$$

Expressed as a ratio to the loss rate from a uniform bunch, the correction factor to account for the Gaussian profile is

$$\mathcal{R}_{\text{unif.}}^{\text{Gauss}} \equiv \frac{P_e^{\text{Gaussian}}}{P_e^{\text{uniform}}} = \frac{1}{Sc\left(\frac{\eta E}{u_{c,0}^*}\right)} \int_{-\infty}^{\infty} \frac{dz}{2\sigma_z} \exp\left(\frac{-z^2}{2\sigma_z^2}\right)$$

$$\times Sc\left(\frac{\eta E}{u_{c,0}^*} \exp\left(\frac{z^2}{2\sigma_z^2}\right)\right). \quad (78)$$

In practice the argument $\eta E/u_{c,0}^*$ will be a big number, like 10 for example, and the value of the $Sc$-function in the denominator will be a very small number, like 0.0001 for example. As $z$ moves away from the origin, the value of the argument of the $Sc$ function in the numerator increases quite strongly. Furthermore the function $Sc(\zeta_0)$ decreases exponentially with increasing $\zeta_0$. This tends to "cut off" the integral, which magnifies the importance of the non-uniformity of the longitudinal charge distribution. One anticipates, therefore, a substantially smaller beamstrahlung loss rate than is obtained assuming uniform longitudinal beam profile.

# 9 C. SIMULATION OF PARAMETER DEPENDENCES OF TUNE SHIFT SATURATION

In a 2002 article published in PRST-AB [5] I described a simulation program designed to an absolute, adjustable-parameter free, calculation of the maximum specific luminosity of an e+e- ring. Though a computer simulation, this code provides an "absolute" calculation in the sense that there are no empirically-adjustable parameters. For the variety of their operating tunes and energies, shown in Table 14 (excerpted from my original paper), the ratio of theoretical to observed was $1.26 \pm 0.45$, suggesting that the model can be trusted within a factor of two.

I have now applied this identical simulation to the design of a Higgs factory. To buttress the claim that there are no free parameters I have not changed the code at all; only the values of the ring-specific parameters exhibited in Eq. (79). $\mu_0$ is the *vertical* betatron phase advance between collision points, and $\mu_x$ and $\mu_s$ are the corresponding horizontal betatron and longitudinal (synchrotron) phase advances. $a_x$ and $a_s$ are the amplitudes in units of the r.m.s. width $\sigma_x$ and bunch length $\sigma_z$. $\delta$ is the "damping decrement" of vertical betatron motion. Eq. (79) is a difference equation calculating the vertical displacement on turn $t + 1$ (time in units of period between collisions) from the two preceding values at $t$ and $t - 1$.

$$y_{t+1} = \frac{1}{1+\delta}\left(2\cos\mu_0 y_t - y_{t-1}(1-\delta)\right. \quad (79)$$
$$- 4\pi\xi \sin\mu_0 \exp\left(-a_x^2 \cos^2\frac{\mu_x(a_x)(t+t_x)}{2}\right)$$
$$\left.\times \sqrt{1 + \left(\frac{\sigma_z}{\beta_y^*}\right)^2 a_s^2 \cos^2(\mu_s(t+t_s)t)} \sqrt{\frac{\pi}{2}} \text{erf}\, \frac{y_t}{\sqrt{2}}\right)$$

The simulation consists of nothing more than checking (repeatedly and ad nauseum) whether the motion described by the difference equation is "stable" or "unstable", and noting the $\xi$-value at the transition. Technical definitions of these terms are in the original paper.

The "physics" of the simulation is that the beam height $\sigma_y$ which would otherwise be negligibly small, is swollen by beam-beam forces. The mechanism is modulation of vertical focusing acting on each particle by its own (inexorable) horizontal betatron and longitudinal synchrotron motion. This modulation provides parametric pumping force with strength inversely proportional to $\sigma_y$ (which is guaranteed to countermand the "negligibly-small" natural single beam, beam height). Amplitude dependence of the vertical tune limits the growth, however, so the resonance causes no particle loss. The growth factor also depends on the distance (in tune space) from the nearest resonance, and diffusive effects are constantly moving particles closer to or away from resonances (which are everywhere). All this is explained in the original paper. The paper analyses the growth mechanism analytically assuming "nearest-resonance" damping, and no resonance overlap. (Fortunately) the simulation automatically accounts for whatever resonances are nearby.

For any particular operating point of any particular ring most of the parameters are taken from appropriate lab reports. Scans over appropriate ranges of $a_x$ and $a_s$ are performed. According to a *saturation principle, the beam height adjusts itself to that value for which the least stable particle is barely stable at small amplitude.* This determines the tune shift parameter $\xi_{\min}^{max}$ (and its corresponding beam height) beyond which further increase of beam current causes the beam height to increase proportionally.

Some of the predictions from the 2002 paper are shown in Table 14. For existing rings these were "post-predictions" since the tunes $Q_x$ and $Q_y$ were given.

For Higgs factory predictions, since the tunes of the Higgs factory are free, what is needed are scans over the tune plane, for various values of damping decrement $\delta$ and vertical beta function $\beta_y$. The results for a six order of magnitude range of $\delta$ are shown in Figure 16. Values of $\xi(Q_x, Q_y)$ can be obtained using the grayscale to scale down the $\xi$-value at $(Q_x, Q_y)$ from the maximum value (for "all" $(Q_x, Q_y)$ pairs) (which is white, with value ximinmax shown at the top of the graph). Here "all" means all of the tune plane shown in the plots. A band of very low and a band of high $Q_y$ values is excluded. High $Q_y$ don't need to be checked since they always give small $\xi$ values. Very low $Q_y$ values usually give large $\xi$-values, but are almost surely excluded operationally.

As the figure shows, only one quarter of the tune plane is scanned. The simulation has symmetries such that all four quadrants of the fractional tune plane are equivalent (as are all integer tunes). As it happens the optimal tunes are almost invariably in near the lower right corner of the fractional tune quadrants. There are always white (maximal $\xi$) regions there. As it happens no storage rings I am aware of has attempted to operate in this region. Or, more likely, there has been some unrelated problem making operation their difficult. For all simulations in this paper I have assumed that the saturation tune shift is less than the maximum. Rather I have found the average value xiav and then, to determine a "typical" good value, have picked a value half way between the average and the maximum, i.e. xityp=(ximinmax+xiav)/2. All simulation results are based on $xi_{\text{typ}}$ values determined in this way. Dependence of $\xi_{\text{typ}}$ on $E_m$, for a particular value of $\beta_y$ and for a plausible scaling of ring radius $R$ with $E_m$, is shown in Figure 15.

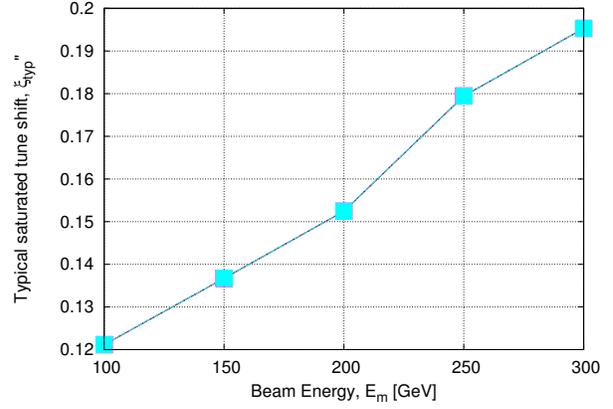

Figure 15: Plot of "typical" saturated tune shift $\xi_{\text{typ}}$ as a function of maximum beam energy $E_m$ for ring radius $R$ scaling as $E_m^{1.25}$. $\beta_y = \sigma_z = 5$ mm.

The main independent variables in the simulation are the damping decrement $\delta$ (DEL in the code), and $\beta_y^*$ (BETYST in the code, simply $\beta_y$ in this paper) the vertical beta function at the intersection point. The main output is $\xi_{\min}^{max}$ (XIMINMAX or ximinmax in the code) which is the quantity plotted in the page of density plots making up Figure 16. For these plots the damping decrement (shown as delta) ranges over six orders of magnitude. The saturation tune shifts can be read off (roughly) using the grayscale.

Figure 17 plots the quantity printed at the top of the plots in Figure 16, for a series of values $\beta_y^* = 0.001, 0.002, 0.005, 0.01, 0.02$ m. For very small values of damping decrement $\delta$ the saturation tune shift is independent of $\delta$. But for values of $\delta$ exceeding 0.01, as will be the case for a Higgs factory, there is an appreciable increase in saturated tune shift with increasing damping decrement. Figure 19 plots the same data, but in the form $\xi_{\min}^{max}\sqrt{\sigma_z/\beta_y^*}$; this provides the simulation code input to luminosity formula (93).

Figure 20 plots the $\xi_{\min}^{max}$ as a function of number of intersection points $N^*$. This is somewhat redundant of the previous plots since the only quantity changing is $\delta_{N^*} = \delta/N^*$ where $\delta_{N^*}$ is the damping decrement from one collision point to the next. For Higgs factory simulations we distill simulation results shown in the following graphs into the form

$$\xi^{\text{sat.}}(\delta_i, \beta_y, \sigma_z) = \xi^{\text{sat.}}(\delta_i, 0.01, 0.01)\sqrt{\frac{\beta_y}{\sigma_z}}$$
$$\equiv \sqrt{r_{yz}}\,\xi^{\text{sat.}}(\delta_i, 0.01, 0.01), \qquad (80)$$

(with all lengths in meters). Values of $\xi^{\text{sat.}}(\delta_i, 0.01, 0.01)$ (with $\sigma_z = \beta_y = 1$ cm) are obtained at each operating point $i$ (with the appropriate damping decrement $\delta_i$) and substituted into Eq. (80) to extrapolate to the actual values of $\sigma_z$ and $\beta_y$. Fits of this form are shown in the appendix. They give $\pm 50\%$ accuracy in the range from $\beta_y \approx 2$ cm down to $\beta_y > 2$ mm, but deviate strongly for $\beta_y < 2$ mm. There are corresponding restrictions on $\sigma_z$. For greater accuracy

Table 14: Parameters of some circular, flat beam, e+e- colliding rings, and the saturation tune shift values predicted (with no free parameters) by the simulation.

| Ring IP's | $Q_x$/IP | $Q_y$/IP | $Q_s$/IP | $\sigma_z$ | $\beta_y^*$ | $10^4 \delta_y$ | $\xi_{\text{th.}}$ | $\Delta Q_{y,\text{exp.}}$ | th/exp |
|---|---|---|---|---|---|---|---|---|---|
| VEPP4 1 | 8.55 | 9.57 | 0.024 | 0.06 | 0.12 | 1.68 | 0.028 | 0.046 | 0.61 |
| PEP-1IP 1 | 21.296 | 18.205 | 0.024 | 0.021 | 0.05 | 6.86 | 0.076 | 0.049 | 1.55 |
| PEP-2IP 2 | 5.303 | 9.1065 | 0.0175 | 0.020 | 0.14 | 4.08 | 0.050 | 0.054 | 0.93 |
| CESR-4.7 2 | 4.697 | 4.682 | 0.049 | 0.020 | 0.03 | 0.38 | 0.037 | 0.018 | 2.06 |
| CESR-5.0 2 | 4.697 | 4.682 | 0.049 | 0.021 | 0.03 | 0.46 | 0.034 | 0.022 | 1.55 |
| CESR-5.3 2 | 4.697 | 4.682 | 0.049 | 0.023 | 0.03 | 0.55 | 0.029 | 0.025 | 1.16 |
| CESR-5.5 2 | 4.697 | 4.682 | 0.049 | 0.024 | 0.03 | 0.61 | 0.027 | 0.027 | 1.00 |
| CESR-2000 1 | 10.52 | 9.57 | 0.055 | 0.019 | 0.02 | 1.113 | 0.028 | 0.043 | 0.65 |
| KEK-1IP 1 | 10.13 | 10.27 | 0.037 | 0.014 | 0.03 | 2.84 | 0.046 | 0.047 | 0.98 |
| KEK-2IP 2 | 4.565 | 4.60 | 0.021 | 0.015 | 0.03 | 1.42 | 0.048 | 0.027 | 1.78 |
| PEP-LER 1 | 38.65 | 36.58 | 0.027 | 0.0123 | 0.0125 | 1.17 | 0.044 | 0.044 | 1.00 |
| KEK-LER 1 | 45.518 | 44.096 | 0.021 | 0.0057 | 0.007 | 2.34 | 0.042 | 0.032 | 1.31 |
| BEPC 1 | 5.80 | 6.70 | 0.020 | 0.05 | 0.05 | 0.16 | 0.068 | 0.039 | 1.74 |

(within the limitations of the model) the full simulation would have to be run at each data point.

## 10  D. LUMINOSITY FORMULAS

### 10.1  Aspect Ratios and Hourglass Correction

To (artificially) reduce the number of free parameters, while retaining the more important dependencies we define two ratios:

$$a_{xy} = \frac{\sigma_x}{\sigma_y} = \text{``transverse aspect ratio''} \gg 1, \tag{81}$$

$$r_{yz} = \frac{\beta_y}{\sigma_z} = \text{ratio of ``Rayleigh length'' } \beta_y \text{ to } \sigma_z \ll 1. \tag{82}$$

We will tend to keep $a_{xy}$ and $r_{yz}$ more or less constant while allowing $\sigma_y$ to vary over significantly large ranges. Built into all luminosity formulas is the flat beam assumption, $\sigma_y \ll \sigma_x$. A nominal value for the width-by-height ratio will be $a_{xy} = 15$. This is not necessarily optimal, but the "flat beam condition" $a_{xy} \gg 1$ is always assumed to be true. It will be shown shortly that the condition $r_{yz} \ll 1$ is highly favorable for reducing the importance of beamstrahlung radiation. Again, though not necessarily optimal, we adopt $r_{yz} = 0.5$ as the nominal value.

When the bunch length $\sigma_z$ is allowed to be long compared to $\beta_y$ one has to reduce the luminosity by an "hourglass factor" [26]

$$H(r_{yz}) = \frac{r_{yz}}{\pi} \exp\left(\frac{r_{yz}^2}{2}\right) K_0\left(\frac{r_{yz}^2}{2}\right), \tag{83}$$

where $K_0$ is a Bessel function. This is plotted in in Figure 21.

### 10.2  RF-Power-Dominated Luminosity

As modified from reference [4] to account for multiple bunches and multiple interaction points, the luminosity summed over $N^*$ interaction points is

$$\mathcal{L}_{\text{summed}} = N^* \frac{N_{\text{tot}}^2}{N_b} \frac{f}{4\pi} \frac{1}{a_{xy} \sigma_y^2}, \tag{84}$$

where the revolution frequency is $f = c/C$. The product $N_b N_p = N_{\text{tot}}$ is the total number of particles stored in each beam.

All numerical values required are in Table 1 and its continuations. The luminosity will always be RF power limited—if not, the number of bunches $N_b$ can be increased, with proportional increase in luminosity. The column labelled $n_1$ in Table 1 gives the number of particles $N_{\text{tot}}$ per MW of RF power. This is obtained, using

$$P_{\text{rf}}[\text{MW}] = I[\text{A}]V[\text{MV}] = f N_{\text{tot}} e U_1[\text{MeV}], \quad \text{or} \tag{85}$$

$$n_1 = \frac{N_p N_b}{P_{\text{rf}}[\text{MW}]} = \frac{10^{-3}}{f e U_1[\text{GeV}]}. \tag{86}$$

For given $P_{\text{rf}}$, $N_{\text{tot}} = n_1 P_{\text{rf}}$ can be treated as known. The best chance for saturating the beam-beam tune shift is with just one bunch, $N_b = 1$, in which case $N_{\text{tot}} = N_p$. But for $N^* = 4$ intersection points, at least $N_b = 2$ bunches are required.

From Eq. (4), including the "hourglass-effect correction factor" $H(r_{yz})$,

$$\boxed{\mathcal{L}_{\text{summed}}^{\text{RF-pow}} = \frac{N^*}{N_b} H(r_{yz}) \frac{n_1^2}{a_{xy}} \frac{f}{4\pi} \left(\frac{P_{\text{rf}}[\text{MW}]}{\sigma_y}\right)^2.} \tag{87}$$

This is the first of four luminosity formulas. The other luminosity formulas, giving only upper limits, are needed to check that beamstrahlung levels are tolerable and that the beam-beam tune shift being assumed does not exceed the saturation level. For any given choice of parameters it will be the lowest of the four luminosity formulas that has to be accepted. But under optimal conditions all four luminosity formulas should give approximately the same values and there is likely to be a nearby parameter set giving better luminosity than the lowest one.

### 10.3  Tune-Shift-Saturated Luminosity

From Eq. (53) we can calculate the beam-beam tune shift,

$$\xi = \frac{1}{2\pi} \frac{1}{\gamma} N_p r_e \frac{\beta_y}{\sigma_x \sigma_y}. \tag{88}$$

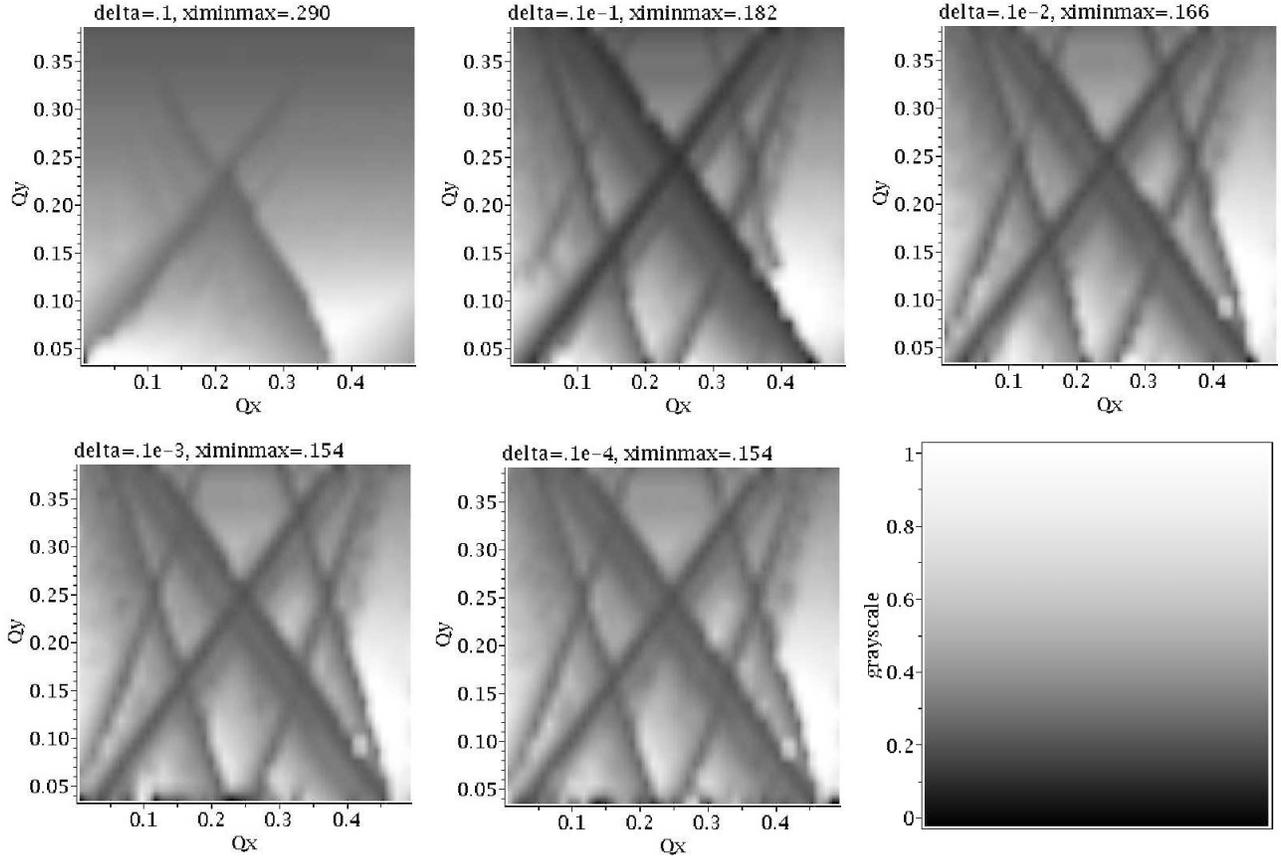

Figure 16: Density plots of $\xi^{\max}_{\min}$ over (most of) one quarter of the $(Q_x, Q_y)$ tune plane for a five order of magnitude range of damping decrements. The symmetry of the simulation is such that the density plots for the other three quadrants of the fractional tune plane are identical to this quadrant. Starting from the upper left, in sequence left to right, the damping decrements are $\delta = 10^{-1}, 10^{-2}, 10^{-3}, 10^{-4}, 10^{-5}$. The $\delta = 10^{-6}$ tune plot (which is indistinguishable from the $\delta = 10^{-5}$ plot) is replaced by the grayscale reference. The highest $\xi^{\max}_{\min}$ value is plotted above each plot. Within each plot the $\xi^{\max}_{\min}(Q_x, Q_y)$ can be inferred from this maximum value multiplied by the (lower right) grayscale value.

The appearances of these density plots are absolutely typical. The hundreds of plots made all look very much like this, including the washing out of resonance lines when the damping decrement is large, and the optimal performance near the lower right corner. Only one quadrant is displayed since the model is unaffected by half integer shifts in either tune.

With unlimited beam power, the maximum value of $\xi$ would be the "saturated value" $\xi^{\text{sat.}}$. Rearranging this equation in this case gives, for the beam "area",

$$A^{(y)}_{\beta_y} = \pi \sigma_x \sigma_y = \frac{N_p r_e}{2\gamma} \frac{1}{(\xi^{\text{sat.}}/\beta_y)}. \tag{89}$$

But we also have to limit $\xi_x$, for which the formula is

$$\xi_x = \frac{1}{2\pi} \frac{1}{\gamma} N_p r_e \frac{\beta_x}{\sigma_x^2} = \frac{1}{2\pi} \frac{1}{\gamma} N_p r_e \frac{\beta_x}{a_{xy} \sigma_x \sigma_y}. \tag{90}$$

which gives an $\xi_x$-limiting area

$$A^{(x)}_{\beta_y} = \frac{N_p r_e}{2\gamma} \frac{\beta_x/\beta_y}{a_{xy}(\xi^{\text{sat.}}/\beta_y)}. \tag{91}$$

To make these areas equal we adjust $\beta_x$ to the value

$$\beta_x = a_{xy} \beta_y. \tag{92}$$

Based on the simulation model, the area denominator factor $(\xi^{\text{sat.}}/\beta_y)$ can be obtained from Figure 18. Note, though, that this figure arbitrarily assumes $\sigma_z = 0.01$ m, which is by no means optimal. More useful is to use Figure 5, which gives $(\xi^{\text{sat.}}/\beta_y)$ constrained by $r_{yz} = 1$; i.e. $\sigma_z = \beta_y$.

But, especially at high energies, the area $A$ may be unachievably small. In this case the actual $\xi$-value may be less because the beam currents are too small to saturate $\xi$. Conversely, in actual operations, especially at low energy, it may be possible to push the luminosity to higher values by pushing the beam currents to higher values than are required to saturate the tune shift.

Substituting into Eq. (4) yields

$$\boxed{\mathcal{L}^{\text{bb-sat.}}_{\text{summed}} = N^* N_{\text{tot}} f \frac{\gamma}{2r_e} H(r_{yz})(\xi^{\text{sat.}}/\beta_y).} \tag{93}$$

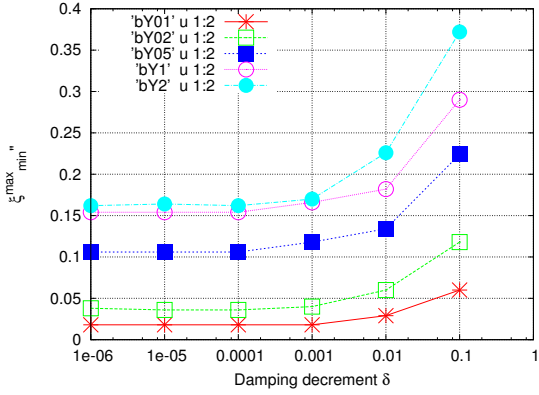

Figure 17: Plot of max/min tune shift value XIMINMAX, $\xi_{\min}^{\max}$ versus damping decrement DEL ($\delta$). For a range of values of $\beta_y$; $\beta_y = 0.1, 0.2, 0.5, 1, 2$ cm. In all cases here, $\sigma_z = 0.01$ m.

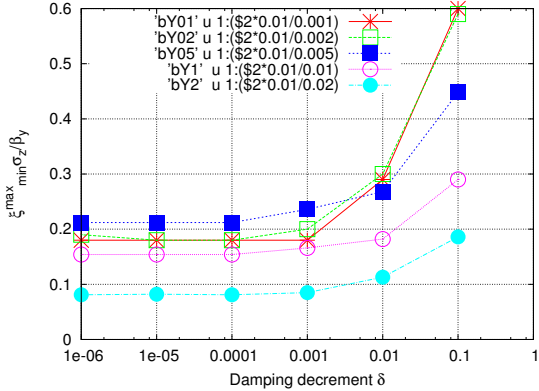

Figure 18: Plot of weighted saturated tune shift value $\xi_{\min}^{\max} \sigma_z/\beta_y$ versus damping decrement DEL$\equiv \delta$. Values from this plot are especially simple (e.g. $\xi/\beta_y \approx 0.19/\sigma_z = 19$ /m for $\delta = 0.01$) for evaluating the luminosity using Eq. (93), which gives luminosity proportional to $\xi_{\min}^{\max}/\beta_y$ when the luminosity is limited by RF power (which is to say, almost always, in a Higgs factory). In all cases here, $\sigma_z = 0.01$ m.

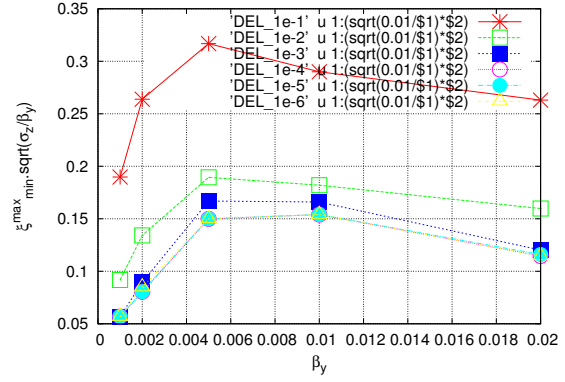

Figure 19: This is the same data as in Figure 17, but the saturated tune shift value is weighted by $\sqrt{\sigma_z/\beta_y}$ and plotted as a function of $\beta_y$ with DEL held constant, instead of the other way round. In all cases here, $\sigma_z = 0.01$ m.

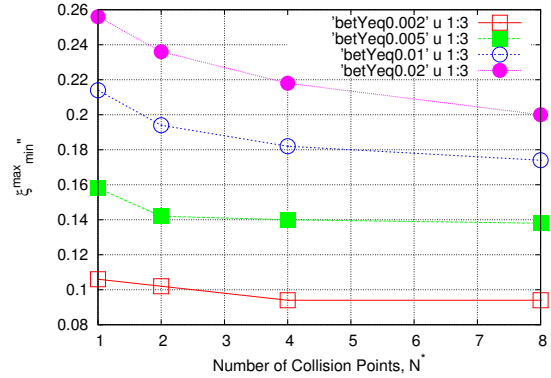

Figure 20: Plot of maximum/minimum tune shift value XIMINMAX $\xi_{\min}^{\max}$ versus number of collision points $N^*$. The damping decrement and synchrotron tune vary as $\delta = \delta_0/N^**$ and $Q_s = 0.03/N^**$. The saturated tune shift depends only weakly on the number of intersection points.

This agrees with Telnov's Eq. (15), except for our extra factor of $N^*$, our inclusion of hourglass correction, and our intention to obtain ($\xi^{\text{sat.}}/\beta_y$) from simulation. $\mathcal{L}_{\text{summed}}^{\text{bb-sat.}}$ is the beam-beam saturated luminosity. With $\xi^{\text{sat.}}/\beta_y$) given by simulation, and $N_{\text{tot}}$ known, Eq. (93) fixes the summed luminosity predicted by the simulation, but only if $N_p$ is large enough to "saturate" the tune shift.

As commented earlier (according to the simulation model) both factors in the product $H(r_{yz})\xi^{\text{sat.}}$ appearing in Eq. (93) are proportional to the ratio $\sqrt{r_{yz}} = \sqrt{\beta_y/\sigma_z}$. See Figure 21. The resulting proportionality to $1/\sigma_z$ tends to frustrate the benefit of decreasing beamstrahlung by increasing bunch length $\sigma_z$.

If the beam currents are unnecessarily large for saturating $\xi$, then $\mathcal{L}_{\text{summed}}^{\text{bb-sat.}}$ can be the actual summed luminosity, but only if the beams are split into the appropriate number of bunches. For determining optimal parameters the important difference between $\mathcal{L}_{\text{summed}}^{\text{bb-sat.}}$ and $\mathcal{L}_{\text{summed}}^{\text{RF-pow}}$ is their different dependencies on the bunch number $N_b$. By appropriate choice of $N_b$ these luminosities could be made exactly equal except for the fact that $N_b$ has to be an integer. In the tables, to obtain smoother variation, this integer requirement for $B_b$ is not imposed. The importance of this defect can be estimated by the degree to which $N_b$ deviates from being an integer or, for $N* = 4$, and even integer.

A significant parameter to be determined is the vertical beta function $\beta_y$. According to Figure 16, $\xi/\beta_y$ is maximal near $\beta_y = 5$ mm, and approximately constant over the range from 1 to 10 mm. To the extent $\xi/\beta_y$ can be treated as independent of $N_b$ and $N_p$, the $N^*N_{\text{tot}}$ factor in Eq. (93) exhibits the expected proportionality to the number of intersection points and to the total circulating charge.

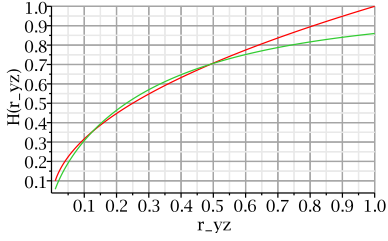

Figure 21: Plot of the "flat beam hourglass correction factor" $H(r_{yz})$, with $r_{yz} \equiv \beta_y/\sigma_z$. (Fortuitously) the function $\sqrt{r_{yz}}$ shown plotted red (and passing through $r_{yz} = 1$) provides an excellent fit to the hourglass function.

## 10.4 Beamstrahlung-Limited Luminosity

As long as the beams remain flat and with constant width through the intersection region, the vertical beam-beam deflection is approximately independent of $z$, the longitudinal deviation from the intersection point. Unlike the differential luminosity, which falls off rapidly as a function of $z$, there is no significant hourglass correction to either the beam-beam tune shift or the beamstrahlung, (but the beamstrahlung penalty is strongly reduced by increasing $\sigma_z$).

The critical energy $u_c^*$ for beamstrahlung radiation is given by Eq. (61);

$$u_c^* = 28.0\,\text{GeV.m}\,\sqrt{2/\pi}\,\frac{r_{yz}}{\beta_y^2}\,\widetilde{E}^3\,(\xi\sigma_y), \qquad (94)$$

and from Eq. (88),

$$\xi\sigma_y = \frac{1}{N_b}\,\frac{1}{2\pi}\,\frac{1}{1.96\times 10^5 \widetilde{E}}\,n_1 r_e\,\frac{\beta_y}{a_{xy}}\,\frac{P_\text{rf}[\text{MW}]}{\sigma_y}. \qquad (95)$$

Substituting this into Eq. (94) produces

$$u_c^* = \frac{1}{N_b}\,\frac{28.0\,\text{GeV.m}\,\sqrt{2/\pi}}{2\pi \times 1.96 \times 10^5}\,\frac{r_{yz}}{a_{xy}\beta_y}\,\widetilde{E}^2\,n_1 r_e\,\frac{P_\text{rf}[\text{MW}]}{\sigma_y}. \qquad (96)$$

Finally, this equation can be used to express the luminosity in terms of the critical beamstrahlung energy by substitution into Eq. (87),

$$\mathcal{L}_\text{summed}^\text{bs-limited} = \qquad (97)$$
$$N^* N_b\,\frac{H(r_{yz})}{r_{yz}^2}\,a_{xy}\beta_y^2\,f\left(\frac{\sqrt{\pi}\,1.96\times 10^5}{28.0\,\text{GeV.m}\,\sqrt{2/\pi}}\right)^2\,\frac{1}{r_e^2 \widetilde{E}^4}\,u_c^{*2}.$$

Clearly it is advantageous to make $u_c^*$ as large as possible, consistent with keeping the beamstrahlung loss rate acceptably low. Increasing the number of bunches is also helpful but, at highest energies this avenue is closed.

There are two ways the radiation of one or a few hard $\gamma$-rays can cause the radiating electron to be lost—transversely or longitudinally.

In Telnov's calculation of beamstrahlung-limited luminosity, $u_c^*$ is compared with the tolerable fractional energy acceptance $\eta E$. The beam decay rate ascribable to beamstrahlung is the time rate of beamstrahlung emissions per electron, multiplied by the probability per beamstrahlung photon that its energy exceeds $\eta E$. Substituting for $u_c^*$ from Eq. (71), the transverse-limited luminosity corresponding to beamstrahlung lifetime $\tau_\text{bs}$ is

$$\mathcal{L}_\text{trans}^\text{bs} = N^* N_b\,\frac{H(r_{yz})}{r_{yz}^2}\,a_{xy}\beta_y^2\,f\left(\frac{\sqrt{\pi}\,1.96\times 10^5}{28.0\,\text{m}\,\sqrt{2/\pi}}\right)^2$$
$$\times \frac{1}{r_e^2 \widetilde{E}^2}\left(\frac{91\eta}{\ln\left(\frac{1/\tau_\text{bs}}{f\,n_{\gamma,1}^*\,\mathcal{R}_\text{unif.}^\text{Gauas}}\right)}\right)^2. \qquad (98)$$

This form of calculation is valid for a low energy storage ring, with one RF cavity, with a synchrotron damping decrement per turn much less than 1. With insufficient damping to pull the particle energy back toward equilibrium, a particle's energy deviation remains outside the lattice energy acceptance for many turns. An electron emitting an anomalously-hard beamstrahlung photon can easily be lost, even if there is ample RF overvoltage to keep it in a stable RF bucket.

In a Higgs factory the situation will be qualitatively different. There will still be ample RF overvoltage, except at the very highest beam energy, where the luminosity will necessarily drop precipitously. But there will be multiple RF cavities distributed at equal intervals around the ring. Furthermore these cavities will be routinely restoring a significant fraction (more than one tenth for example) of the beam energy each turn.

In this circumstance it seems appropriate to compare $u_c^*$ to $eV_\text{excess}$, the excess energy the RF cavities are capable of restoring each turn, rather than to $\eta E$. Values of $eV_\text{excess}$ are given in Table 1. When $eV_\text{excess}$ is negative, the luminosity obviously vanishes. But, especially for beam energies less than $250\,\text{GeV}$, $eV_\text{excess}$ is many GeV. Even a 10 or 20 GeV beamstrahlung emission will typically not knock an electron out of its stable bucket. This luminosity further assumes (and it must be checked) that the distributed RF will save the particle as long as it remains within the stable RF bucket.

It has been seen earlier that the probability for the energy of a single beamstrahlung photon to exceed $10u_c^*$ is about $10^{-4}$. (Multiplied by the emission rate, this would give an unacceptably large loss rate.) The probability for the same electron to emit two photons with energy summing to $20u_c^*$ will be about $10^{-12}$. (Multiplied by a total number of emissions in one damping time, this gives the probability of energy loss in excess of $20u_c^*$, which would cause the particle to be lost.) This probability is small enough to make beamstrahlung loss small compared to other loss mechanisms. By this estimate, the requirement on the critical energy will be

$$u_c^* < \frac{eV_\text{excess}}{20}. \qquad (99)$$

Accepting this as an equality, and substituting into Eq. (97)

$$\boxed{\begin{aligned}\mathcal{L}_{\text{longit}}^{\text{bs}} &= N^* N_b \frac{H(r_{yz})}{r_{yz}^2} a_{xy} \beta_y^2 \\ &\times f\left(\frac{\sqrt{\pi}}{28.0\,\text{GeV.m}} \frac{1.96 \times 10^5}{\sqrt{2/\pi}}\right)^2 \frac{1}{r_e^2 \widetilde{E}^4} \left(\frac{eV_{\text{excess}}}{20}\right)^2.\end{aligned}}$$
(100)

This formula is deceptively simple because it ascribes all luminosity restriction to beamstrahlung. Though it is sensitive to the bunch aspect ratios, it has the curious property of being independent of the overall bunch scale. Only after $\sigma_y$ has been fixed are the absolute bunch dimensions determined. Furthermore the formula suggests, unrealistically, that the luminosity can be made arbitrarily large by choosing $a_{xy}$ large (wide beam) and $r_{yz}^2$ small (long beam). Counter-intuitively, the formula also favors large $\beta_y$. But these factors are constrained by other design considerations. Only realistically achievable parameters can be used, and other constraints, such as whether the assumed aspect ratios are achievable, have to be checked.

For all these reasons $\mathcal{L}_{\text{summed}}^{\text{bs-limited.}}$ will always be treated as an *upper limit* on the luminosity that can be achieved with beamstrahlung still being negligible.

### 10.5 Horizontal Beam Conditions

The r.m.s. horizontal beam width at the crossing points is given by $a_{xy}\sigma_y^*$ and the arc-determined horizontal emittance is given by Eq. (49). This value of emittance is known (from the lattice design) to give an acceptably small horizontal beam size in the arcs of the ring, $\sigma_x^{\text{arc}} = \sqrt{\beta_x^{\text{arc}}\epsilon_x}$. The discussion of beam-beam saturated operation showed that $\beta_x^*$ has to be determined by the condition that $x$ and $y$ beam-beam tune shifts are approximately equal. The two determinations of $\beta_x$ will not, in general, be equal. If this is the case we have to suppose that the lattice will be modified to make the arc-determined horizontal emittance give the required beam aspect ratio at the IP. The horizontal emittance is then given by

$$\epsilon_x = \frac{\sigma_x^2}{\beta_x^*}. \qquad (101)$$

One thing that can go wrong with $\beta_x$, and very hard to estimate, is whether the intersection region optics can actually be designed. A possible failure mechanism is for $\sigma_{x,\text{max}}^{\text{arc}}$ to be insufficiently small compared to the separation of the counter-circulating beams. These concerns have been largely allayed by preliminary lattice design by the LBNL/SLAC group [17].

## 11 E. DECONSTRUCTING YUNHAI CAI'S IR OPTICS

This section contains an elementary discussion of chromaticity correction of a circular e+e- Higgs factory, with emphasis on CepC. To achieve high luminosity at the intersection point requires the beam to be focussed down to a "point" or a "ribbon" or (given Liouville's phase space requirements) a flattened ellipse, whose dimensions are characterized by the (Twiss) beta-functions $\beta_x$ and $\beta_y$.

For various reasons the entire ring, including the intersection region, has to be approximately "achromatic", meaning independent of fractional momentum offset $\delta$. Regrettably magnetic lenses are inherently chromatic since their focal lengths are proportional to particle momentum. The only way known to reduce the chromaticity is to place nonlinear elements, namely sextupoles (loosely speaking they are lenses with focal length proportional to radial position) at locations in the lattice at which there is non-vanishing dispersion (radial position proportional to momentum). This is routine, since higher momentum particles tend to circulate more toward the outside of the vacuum chamber.

In early storage rings, up to and including LEP, the IP chromaticity was cancelled "globally" using sextupoles distributed more or less uniformly in the arcs of the ring. Starting with the B-factories, and influenced by ILC "final focus optics" studies, schemes for distributing the chromatic compensation "locally" in intersection region (IR) have been developed. Yunhai Cai's intersection region optics design is local in this sense.

Local chromatic compensation has been critical to obtaining the astonishingly large B-factory luminosities. Unlike B-factories, at the Higgs production threshhold energy and higher energies, the Higgs factory will be RF-dominated (as explained in the rest of the paper). This is a sufficiently large difference to call into question the superiority of local over global chromaticity correction. But this appendix *does not* address this question. Rather it assumes that the local route will be taken.

Large values of either of the lattice beta-functions, $\beta_x$ or $\beta_y$ implies operational sensitivity to lattice imperfections. It will become obvious that achieving small $\beta_y^*$ at the IP (as needed for high luminosity) implies a very large value of $\beta_y$ peak near the IP. And, for the same reason, a smallish value of $\beta_x^*$ implies a largish value of $\beta_x$.

Primary emphasis is therefore on preserving the capability to adjust the vertical beta function $\beta_y^*$ at the crossing points, without changing in quadrupole locations. As Cai has explained, this precise capability is automatically built into his lattice design. This is important to retain the option of running with large numbers of bunches $N_b$ and small $\beta_y^*$ on the one hand, or small $N_b$ and larger $\beta_y^*$ on the other.

In section "Scaling Law Dependence of Luminosity on Free Space $L^*$" this capability is used to investigate the dependence of luminosity on $L^*$. (This is of special importance for particle physics experimentalists who have to compromise between their desire for long free space for their detection apparatus and their desire for high luminosity.) Using dimensional analysis a scaling relation between $L^*$ and the maximum value of $\beta_y^{\text{max}}$ (and hence dynamic aperture) is obtained.

Copied from Yunhai Cai's $\beta_y^* = 1$ mm design, the main chromaticity compensation is performed by matched sextupole pairs, separated by the -I, $\pi$-phase advance sections, needed to cancel their nonlinear kicks. Tunability of these sections, for example to reduce maximum beta function values compared to the $\beta_y = 1$ mm design, has to be demonstrated.

To obtain closed-form, analytic solutions, in this section only thin quadrupoles are used. The solutions are obtained using MAPLE. There is no numerical equation solving, for example using MAD. But eventual precise parameter fitting will require thick element analysis and numerical fitting, using MAD for example.

An example in applying the formulas will be to reduce and equalize the beta function peaks. For Yunhai, using MAD this would be a few hours work. We will take longer and be less successful, but only as an exercise.

### 11.1 Waist-to-Waist Lattice Matching

In terms of Twiss parameters, the transfer matrix from arbitrary lattice point 1 (e.g. the IP) to arbitrary lattice point 2 (e.g. the start of the FODO arc) is

$$\mathbf{M}(2 \leftarrow 1) = \begin{pmatrix} \sqrt{\frac{\beta_2}{\beta_1}}(C + \alpha_1 S) & \sqrt{\beta_1 \beta_2} S \\ \frac{-S(1+\alpha_1\alpha_2)+C(\alpha_1-\alpha_2)}{\sqrt{\beta_1\beta_2}} & \sqrt{\frac{\beta_1}{\beta_2}}(C - \alpha_2 S) \end{pmatrix}, \quad (102)$$

where $S \equiv \sin\phi$, $C \equiv \cos\phi$, and where $\phi$ is the betatron phase advance through the section from 1 to 2.

By symmetry at the IP, $\alpha_1 = 0$ in both planes, and this remains true independent of momentum offset $\delta$. By starting and ending the regular arc with a half-quad one can, without loss of generality, assume that $\alpha_2 = 0$ where the IR section joins the arc, which we assume to be FODO. (This equality will break down for $\delta \neq 0$ however.) The full on-momentum transfer matrix $\mathbf{M}$ from IP to arc start then has the form

$$\begin{pmatrix} \sqrt{\frac{\beta_{x2}}{\beta_{x1}}} C_x & \sqrt{\beta_{x1}\beta_{x2}} S_x & 0 & 0 \\ -\frac{1}{\sqrt{\beta_{x1}\beta_{x2}}} S_x & \sqrt{\frac{\beta_{x1}}{\beta_{x2}}} C_x & 0 & 0 \\ 0 & 0 & \sqrt{\frac{\beta_{y2}}{\beta_{y1}}} C_y & \sqrt{\beta_{y1}\beta_{y2}} S_y \\ 0 & 0 & -\frac{1}{\sqrt{\beta_{y1}\beta_{y2}}} S_y & \sqrt{\frac{\beta_{y1}}{\beta_{y2}}} C_y \end{pmatrix}. \quad (103)$$

This same form can be used for matching between any sectors that have the property that there are simultaneous waists ($\alpha_x = \alpha_y = 0$) in both planes at both ends. usually there is a maximum or minimum of one or both beta functions at any given quadrupole location. Both $\alpha$-functions change sign in these cases, and for these quads there is a "waist" in both planes in the quad interior. Especially for thin quads it is usually valid to approximate these waist locations as coinciding. In terms of the beta functions at both ends of a simultaneous waist, the following equations can be derived by combining elements of the partitioned sub-matrices $\mathbf{M}$;

$$\begin{aligned} M_{11}\beta_{x1} - M_{22}\beta_{x2} &= 0, \\ M_{33}\beta_{y1} - M_{44}\beta_{y2} &= 0, \\ M_{12} + \beta_{x1}\beta_{x2}M_{21} &= 0, \\ M_{34} + \beta_{y1}\beta_{y2}M_{43} &= 0. \end{aligned} \quad (104)$$

Other equations, such as

$$\begin{aligned} \beta_{x1} M_{11}M_{21} + M_{12}M_{22}/\beta_{x1} &= 0, \\ M_{11}M_{12}/\beta_{x2} + \beta_{x2} M_{12}M_{22} &= 0, \end{aligned} \quad (105)$$

plus the corresponding $y$ equations can be derived but, being quadratic in the matrix elements, and hence giving higher order polynomials in the unknowns, they are less amenable to solution by polynomial solvers such as MAPLE, which is what I use. But they can be used for checking purposes.

Here we discuss transfer matrices between lattice points which can, potentially, have waists in both planes. Just because there *can* be waists at both ends does not, however, guarantee that there *will* be. But the transfer matrix through a sector can be designed so that a waist at one end will guarantee a waist also at the other end; this is not unlike requiring a differential equation to satisfy periodic boundary conditions. For lattice design this can be a useful approach since some waists have to exist by symmetry and some others can be required to exist by design constraints which simply *must not* be compromised.

We cannot afford to assume there is zero dispersion through the IR sector because this would make it impossible to compensate chromaticity locally. Neglecting the tiny radial focusing that occurs in bend elements, the transverse optics is determined by the drifts and quadrupoles. Then the elements $M_{ij}$ are polynomial functions of the quad strengths $q_i$ and drift lengths $L_i$. In principle, if the four parameters $\beta_{x1}$, $\beta_{x2}$, $\beta_{y1}$, and $\beta_{y2}$ are given, and four of the $q_i$ and $L_i$ are free variables, then the variables can be determined to match the betas. In practice it is not so simple, as there may be multiple solutions, or worse, none at all. Some or all of the matches may have complex strengths or complex or negative lengths. Nevertheless, with a certain amount of trial and error, Eqs. (104) can be used to derive closed form, analytically-exact matched solutions. To obtain unique solutions it may be necessary to introduce intermediate points and to complete the matching sector by sector.

Supposing that the $q_i$ and $L_i$ of the IP-to-regular-arc-region have been determined, the matrix $\mathbf{M}$ is known, and satisfies Eqs. (104) and (105). It would be nice if these equations could be satisfied identically in $\delta$. But this is clearly impossible. (See, for example, in Steffen [27] for proof of the impossibility of designing fully momentum-independent bend-free sections.) One sometimes concentrates on compensating just the "global chromaticities" $\chi_x \equiv Q'_x \equiv dQ_x/d\delta$ and $\chi_y \equiv Q'_y \equiv dQ_y/d\delta$. Typically the chromaticities due to the IR section are comparable to or, especially after "$\beta$ squeeze" greater than the chromaticities due to the rest of the ring.

Various considerations have always made it important for the dispersion to vanish at the intersection point (IP). For an e+e- Higgs factory there us another new important reason; beamstrahlung radiation. The simplest way to cancel the ring chromaticities is to uniformly increase the strengths of the arc sextupoles. Even with no intersection regions the presence of arc sextupoles limits the dynamic aperture of the ring.

But this limitation can be quite harmless and, for a Higgs factory there are so many arc sextupoles that the arc-dominated dynamic aperture would probably remain acceptably large by simply increasing the arc sextupole strengths uniformly to cancel the IR-induced chromaticity. Though this easily adjusts the chromaticities to zero, this approach proves to be unsatisfactory.

Unfortunately it is the chromatic mismatch between arcs and IR sections that dominates the off-momentum dynamic aperture. For a Higgs factory it is critically important to maximize the off-momentum dynamic aperture. It is these considerations that makes achromatic IR design attractive. One (unfortunate) consequence of local chromaticity control is that bend elements need to be built into the IR design in order to produce and control the momentum dispersion.

It is usually assumed therefore that the dominant phenomenon limiting off-momentum acceptance is "chromatic beta-mismatch". Because low-beta IR's are "highly tuned", the $q_i/(1+\delta)$ momentum dependence of the IR region quadrupoles causes a mismatch which launches a $\delta$-dependent "beta-wave" which seriously reduces the off-momentum aperture.

## 11.2 Achromatic Higgs Factory IR Optics

The entries in Table 15 and its attached figure describe (an approximation to) the Cai IR design, with waist locations identified by letters A,B,C, ..., I. With no $x, y$ coupling the waist-to-waist transfer matrix of Eq. (103) is block diagonal with blocks of the form,

$$\mathbf{M}(\mu) = \begin{pmatrix} \sqrt{\frac{\beta_2}{\beta_1}} \cos\mu & \sqrt{\beta_1\beta_2} \sin\mu \\ -\frac{1}{\sqrt{\beta_1\beta_2}} \sin\mu & \sqrt{\frac{\beta_1}{\beta_2}} \cos\mu \end{pmatrix}. \quad (106)$$

The following special cases of the form (106) are useful for designing chromaticity correcting sectors, such as the matched sectors BC and CD shown Figure 23:

$$\mathbf{M}(0) = \mathbf{M}(2\pi) = \begin{pmatrix} \sqrt{\frac{\beta_2}{\beta_1}} & 0 \\ 0 & \sqrt{\frac{\beta_1}{\beta_2}} \end{pmatrix} = \text{``}0\text{''},$$

$$\mathbf{M}(\pi/2) = \begin{pmatrix} 0 & \sqrt{\beta_1\beta_2} \\ -\frac{1}{\sqrt{\beta_1\beta_2}} & 0 \end{pmatrix} = \text{``}\pi/2\text{''},$$

$$\mathbf{M}(\pi) = \begin{pmatrix} -\sqrt{\frac{\beta_2}{\beta_1}} & 0 \\ 0 & -\sqrt{\frac{\beta_1}{\beta_2}} \end{pmatrix} = \text{``}\pi\text{''},$$

$$\mathbf{M}(3\pi/2) = \begin{pmatrix} 0 & -\sqrt{\beta_1\beta_2} \\ \frac{1}{\sqrt{\beta_1\beta_2}} & 0 \end{pmatrix} = \text{``}3\pi/2\text{''}. \quad (107)$$

As an example, consider sections represented by the matrix product

$$M_{BC} = Q(q3) * D(1) * Q(q2) * D(1) * Q(q1),$$
$$M_{CD} = Q(q5) * D(1) * Q(q4) * D(1) * Q(q3), \quad (108)$$

with matrix multiplication indicated by asterisks. Here $Q(q)$ is the 2x2 transfer matrix for a thin quad of strength $q$ and $D(1)$ is the 2x2 transfer matrix for a drift (with length taken to be 1 to simplify the formulas). These sections are to be combined to form the full chromatic adjustment sectors. The quadrupoles strengths can be adjusted so that

$$M^{BC} = \begin{pmatrix} 0 & \sqrt{\beta^B\beta^C} \\ -\frac{1}{\sqrt{\beta^B\beta^C}} & 0 \end{pmatrix},$$

$$M^{CD} = \begin{pmatrix} 0 & \sqrt{\beta^C\beta^D} \\ -\frac{1}{\sqrt{\beta^C\beta^D}} & 0 \end{pmatrix}, \quad (109)$$

in one or the other of the horizontal and vertical planes. Tuning curves for the quadrupole strengths accomplishing this as a function of the product $\sqrt{\beta^C\beta^D}$ are shown in Figure 22. Because the drift lengths in the line have been taken to be 1, the beta functions are in units of the drift length and the quad strengths are in units of 1/drift_length. One sees that the beta function $\beta^C$ at the sextupole locations can be varied over a large range by controlling the quadrupole strengths and the matched beta functions $\beta^B$ and $\beta^D$ at the start and finish. The concatenated matrix product is

$$M^{CD}M^{BC} = \begin{pmatrix} -\sqrt{\beta^D/\beta^B} & 0 \\ 0 & -\sqrt{\beta^B/\beta^D} \end{pmatrix}. \quad (110)$$

To obtain the desired relation $M^{CD}M^{BC} = -I$ the quadrupole strengths have to be symmetric about the center and $\beta^B = \beta^D$.

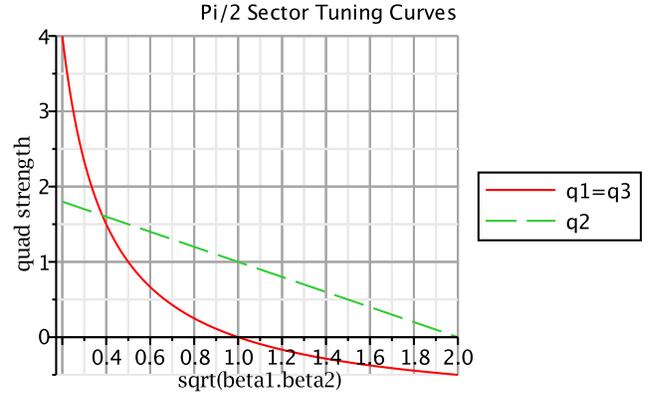

Figure 22: Tuning curves for adjusting $\beta$-functions at sextupole locations. Beta functions are in units of drift length and quad strengths are in units of 1/drift_length.

Ideally one would have both $M_x^{CD}M_x^{BC} = -I$ and $M_y^{CD}M_y^{BC} = -I$. Unfortunately no such solution exists. The best one can do is to satisfy the more important of these two conditions—the $y$-relation for the $\chi_y$ compensation module and the $x$-relation for the $\chi_x$ compensation module.

The impossibility of simultaneously satisfying the $-I$ condition in both planes has only been demonstrated for thin quadrupoles. But it would be very curious for such a condition to be possible for thick element but not for thin elements. Erratic behavior frequently reported for MAD Higgs factory simulations can be due to striving for an unachievable result.

What can be done, and has been done in the design, is to arrange for the sextupole locations to approximately coincide with minimum beta locations of the "insensitive coordinate"; for example there are minima of $\beta_x$ near locations C and E, where the y-chromaticity is being compensated. Since these points are not "$\pi$-separated" in $x$, their nonlinear horizontal kicks do not cancel but, because $\beta_x$ is so small at these locations, their nonlinear horizontal effect is expected to be small.

All the waist-to-waist sections are broken out, with parameters defined, in the following series of figures. The main "sacred" parameters, that will not be allowed to change (after having been negotiated between detector and accelerator groups) are $L^* = 4$ m, the free length in the IR, and $lIR = 20$ m, the distance between same-sign quadrupoles throughout the IR region. Of course even these dimensions can change in the early design days, but they need to be frozen fairly early in the design process.

## 11.3 Tunability of the CepC Intersection Region

The preceeding discussion of the CepC intersection region optics suggests that there is some tunability of the IR optics, even with all the labeled waist locations held in fixed locations, and some ability to alter the heights of the high beta peaks needed for chromaticity compensation. Minor changes to increase the flexibility will be suggested below. It is unnecessary to discuss the bend elements initially since, to a good approximation, they do not affect the linear optics. However they are critical for the chromaticity compensation. Also they represent serious sources of synchrotron radius from which the detectors will need to be shielded.

It is often (even usually) convenient in a lattice model to represent a ring quadrupole by two thin quadrupoles, so that the waist location occurs midway between them, even when there is just one physical quadrupole.

To assure symmetry of the y-chromatic adjustment centered at D, $\beta_y$ has to have the same values at waist locations B, D, and F. Then, for the same reason, the $\beta_x$ values at F, H, and J must also be the same, though not necessarily with the same values as $\beta_y$ at B, D, and F. One eventually has to check that the bend elements are satisfactorily positioned to be able to flexibly provide the required dependence of the dispersion $D(s)$.

These assumptions heavily constrain the IR section optics, but there is still considerable freedom. The height of the first peak at C of the y-chromaticity module can be adjusted using quad strengths qBC1=q1, qBC2=q2, and qBC3=q1, as governed by the tuning curves of Figure 22 including qBC3=qBC1. By adjusting the beta functions at B the heights of the $\beta$ peaks (especially $\beta_y^C$ which is large) can be adjusted. The relations qCD1=q1, qCD2=q2, and qCD3=q1 are then imposed by symmetry. This establishes the $\chi_y$ compensation centered at D. Quad strengths for the peak at E can be copied: qDE1=q1, qDE2=q2, qDE3=q1, qEF1=q1, qEF2=q2, qEF3=q1.

The x-chromatic adjustment peaks follow the same pattern. The same values of quad strength magnitudes, but with reversed signs, qFG1=qFG3=-q1 and qFG2=-q2, establish $\pi$-separated $\beta_x$ peaks in the $\chi_x$-compensating module. Even so, the equal $\beta_x$ peak heights at D and I are not necessarily equal to the earlier $\beta_y$ peaks.

There is still freedom in adjusting section AB, which can usefully be thought of a forming a complex triplet with the mirror-symmetric section on the other side of the IP and with the beam-beam focusing force forming the central element. Since the beam-beam strength is variable, and can be strong, one can try to keep this section matched independent of beam current. Then all the other quads can track, to maximize the dynamic aperture at every beam current.

Part of the lore of beam-beam interaction effects is that the beam-beam tune shift parameters $\xi_x$ and (especially) $\xi_y$ are more usefully thought of as measures of the nonlinearity of the beam-beam interaction than as their technical accurate definitions as quantitative shifts of the betatron tunes. Nevertheless, it still makes sense to incorporate the linear part of this focusing into the linear lattice model. Though the beam-beam focusing is far from being linear, the leading nonlinearity is octupole. This leads to amplitude-dependent detuning which is not automatically undesirable, especially in conjunction with the extremely strong betatron damping at a Higgs factory.

The strong damping at a Higgs factory also motivates looking into the possibility of avoiding the so-called "beta-squeeze" or beam separation cycling between injection and data collection operational states. One should strive to determine optics that can evolve adiabatically as the beam currents are gradually increased during initial injection. With topping-off injection the beam-beam tune shifts need not change by more than a percent or so between injection cycles. Except for this change, the beam-beam force is just a linear element, focusing in both planes.

One can further strive to design turn-by-turn, bumper-free, kicker-free, injection. If this injection process can be designed to be continuous and consistent with simultaneous data collection, then there will be no need for frequent cycling between injection and data collection. This holds out the possible goal of day-long, or longer, steady-state data collection runs.

## 11.4 Modular Chromatic Adjuster Design

**Unique Parameter Determinations.**

*The y-Chromaticity Correction Module*  Because the IP beta functions satisfy $\beta_y^* < \beta_x^*$ the vertical chromaticity due to the IP is greater than the horizontal. This makes it appropriate to concentrate first on compensating the vertical chromaticity and to place its compensating module as close to the IP as possible. The transfer matrix through an arbitrary sector has the partitioned form

$$\mathbf{M}^{BC} = \begin{pmatrix} \mathbf{M}_x^{BC} & 0 \\ 0 & \mathbf{M}_y^{BC} \end{pmatrix}. \quad (111)$$

To simplify notation the initial and final lattice locations are labeled B and C, but the same constraints will be applied to

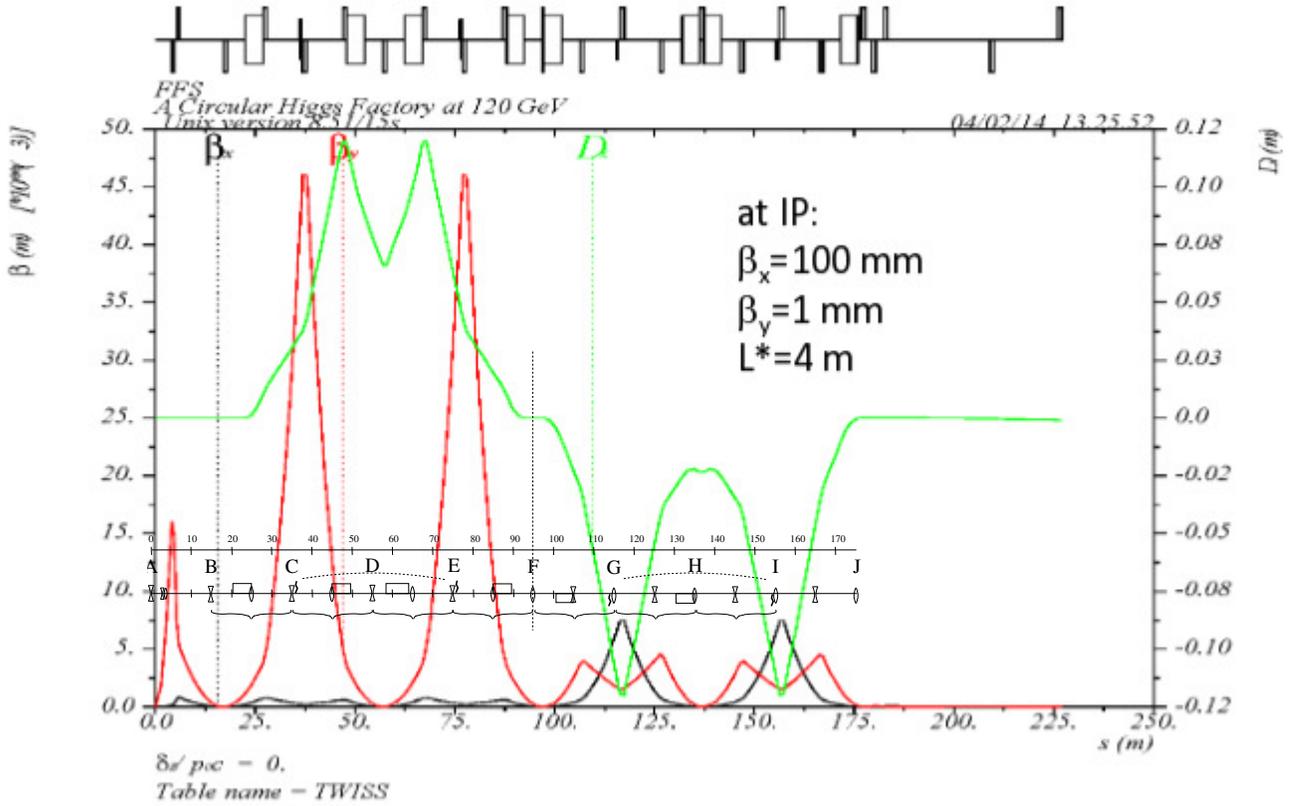

Figure 23: Yunhai Cai IR design, with significant waist locations identified as A, B, C, ….

all subsequent sections. It will be shown that the following relations unambiguously determine the entire *matched y-chromaticity correcting module*:

for $y$, $M_{y11} = 0$, $M_{y22} = 0$, $M_{y12} = \sqrt{\beta_y^B \beta_y^C}$,

and for $x$, $M_{x12} = 0$. (112)

Notice that these equations are consistent with $y$ having waists at both B and C. But they do not allow $x$ to have waists at both positions. These constraints leave considerable flexibility to choose necessary adjustment parameters. From the $y$ constraints (along with the unit determinant requirement) one sees that the precise "$\pi/2$" pattern of Eq. (107) will have been achieved for $\mathbf{M}_y$. From the $x$ constraint one sees that part of the "$\pi$" pattern will have been achieved, $M_{x12} = 0$; but the $M_{x21} = 0$ condition will not have been met. How to overcome this defect will be discussed later. With these constraints the full transfer matrix has the form

$$\mathbf{M}^{BC} = \begin{pmatrix} -1 & 0 & 0 & 0 \\ \frac{\alpha_x^C - \alpha_x^B}{\beta_x^B} & -1 & 0 & 0 \\ 0 & 0 & 0 & \sqrt{\beta_y^B \beta_y^C} \\ 0 & 0 & -\frac{1}{\sqrt{\beta_y^B \beta_y^C}} & 0 \end{pmatrix}. \quad (113)$$

In this form $\beta_x^B$ is free, and $\beta_x^C$ has the same value. Also $\beta_y^B$ is free, but the product $\beta_y^B \beta_y^C$ is fixed. To make $\beta_y^C$ large one will make $\beta_y^B$ small. The $y$ phase advance is exactly $\pi/2$ as required for nonlinearity cancellation, but, as mentioned earlier, the $x$ propagation is not quite matched—i.e. there is no $x$-waist at C even if there is an $x$-waist at B.

*Exploiting Quadrupole Polarity Reversal.* It was stated above that the solution to Eqs. (112) is unique, and that is correct, but these equations have a natural symmetry that guarantees the existence of a solution to a related set of equations. This symmetry is associated with reversing the polarities of all the quadrupoles. This interchange can be associated with interchanging the $x$ and the $y$ constraint equations. This means that the design of the $x$-chromaticity module can be (in fact has to be) identical to that of the $y$-module. In both cases the height of the beta peaks at sextupole locations are controlled by the beta functions at the cell ends. To obtain high $\beta_y$ peaks the input $\beta_y$ value is low. This choice also affects $\beta_x$ at the same locations, but the values of $\beta_x$ are low (and hence negligible) at those locations.

Table 15: Twiss function values eye-balled from the Yunhai Cai Higgs factory final focus optics shown above for $\beta_y^* = 1$ mm at the IP (interaction point). Columns on the right give target values for a more conservative $\beta_y^* = 10$ mm design that would use the same or almost the same element locations.

| waist label | waist (or other) identifier | location m | Yunhai Cai $\beta_x$ | $\beta_y$ | IR $\eta$ | design $\beta_x$ | targets $\beta_y$ |
|---|---|---|---|---|---|---|---|
| A | collision point | 0 | 50 mm | 1 mm | 0 | 1 m | 10 mm |
|   | first quad | $L^*/2 + lQ/2$ |  | 16 km | 0 |  | < 2 km |
|   | second quad | +1.5 | 0.6 km |  | 0 |  |  |
| B | begin y chrom-comp. | 15 |  |  | 0 |  |  |
| C | sextupole twin SIR1-1 | 35 |  | 46 km | 35 mm |  | < 2 km |
| D | center of y chr-comp. | 55 |  |  | 55 mm |  |  |
| E | sextupole twin SIR1-2 | 75 |  | 46 km | 35 mm |  | < 2 km |
| F | symmetry reversal | 95 |  |  | 0 |  |  |
| G | sextupole twin SIR2-1 | 115 | 7.5 km | 1.5 km | -110 mm | < 1 km |  |
| H | center of x chr-comp. | 135 | 148 mm | 18 mm | -40 mm |  |  |
| I | sextupole twin SIR2-2 | 155 | 7.5 km | 1.5 km | -110 mm | < 1 km |  |
| J | begin DS&optics match to arc | 175 |  |  | 0 |  |  |

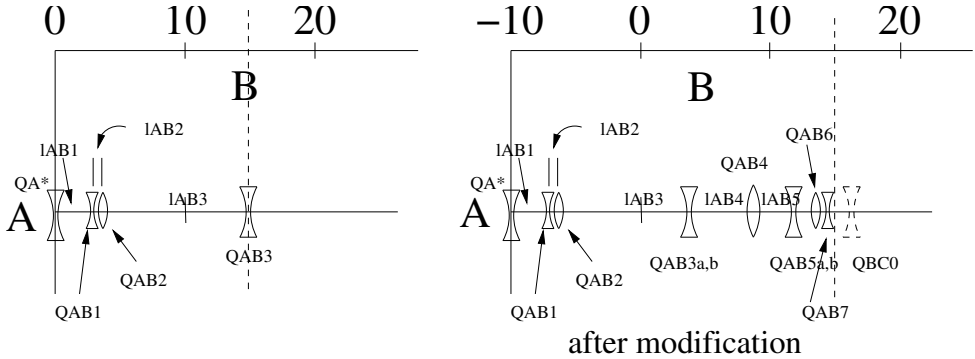

Figure 24: The AB "demagnifying" section adjacent to the IP, shown before and after extensive redesign, lengthening the IR region by 10 meters. The beam-beam focusing element QA$^*$ represents half of the beam-beam focusing, with the other half provided by the mirrored section on the other side of the IP. The linear focusing part can be designed into the basic lattice design, even adapting the full ring optics as the strength of the beam-beam force varies with beam currents. Though shown as horizontally defocusing the beam-beam force actually *focuses* in both planes.

One can now understand the function of the symmetry reversal (i.e. quadrupole polarity reversal) starting at location F. For full self-consistency the quadrupole at F has to be turned off to preserve this symmetry. That is $qFG1 = -qEF3$. As one consequence of symmetry, the beta function pattern will be the same in the $x$-module as in the $y$-module, with the exception that horizontal and vertical beta functions are reversed. In particular the $\beta_y$ peaks can be large in the $y$-module and the $\beta_x$ peaks can be large in the $x$-module. Conservation of both beta function values at (every second waist) locations B, D, F, H, and J, makes it possible to tune the heights differently in the $y$ and $x$ modules by adjusting these *conserved* $\beta_x$ and $\beta_y$ values. This degree of tunability makes it practical to use the same lattice layout over a large range of chromaticity adjustment, even including large differences between $\chi_x$ and $\chi_y$.

**Resulting Lattice Parameters.** Referring to the (completely general) form in Eq. (102), one sees that the following phase-advance-dependent determinations have been made:

$$S_x^{BC} = 0, \quad C_x^{BC} = -1, \quad S_y^{BC} = 1, \quad C_y^{BC} = 0. \quad (114)$$

Also determined have been

$$\beta_y^B = \beta_y^C, \quad \alpha_y^B = \alpha_y^C = 0, \quad \text{and}$$
$$\beta_x^B = \beta_x^C, \quad \text{but} \quad \alpha_x^B \neq \alpha_x^C. \quad (115)$$

The initial beta functions $\beta_x^B$ and $\beta_y^B$ remain free.

The numerical values of the elements in $\mathbf{M}^{BC}$ depend on a master IR length scale, which is being held frozen here, and on the quadrupole strengths $qBC1 = qBC3 = -3/40 = -0.075\,/\text{m}$ and $qBC2 = 1/5 = 0.2\,/\text{m}$, as determined by the conditions (112). The resulting transfer matrix is

$$\mathbf{M}^{BC} = \begin{pmatrix} -1 & 0 & 0 & 0 \\ -7/20 & -1 & 0 & 0 \\ 0 & 0 & 0 & 40 \\ 0 & 0 & 1/40 & 0 \end{pmatrix}. \quad (116)$$

The reason these elements are so thoroughly numerical, is that the basic (half-cell) length $lIRh$ has been specified to be 10 m. The entries are not approximate, they are exact. For different values of $lIRh$ the entries can simply be scaled on the basis of dimensional analysis. A numerical consequence of the $y$ entries is that, with $\beta_y^B = 1$ m, the height of the $\beta_y^C$ peak is given by $\beta_y^C = 40^2/\beta_y^B \stackrel{\text{e.g.}}{=} 1600$ m,

**Concatenating Successive Lattice Sections.** The transfer matrices from B to D are given by

$$\mathbf{M}_x^{BD} = \begin{pmatrix} -1 & 0 \\ \frac{\Delta \alpha_x^{BC}}{\beta_x^B} & -1 \end{pmatrix} \begin{pmatrix} -1 & 0 \\ \frac{\Delta \alpha_x^{BC}}{\beta_x^B} & -1 \end{pmatrix} = \begin{pmatrix} -1 & 0 \\ -\frac{2\Delta \alpha_x^{BC}}{\beta_x^B} & -1 \end{pmatrix},$$

$$\mathbf{M}_y^{BD} = \begin{pmatrix} 0 & \sqrt{\beta_y^B \beta_y^C} \\ -\frac{1}{\sqrt{\beta_y^B \beta_y^C}} & 0 \end{pmatrix} \begin{pmatrix} 0 & \sqrt{\beta_y^B \beta_y^C} \\ -\frac{1}{\sqrt{\beta_y^B \beta_y^C}} & 0 \end{pmatrix}$$

$$= \begin{pmatrix} -1 & 0 \\ 0 & -1 \end{pmatrix}. \quad (117)$$

Defining $\Delta_x^{BC} = \alpha_x^C - \alpha_x^B$, the transfer matrices from B to F are

$$\mathbf{M}_x^{BF} = \begin{pmatrix} 1 & 0 \\ \frac{4\Delta \alpha_x^{BC}}{\beta_x^B} & 1 \end{pmatrix}, \quad \mathbf{M}_y^{BF} = \begin{pmatrix} 1 & 0 \\ 0 & 1 \end{pmatrix}. \quad (118)$$

It can be seen that the magnitude of the matrix element $M_{x21}$ "defect" accumulates by the same amount each section. Location F marks the end of the $y$-chromaticity correction module and the beginning of the $x$-chromaticity correction module. To make F be a true waist one can refer back to the starting point at B and impose, as an initial condition,

$$\alpha_x^{B-} = 4\Delta \alpha_x^{BC}. \quad (119)$$

**Checking the Twiss Parameter Evolution.** One can use evolution formulas (129) (which simplify markedly) to perform various checks of the results just obtained:

$$\beta_y^C \stackrel{?}{=} \frac{M_y^{BC}}{\beta_y^B} = \frac{\beta_y^B \beta_y^C}{\beta_y^B} = \beta_y^C, \quad \checkmark \quad (120)$$

$$0 \stackrel{?}{=} \alpha_y^C = M_{y,12}^{BC} M_{y,21}^{BC} \alpha_y^B = 0, \quad \checkmark \quad (121)$$

$$\beta_x^C \stackrel{?}{=} M_{x,11}^{BC} M_{x,22}^{BC} \beta_x^B = \beta_x^B, \quad \checkmark \quad (122)$$

$$\alpha_x^F \stackrel{?}{=} -M_{x,11}^{BF} M_{x,21}^{BF} \beta_x^B + M_{x,11}^{BF} M_{x,22}^{BF} \alpha_x^{B-}$$
$$= -4\frac{\Delta \alpha^{BC}}{\beta_x^B} \beta_x^B + 4\Delta \alpha^{BC} = 0. \quad \checkmark \quad (123)$$

The $x$-chromaticity module from F to J can just be copied from the module from B to F just derived. With the modules joining at the common waist at F, the $x$-correction module simply inherits its input $\beta$ functions from the $y$-correction module.

**Suggested Lattice Alterations.** Figure 25 is almost the same as earlier figures, but suggested modifications are included, primarily by splitting and slightly separating the previously-superimposed quadrupoles at all of the labeled lattice locations, B, C, D, E, F, G, H, I, and J. There are different reasons for doing this in each case and some of the reasons are more important than others.

The most important quadrupole splits are at sextupole locations C, E, G, and I. By placing a sextupole at the exact center of its two matched quadrupoles one makes both its chromatic compensation and its betatron phase location relatively insensitive to lattice errors. Without doubling such a quadrupole, its sextupole sits where the beta function depends strongly on position, making it hypersensitive to lattice imperfections. The effects of such errors are "amplified" by unballencing the sextupole pair cancellation.

Since the quadrupole at F has to be "turned off" to comply with the symmetry reversal, this quadrupole could simply be removed. Better, however, is to split it and separate it to form a (weak) doublet. To the extent both are turned off, this corresponds to no change whatsoever. But, when turned on, such a doublet can be used to alter the $\beta_x$ and $\beta_y$ values in the separate $x$ and $y$ modules. This can be used to produce an independent knob for tuning the $\chi_x$ module relative to the $\chi_y$ module.

Doubling the quadrupoles at B and J is also likely to prove useful. The main defect of the overall design is the non-vanishing $M_{21}$ element in Eq. (113). It will be seen in the next section how this defect is to be handled. This places extra demands on tuning the transitions into and out of the chromatic correction modules. The (newly-available) doublet at F can be involved in this same adjustment.

There may be no great advantage to splitting the quadrupoles as locations D and H, at the exact centers of the $\chi_x$ and $\chi_y$ modules. However there is an important formal advantage to treating these two locations the same as all the others. It is that (until the split quadrupoles are tuned to no longer exact match) the overall transfer matrices can be obtained by concatenating identical transfer matrices. As it happens it is only the quadrupoles with odd indices (e.g. QBC1 and QBC3) that are split. Quadrupoles with even indices (e.g. QBC2) are not split. The individual transfer matrices then depend on only two independent quadrupole strengths, $qBC1 = q1$ and $qBC2 = q2$. As a consequence the overall transfer matrices have the same property.

There is yet another advantage to splitting the quadrupoles as I am proposing. By splitting the quads at the maximum beta locations, quadrupoles what will necessarily be implemented as thick quads in the physical lattice will be being represented by two, slightly-separated, thin quads in the thin quad model. Being at beta function maxima, the overall optics is more dependent on the quality of these quadrupoles than on any of the others. Representing these quads as two thin quads is the first step in the eventual further splitting of these quads (as required to preserve symplecticity in the thick element model) into the enough zero-length quads for faithful representation. As a result the thin element model being described here will more nearly resemble the more nearly faithful thick element model.

This last argument is sufficiently persuasive that I will implement a CepC thin quadrupole model with the A,B,…,J quadrupoles split as I suggest, whether or not they are really split in the eventual physical model. Such a model is certain to be symplectic, and is likely to be have properties quite close to other models of the expected performance (such as dynamic aperture and injection efficiency).

### 11.5 Transfer Matrices for the Modified Lattice

The effects of splitting the quads as suggested have been investigated numerically by repeating the calculations of Sections 11.4 and 11.4. The quad separation distances are always $lq = 1.4$ m, which is approximately the thickness of the quads in the the Yunhai Cai lattice. Eqs. (114) and (115) remain applicable. The matched quadrupole strengths are qBC1=-0.0791785, qBC2=0.200738, qBC3, -0.0791785, and $\sqrt{\beta_y^B \beta_y^C}$=35.9754. Corresponding to Eq. (116) the resulting BC transfer matrix is

$$\mathbf{M}^{BC} = \begin{pmatrix} -1 & 0 & 0 & 0 \\ -0.33025 & -1 & 0 & 0 \\ 0 & 0 & 0 & 35.97544 \\ 0 & 0 & -0.027797 & 0 \end{pmatrix}. \tag{124}$$

The transfer matrices from B to D are

$$\mathbf{M}_x^{BD} = \begin{pmatrix} -1 & 0 \\ 0.66049 & -1 \end{pmatrix},$$

$$\mathbf{M}_y^{BD} = \begin{pmatrix} -1 & 0 \\ 0 & -1 \end{pmatrix}. \tag{125}$$

The transfer matrices from B to F are

$$\mathbf{M}_x^{BF} = \begin{pmatrix} 1 & 0 \\ 1.32100 & 1 \end{pmatrix}, \quad \mathbf{M}_y^{BF} = \begin{pmatrix} 1 & 0 \\ 0 & 1 \end{pmatrix}. \tag{126}$$

All these results are consistent with expectations. As before the only (and unavoidable) blemish is the accumulating $M_{x12}$ elements. The only significant numerical changes (compared to the pre-modification results) is in the $M_{y12}$ element, which has been reduced from 40/m to 35.99/m. As a result the maximum $\beta_y$ value is, for example, $35.99^2$=1295 m compared to the corresponding earlier value of 1600 m (with the $\beta_y^C$ value taken to be 1 m in both cases.) This reduction is presumably the result of moving the end quadrupoles in slightly and representing the quads at C and E more faithfully by two separated zero-length quads than by single zero-length quads. This change has the cosmetic benefit of eliminating unsightly beta function kinks at sextupole locations D, E, G, and I and providing the sextupoles comfortable nests (as emphasized above).

### 11.6 Preliminary IR Lattice Design

The following figures illustrate a tentative IR design based on the principles described above.

### 11.7 Limitation in Matching Achromatic IR Sections to Achromatic Arcs

Suppose that the arc sextupoles have been tuned to make the ring achromatic in the $Q'_x = Q'_y = 0$ sense. To the extent this includes compensating for IR chromaticity this means that the arcs themselves are "over-compensated", so their periodic Twiss functions depend on $\delta$. In particular their "period matched" end values, in both planes, are $\alpha_2(\delta)$ and $\beta_2(\delta)$. It is this limitation that forces the IR chromaticity to be compensated locally. But, to the extent this compensation is only approximate, the same considerations will be applicable.

Fortunately the compensation of excess chromaticity will have only minor effect on the arc optics when spread uniformly over the entire arcs. It is the residual chromatic dependence of the IR sections that really matters (even if this chromaticity has been greatly reduced in the IR design). After some iterations to optimize the overall design, assume that the derivatives with respect to $\delta$,

$$\beta'_{x2}, \beta'_{y2}, \alpha'_{x2}, \alpha'_{y2}, \tag{127}$$

at the ends of the arcs are known. Perfect IR design would exactly compensate these dependencies to make the IP values $\alpha_1(\delta)$ and $\beta_1(\delta)$ independent of $\delta$ at the ends of the IR sectors. We expect this to be impossible, but it might be thought unnecessary to be quite so fussy. We could afford to let $\beta_{x1}(\delta)$ and $\beta_{y1}(\delta)$ continue to depend on $\delta$, but with matching good enough to preserve about the same dynamic ranges as on-momentum. Even if the beta values vary substantially over the momenta present in the beam, the resulting beam distortion at the IP could be acceptable, and might cause only a minor loss of luminosity. However, closure of the lattice off-momentum requires $\alpha_{x1}(\delta)$ and $\alpha_{y1}(\delta)$ to be independent of $\delta$; that is

$$\alpha'_{x1}(\delta) = \alpha'_{y1}(\delta) = 0 . \tag{128}$$

If these relations are satisfied the lattice might be expected to stay approximately matched over a substantial range of $\delta$ even if the $\beta$'s vary with $\delta$. This is a luxury which is made possible for the designers of linear colliders by the fact that, after collision, the unscattered particles do not need to remain captured. Unfortunately this "freedom" will now be shown to be unavailable in a storage ring.

The formulas by which Twiss functions evolve from lattice position 1 to 2 are

$$\beta_2 = M_{11}^2 \beta_1 - 2M_{11}M_{12}\alpha_1 + M_{12}^2 \frac{1+\alpha_1^2}{\beta_1},$$

$$\alpha_2 = -M_{11}M_{21}\beta_1 + (M_{11}M_{22} + M_{12}M_{21})\alpha_1$$
$$- M_{12}M_{22}\frac{1+\alpha_1^2}{\beta_1}. \quad (129)$$

To proceed backwards through a sector such as the demagnifying section we require the inverse matrix,

$$\mathbf{M}^{-1} = \begin{pmatrix} M_{22} & -M_{12} & 0 & 0 \\ -M_{21} & M_{11} & 0 & 0 \\ 0 & 0 & M_{44} & -M_{34} \\ 0 & 0 & -M_{43} & M_{33} \end{pmatrix}. \quad (130)$$

The elements of this matrix can then be substituted into the inverses of Eq. (129);

$$\beta_1 = M_{22}^2 \beta_2 + 2M_{22}M_{12}(-\alpha_2) + M_{12}^2 \frac{1+\alpha_2^2}{\beta_2},$$

$$\alpha_1 = M_{22}M_{21}\beta_2 + (M_{22}M_{11} + M_{12}M_{21})(-\alpha_2)$$
$$+ M_{12}M_{11}\frac{1+\alpha_2^2}{\beta_2}. \quad (131)$$

In this step the signs of $\alpha_2$ have been reversed since the evolution is back through the demagnifying section. A useful special case of Eq. (131) applies when $\alpha_2$ is known to vanish, in which case

$$\beta_1 = M_{22}^2 \beta_2 + M_{12}^2 \frac{1}{\beta_2},$$

$$\alpha_1 = M_{22}M_{21}\beta_2 + M_{12}M_{11}\frac{1}{\beta_2}. \quad (132)$$

On-momentum ($\delta = 0$) and Eqs. (131) are presumably already satisfied (in both planes) because the lattice is assumed to be already matched. It is the first order momentum dependence we are interested in. Furthermore, as explained above, we are only insisting on the $\alpha$ matches. In particular, we demand that conditions Eq. (128) be satisfied, with $\alpha$s substituted from Eq. (131). Since this is two fewer conditions than full achromaticity requires, we can remain hopeful that a bend-free IR could be designed to satisfies them.

All quantities appearing on the left hand sides of Eq. (128) are known. The matrix elements $M_{ij}$ are all known as polynomials in the $q_i$, the $L_i$ and $\delta$. The on-momentum $\alpha_2$'s and $\beta_2$'s are known from the matched lattice design and their slopes are known, according to Eq. (127). The operative word "known" is being used loosely here since, as mentioned already, a certain amount of iteration will be required. The Twiss parameters and their first order momentum derivatives will be known from whatever lattice fitting software is being used.

If the IR section were being matched to general arcs, according to Eqs. (128), formulas for the $\alpha$-functions would also be required. But since we are fitting to a pure FODO arc (neglecting the perturbing influence of the far straight) the arc $\alpha$-functions vanish at the IR boundaries. Conditions Eq. (128) then reduce to

$$\frac{d}{d\delta}\left(\tilde{M}_{22}\tilde{M}_{21}\tilde{\beta}_{x2} + \tilde{M}_{12}\tilde{M}_{11}\frac{1}{\tilde{\beta}_{x2}}\right) = 0,$$

$$\frac{d}{d\delta}\left(\tilde{M}_{44}\tilde{M}_{43}\tilde{\beta}_{y2} + \tilde{M}_{34}\tilde{M}_{33}\frac{1}{\tilde{\beta}_{y2}}\right) = 0. \quad (133)$$

The tildes on the $\tilde{M}_{ij}$, $\tilde{\beta}_{x2}$ and $\tilde{\beta}_{y2}$ indicate that all quadrupole strength parameters $q_i$ have been replaced by $q_i/(1+\delta) \approx q_i(1-\delta)$ in the formulas expressing the matrix elements in terms of the quadrupole strengths and drift lengths; sextupoles do not contribute in leading order. For a match to the regular arc, valid to linear order in $\delta$, the factors $\tilde{\beta}_{x2}$ and $\tilde{\beta}_{y2}$ have to agree with the values in Eq. (127). (Because of the weak chromaticity of the arcs, just treating $\tilde{\beta}_{x2}$ and $\tilde{\beta}_{y2}$ as independent of $\delta$ may be adequate.)

One can note that the validity of Eqs. (133) would imply that Eqs. (105) are satisfied independent of momentum (to leading order.) Also, if points 1 and 2 are reversed in the above argument, one obtains an equation equivalent to the first of Eqs. (105) (in lowest order.) So requiring Eq. (133) amounts to requiring that Eqs. (104) hold not only for $\delta = 0$ but also to linear order in $\delta$. This means that demanding a momentum-independent $\alpha$ match implies also a momentum-independent $\beta$ match. As suggested above, such a match is probably impossible.

It is considerations like these that make it obligatory to compensate the chromaticity locally within the IR regions to make the compensation as local as possible. At a minimum this has required the introduction of bending elements and sextupoles into IR lattice designs, for example to satisfy conditions Eqs. (133). Since the simplest form of achromat uses identical sextupoles separated by phase advance of $\pi$, it is appropriate to use two sections, each with phase advance $\pi/2$.

### 11.8 Arc Chromaticity and its Compensation

*This section will be of little value until it contains a discussion of sextupole families. It is present primarily to evaluate derivatives needed in Eq. (133).*

For simplicity we assume thin lenses everywhere, even though, ultimately, thick lens formulas have to be applied, especially to the quadrupoles adjacent to the IP. Since the sextupole strengths $S_1$ and $S_2$ are determined only implicitly, they have to be determined to adjust the overall chromaticities to zero (or whatever nearby values are called for.) Let us assume that each FODO cell starts and ends with a vertically focusing half quad of strength $q_1$ (which is negative, meaning horizontally defocusing), the middle quad strength is $q_2$ (which is positive), and the half-cell lengths are $\ell$. The horizontal transfer matrix through the first half-cell is

$$\begin{pmatrix} 1 & 0 \\ -q_2 & 1 \end{pmatrix}\begin{pmatrix} 1 & \ell \\ 0 & 1 \end{pmatrix}\begin{pmatrix} 1 & 0 \\ -q_1 & 1 \end{pmatrix}$$
$$= \begin{pmatrix} 1 - q_1\ell & \ell \\ -q_1 - q_2 + q_1 q_2 \ell & 1 - q_2\ell \end{pmatrix}. \quad (134)$$

Momentum dependence can be built into this formula by the replacements $q_i \to q_i/(1+\delta)$. Sextupoles can also be incorporated if they are superimposed on the quadrupoles. Decomposing the horizontal displacement as $x = x_\beta + \eta_x \delta$, the horizontal angular deflection caused by a quadrupole of strength $q_i$ with a sextupole of strength $S_i$ superimposed, is $S_i x^2/2$, or

$$\Delta x' = \left(\frac{q_i}{1+\delta} + S_i \eta_x \delta\right) x + \text{terms to be dropped}$$
$$\approx (q_i + (S_i \eta_x - q_i)\delta) x. \quad (135)$$

The "terms to be dropped" require special discussion. Dropping the term proportional to $\delta^2$ is probably always valid. The term $S_i x^2$, being nonlinear, is always serious. By canceling pairs of $\pi$-separated kicks one can hope to validate dropping these terms. Let us therefore define the dimensionless parameters

$$\tilde{q}_1 = (q_1 + (S_1\eta_{x1} - q_1)\delta)\,\ell, \quad \tilde{q}_2 = (q_2 + (S_2\eta_{x2} - q_2)\delta)\,\ell, \quad (136)$$

which "wrap" or "hide" the functional dependencies on $\delta$, the $S$'s, $\eta_x$ and $\ell$. The horizontal transfer matrix through the full cell is given by

$$\mathbf{M}_x(\delta)$$
$$= \begin{pmatrix} 1 - 2\tilde{q}_1 - 2\tilde{q}_2 + 2\tilde{q}_1\tilde{q}_2 & 2(1-\tilde{q}_2)\ell \\ 2(-\tilde{q}_1 - \tilde{q}_2 + \tilde{q}_1\tilde{q}_2)(1-\tilde{q}_1)/\ell & 1 - 2\tilde{q}_1 - 2\tilde{q}_2 + 2\tilde{q}_1\tilde{q}_2 \end{pmatrix}$$
$$\equiv \begin{pmatrix} \cos\mu_x(\delta) & \beta_x(\delta)\sin\mu_x(p) \\ -\frac{\sin\mu_x(\delta)}{\beta_x(\delta)} & \cos\mu_x(\delta) \end{pmatrix} \quad (137)$$

Here $\mu_x(\delta = 0)$ is the on-momentum horizontal phase advance per cell. The $\beta$-functions, now valid to linear order in $\delta$ are obtained from $-M_{12}/M_{21}$,

$$\beta_{x1} = l\sqrt{\frac{1-\tilde{q}_2}{1-\tilde{q}_1}}\sqrt{\frac{1}{\tilde{q}_1 + \tilde{q}_2 - \tilde{q}_1\tilde{q}_2}},$$

$$\beta_{y1} = l\sqrt{\frac{1+\tilde{q}_2}{1+\tilde{q}_1}}\sqrt{\frac{1}{-\tilde{q}_1 - \tilde{q}_2 - \tilde{q}_1\tilde{q}_2}},$$

$$\beta_{x2} = l\sqrt{\frac{1-\tilde{q}_1}{1-\tilde{q}_2}}\sqrt{\frac{1}{\tilde{q}_1 + \tilde{q}_2 - \tilde{q}_1\tilde{q}_2}},$$

$$\beta_{y2} = l\sqrt{\frac{1+\tilde{q}_1}{1+\tilde{q}_2}}\sqrt{\frac{1}{-\tilde{q}_1 - \tilde{q}_2 - \tilde{q}_1\tilde{q}_2}}. \quad (138)$$

The phase advances are given by

$$\sin^2\mu_x(\frac{\delta)}{2}) = \tilde{q}_1 + \tilde{q}_2 - \tilde{q}_1\tilde{q}_2, \quad \sin^2\mu_y(\frac{\delta)}{2}) = -\tilde{q}_1 - \tilde{q}_2 - \tilde{q}_1\tilde{q}_2. \quad (139)$$

To use relations (136) it is necessary to have formulas for the $\eta_x$ functions;

$$\eta_{x1} = \frac{(1 - q_2\ell/2)l\Delta\theta}{(\sin\frac{\mu^{(x)}}{2})^2}; \quad \eta_{x2} = \frac{(1 - q_1\ell/2)l\Delta\theta}{(\sin\frac{\mu^{(x)}}{2})^2}. \quad (140)$$

For equal tunes these reduce to

$$\eta_{x1} = \frac{(1 + |q|\ell/2)l\Delta\theta}{|q|^2\ell^2}; \quad \eta_{x2} = \frac{(1 - |q|\ell/2)l\Delta\theta}{|q|^2\ell^2}. \quad (141)$$

Since these are already the coefficients of terms first order in $\delta$ it is not necessary to allow for their momentum dependence. This means their values can simply be copied from the output of a lattice program. In fact, since the same comments can be made about the sextupole strengths, it will only be necessary to know the products $S_1\eta_{x1}$ and $S_2\eta_{x2}$. The most rudimentary, most local, form of chromatic correction is to choose

$$S_1 = \frac{q_1}{\eta_{x1}}, \quad S_2 = \frac{q_2}{\eta_{x2}}, \quad (142)$$

though, of course, this compensates only the arc chromaticity. (In fact. replacement (142) does not even exactly cancel the arc chromaticities since it does not allow for differences of the beta functions at the sextupole locations.) Since $\eta_{xi}$ is (almost) always positive, $S_i$ will normally have the same sign as $q_i$. The following derivatives enter Eqs. (133):

$$\frac{d\tilde{q}_1}{d\delta} = (S_1\eta_{x1} - q_1)\,\ell, \quad \frac{d\tilde{q}_2}{d\delta} = (S_2\eta_{x2} - q_2)\,\ell. \quad (143)$$

In these formulas the coefficients $S_1\eta_{x1} - q_1$ and $S_2\eta_{x2} - q_2$ can be regarded as the excess due to compensating also the IR chromaticity.

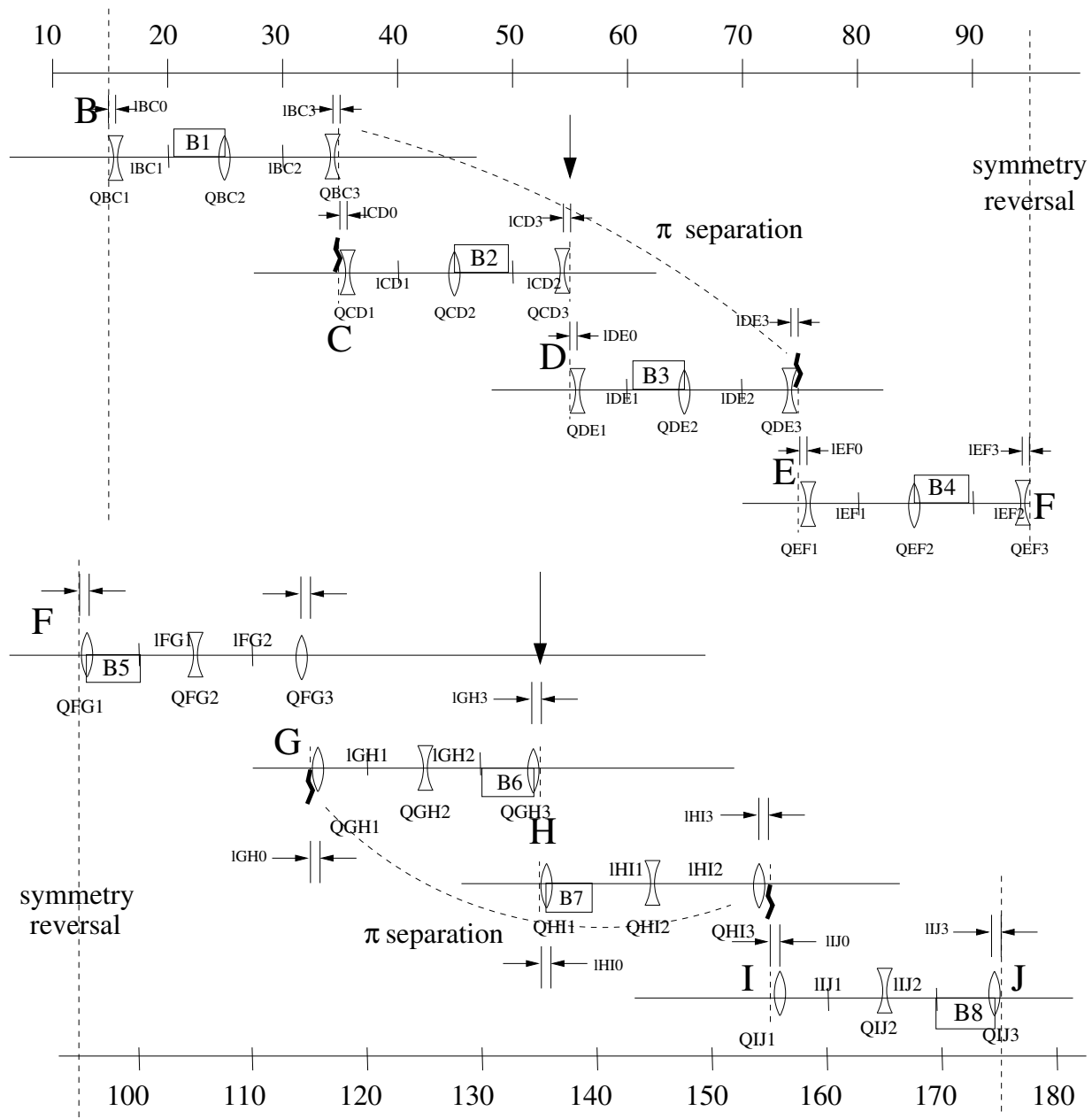

Figure 25: Suggested modifications to improve lattice tunabiity. The changes amount to cutting the quadrupoles in half at all labeled locations B through I and separating them symmetrically. This permits the compensating sextupoles to be optimally situated, and other advantages are given in the text.

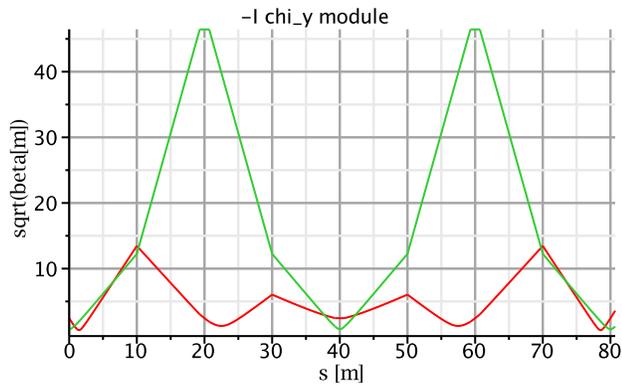

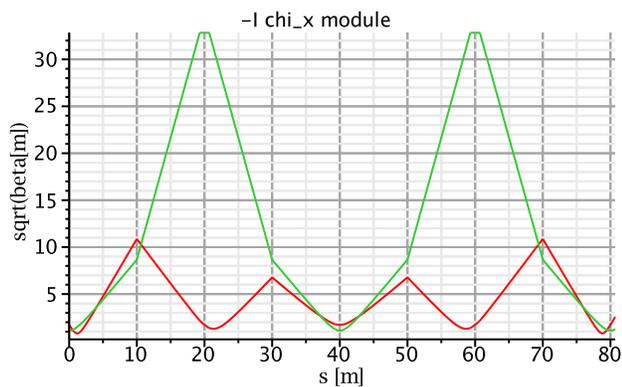

Figure 26: Chromaticity tuning modules, vertical above, horizontal below.

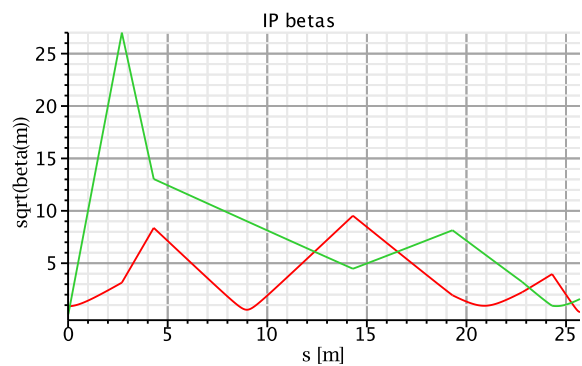

Figure 28: Tentative fit to full CepC beamline from the IP to the beginning of the chromaticity correcting sections. This section is probably overly complicated (to simplify the match to the chromatic module) and it has not been optimized using MAD or any other fitting program.

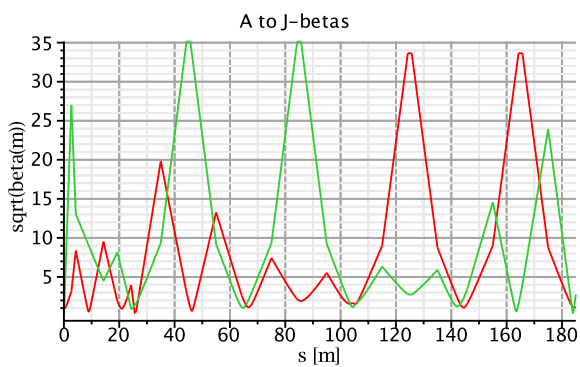

Figure 27: Beta function plot for matched fit to full CepC beamline, with $\beta_y^* = 10$ mm, from the IP to the end of the chromaticity correcting modules. Relative to the Yunhai IR, the demagnifying section is 10 m longer which shifts the remaining elements 10 m towards larger $s$. The variation of the "non-peaking" beta function through the chromatic modules cannot be eliminated without disrupting the required **-I** compensation condition. However the **-I** preservation is extremely robust.